%% file: main_v3.tex
\documentclass[a4paper, 11pt]{article}
\pdfoutput=1
\usepackage{jheppub}
\input{packages}

\input{definitions}
\newcommand{\update}[1]{#1}
\newcommand{\updatetwo}[1]{#1}
\title{Parton distributions in the shockwave formalism}

\author[a]{Shohini Bhattacharya,}
\author[b,e]{Chuan-Qi He,}
\author[b,c,d]{Zhong-Bo Kang,}
\author[b,c]{Diego Padilla,}
\author[b,c]{and Jani Penttala}

\affiliation[a]{Department of Physics, University of Connecticut, Storrs, CT 06269, USA}
\affiliation[b]{Department of Physics and Astronomy, University of California, Los Angeles, CA 90095, USA}
\affiliation[c]{Mani L. Bhaumik Institute for Theoretical Physics, University of California, Los Angeles, CA 90095, USA}
\affiliation[d]{Center for Frontiers in Nuclear Science, Stony Brook University, Stony Brook, NY 11794, USA}
\affiliation[e]{State Key Laboratory of Nuclear Physics and Technology, Institute of Quantum Matter, South China Normal University, Guangzhou 510006, China}

\emailAdd{shohinib@uconn.edu}
\emailAdd{legend\_he@m.scnu.edu.cn}
\emailAdd{zkang@physics.ucla.edu}
\emailAdd{dpadi022@g.ucla.edu}
\emailAdd{janipenttala@physics.ucla.edu}

\abstract{
In this work, we calculate a broad class of parton distributions---including parton distribution functions (PDFs), transverse-momentum-dependent distributions (TMDs), generalized parton distributions (GPDs), generalized transverse-momentum-dependent distributions (GTMDs), and diffractive parton distributions---directly from their operator-level definition in the shockwave approximation for the target nucleon.
This approximation is valid in the high-energy limit of scattering, corresponding to the small-$x$ regime.
The shockwave framework allows us to employ the eikonal approximation and express the parton distributions in terms of Wilson-line correlators, naturally formulated within the color-glass condensate effective field theory.
We present  a comprehensive set of Feynman rules for evaluating parton distributions in this limit, and demonstrate how they can be systematically applied to calculate all phenomenologically relevant leading-twist parton distributions at leading order.
This work establishes a unified starting point for future studies that aim to bridge the color-glass condensate approach with the partonic description of the nucleon. 

}

\begin{document}

\maketitle

\section{Introduction} 
\label{sec:intro}

A central frontier in hadronic physics is the multi-dimensional imaging of strongly interacting systems~\cite{Achenbach:2023pba}. The conventional tool for describing hadron structure has been the parton distribution function (PDF), which characterizes how quarks and gluons share the longitudinal momentum of the hadron. Over the past two decades, significant effort has been devoted to moving beyond this one-dimensional description toward richer, multi-dimensional frameworks that resolve the spatial and momentum distributions of partons inside nucleons; for a comprehensive recent review, see Ref.~\cite{Lorce:2025aqp}.

Two key generalizations of PDFs have emerged: generalized parton distributions (GPDs) and transverse-momentum dependent distributions (TMDs). GPDs extend PDFs by incorporating information on the transverse spatial distribution of partons, thereby enabling a genuine three-dimensional imaging of hadrons in mixed longitudinal momentum and transverse position space~\cite{Ji:1996ek,Radyushkin:1996nd,Burkardt:2000za,Diehl:2003ny}. Through Ji's sum rule, specific GPDs provide access to the total angular momentum carried by quarks and gluons~\cite{Ji:1996ek}, making them an important ingredient in addressing the nucleon spin puzzle~\cite{Ji:1996ek}. TMDs, on the other hand, resolve the transverse momentum of partons, providing information on spin–momentum correlations and effects arising from initial- and final-state interactions in semi-inclusive processes~\cite{Collins:2011zzd,Boussarie:2023izj,Collins:2002kn,Belitsky:2002sm,Ji:2004wu,Bomhof:2006dp}.  

At the most general level, both GPDs and TMDs can be obtained as limiting cases of the generalized transverse-momentum dependent distributions (GTMDs)~\cite{Meissner:2009ww,Lorce:2013pza}. GTMDs thus constitute the most complete characterization of partonic structure, earning them the designation of ``mother distributions". Their Fourier transforms correspond to Wigner functions~\cite{Ji:2003ak,Belitsky:2003nz}, the quantum-mechanical analogues of classical phase-space distributions, which encode simultaneously the longitudinal momentum, transverse momentum, and transverse position of partons. GTMDs, in contrast to GPDs, provide a more direct connection to the orbital angular momentum of partons inside nucleons~\cite{Lorce:2011kd,Hatta:2011ku,Ji:2012sj,Bhattacharya:2024sck,Bhattacharya:2023hbq,Bhattacharya:2022vvo,Bhattacharya:2017bvs,Bhattacharya:2018lgm,Kovchegov:2024wjs,Manley:2024pcl,Kovchegov:2023yzd}. This distinctive feature highlights their central role in addressing the nucleon spin problem while simultaneously revealing the complete phase-space structure of hadrons.

Although the operator definitions of these distributions are well established, carrying out practical calculations remains highly challenging (see, e.g., Ref.~\cite{Ji:2013dva}). Moreover, the conventional partonic picture becomes inadequate in the high-energy (small-$x$) regime of QCD, where gluon densities grow large and collective dynamics dominate. In this domain, an alternative yet complementary description emerges: the shockwave, or the color-glass condensate (CGC), formalism. Here the hadron is no longer represented by individual partons but by dense gluon fields, with scattering processes encoded in Wilson-line correlators~\cite{Iancu:2003xm}. Perturbative calculations in this framework follow the shockwave approximation, in which the Lorentz-boosted target interacts with the projectile instantaneously in light-cone time. The scattering off the target factorizes from the rest of the process, and the corresponding dynamics are described by Wilson lines that connect the fields before and after the interaction. 

It is important to emphasize that, despite their conceptual differences, the partonic (PDF, GPD, TMD, GTMD) and CGC approaches are both founded on the principle of factorization: the separation of perturbative hard dynamics from the nonperturbative physics of the target. It is therefore natural to expect that these descriptions should be connected when their domains of validity overlap---namely, at very high energies where the CGC is justified, and in processes with a hard transverse momentum scale where collinear or TMD factorization applies. Establishing this matching is crucial: it unifies the physics of small-$x$ gluon saturation with the multi-dimensional imaging of hadrons, providing a common language that connects two seemingly different pictures of QCD dynamics. On the one hand, it allows distributions such as GPDs, TMDs, and GTMDs to be computed directly from Wilson-line correlators in the CGC formalism. On the other hand, it ensures that observables derived in the partonic framework remain consistent with the dynamics of dense gluon fields at high energies. In this way, the matching not only strengthens the theoretical connection between small-$x$ saturation physics and partonic physics, but also makes these ideas directly applicable to collider phenomenology. By providing a unified framework, it allows multi-dimensional parton distributions—such as GPDs, TMDs, and GTMDs—to be computed in the kinematic domains relevant for experiments, where both high energies and hard momentum scales are present. This is essential for interpreting observables at facilities like the Electron–Ion Collider (EIC)~\cite{Boer:2011fh,Accardi:2012qut} and the Large Hadron Collider (LHC), where signals of gluon saturation and partonic imaging naturally coexist. 

Recognizing this connection, substantial progress has already been achieved at leading order, where various parton distributions have been expressed in terms of Wilson-line operators. Examples include the gluon PDF~\cite{Baier:1996sk,Mueller:1999wm,Mueller:2001fv}, GPDs~\cite{Hatta:2016dxp,Hatta:2017cte}, TMDs~\cite{Dominguez:2011wm,Marquet:2009ca}, GTMDs~\cite{Hatta:2016dxp,Boussarie:2018zwg}, and diffractive distributions~\cite{Hatta:2022lzj,Hauksson:2024bvv}. Additionally, gluon TMDs have also been considered at next-to-leading order~\cite{Xiao:2017yya,Zhou:2018lfq}. 

However, these results have so far been obtained only on a case-by-case basis, and a general, systematic framework with transparent Feynman rules has been lacking. In this work, we address this gap by formulating a simple and comprehensive set of Feynman rules that directly connect the operator definitions of parton distributions to their realization in the CGC shockwave formalism. Within this framework, we present for the first time complete leading-order results for the quark PDF and GPD, including their finite parts, as well as TMDs and GTMDs with arbitrary gauge-link structures. This unified approach not only consolidates previous findings but also establishes a foundation for systematic higher-order studies. 

Beyond its theoretical significance, this framework further strengthens the phenomenological implications discussed above. By enabling the computation of multi-dimensional parton distributions in the high-energy limit, it provides new means to study spin–orbit correlations, orbital angular momentum, and the phase-space structure of nucleons. These aspects are central to the physics program of the upcoming EIC, where multi-dimensional imaging of hadrons will be a primary goal. Moreover, connections between small-$x$ dynamics and parton distributions are also highly relevant for interpreting results from the LHC and other high-energy facilities, where gluon saturation effects play an increasing role. In this sense, our results bridge two complementary pictures of hadronic structure and open new avenues for both theory and experiment.  

The article is organized as follows. In Sec.~\ref{sec:theory}, we describe the operator-level starting point and derive the Feynman rules for the matching. We then apply the framework to compute specific parton distributions: GTMDs in Sec.~\ref{sec:GTMD}, GPDs in Sec.~\ref{sec:GPD}, TMDs and PDFs in Sec.~\ref{sec:TMDPDF}, and diffractive distributions in Sec.~\ref{sec:diffraction}. Finally, in Sec.~\ref{sec:Conclusion}, we summarize our results and discuss future directions.

\section{Theoretical framework}
\label{sec:theory}

In this Section, we present the overall framework for computing parton distributions in the shockwave approximation used widely in small-$x$ calculations.
This allows us to perform computations starting from the operator-level definition, without considering a specific observable.
We will work in a frame where the target moves with large $p^+$ (i.e. along the $+z$ direction). This is the standard choice in computations with parton distributions but differs from the convention adopted in many small-$x$ works. The main reason for this choice is to make comparisons to the standard literature for parton distributions easier.

Note, however, that while our main motivation is to be able to compute parton distributions at small $x$, we wish to present our results in a general shockwave approximation without approximating the relevant $x$ variable to be small.
While it will turn out that small $x$ is a \textit{requirement} for nucleon targets for the shockwave approximation to be valid,
the small-$x$ approximation and the shockwave limit correspond to a power counting in terms of different parametric variables,  allowing us to treat these as separate approximations.
We hope that making a clear distinction between the shockwave limit and small $x$ will clarify the approximations done in the calculations.
% The small-$x$ limit can then be taken at the end, although the final expressions are expected to hold for any $x$ when the shockwave approximation is valid.

Let us remark on the connection between small $x$ and the shockwave limit.
Working in the frame where the target is moving in the plus direction,
the shockwave approximation is valid when the plus momentum $k^+$ of the particle scattering off the target is small compared to the width of the target in the minus direction $L^-$.
The relevant $x$-variable describing the small-$x$ power counting is defined as $x = k^+ / p^+$ where $p^+$ is the plus momentum of the target.
We can then see that the shockwave approximation is valid when $k^+ L^- = x p^+ L^- \ll 1$, and
working at a small enough $x$ guarantees that the shockwave approximation is also satisfied.
In general, however, we see that the proper power counting for the shockwave approximation depends on the constant $p^+ L^- = M R$, where $M$ and $R$ are the mass and the radius of the target,
which depends on the specific target.
For example, for protons we can estimate $M_p R_p \approx 5$, meaning that the shockwave approximation requires $x \ll 1/(M_p R_p) \approx 0.2$.
For heavy nuclei, the $x$-variable is typically defined with respect to the average momentum of the nucleon, and we find
 $(M_A /A) \times R_A \approx A^{1/3} \times 5$.
 This means that we generally need a slightly lower value of $x$
for heavy nuclei compared to protons for the shockwave approximation to be valid, which can be understood from the increase in the width of the shockwave due to multiple nucleons.

This requirement of a small $x$ variable can also be understood from the uncertainty principle:
because the $x$ variable corresponds to the plus momentum fraction exchanged with the target, in the parton picture we can relate it to the plus momentum fraction carried by the parton interacting with the projectile.
By momentum conservation, the plus momentum distribution of the parton is restricted to values $0 \leq k^+ \leq p^+$.
In position space, this corresponds to a distribution of the parton in minus direction, and the variance of this distribution can be interpreted as the width of the shockwave $L^-$.
The uncertainty principle then gives us the estimate $ p^+ L^- \gtrsim 1$, meaning that this constant cannot be too small.
Hence, the shockwave approximation in general requires a small $x$.

\update{
For a consistent matching to the parton distributions, however, it is useful to consider the expansion around the relevant $x$-variable (corresponding to small $x$) and the expansion around $x p^+ L^-$ (corresponding the shockwave limit) as separate.
As it turns out, \textit{only} expansion around the parameter $x p^+ L^-$ is required for the calculations of the parton distributions without further expansion around $x$.
By considering the shockwave limit with the expansion in $x p^+ L^-$ we are then making the minimal amount of approximations for the matching, which should clarify the relation between the shockwave limit and the parton distributions.
% While the distinction between these two expansions is not highly relevant for this work, where we only work in the eikonal limit and keep the leading term in the expansion around the variable  $x p^+ L^-$, we expect it to be important for subeikonal corrections where a well-defined power counting is required.
}

\subsection{Relation between parton distributions and the shockwave approximation}
\label{sec:distributions}

Let us now explain the main procedure for computing parton distributions in the shockwave limit.
In general, parton distributions can be written as matrix elements such as~\cite{Diehl:2003ny}
\begin{equation}
\label{eq:parton_distribution1}
    F^{ij}(\kt, \Deltat,x, \xi) = (P^+)^n \int \frac{\dd{r^-} \dd[2]{\rt}}{(2\pi)^3} 
    e^{ ixP^+ r^- - i \kt \vdot \rt}
    \mel{p'}{\mathcal{T} \Phi^i \qty(-\frac{r}{2}) \mathcal{W}^R_C\qty[- \frac{r}{2},\frac{r}{2}] \Phi^j \qty(\frac{r}{2})}{p}
   ,
\end{equation}
where $p$ and $p'$ describe the incoming and outgoing target states, $\Phi$ is a field operator corresponding to the particle of interest, and the factors of the momentum $(P^+)^n$ are added to make the matrix element boost invariant. 
Here, $\mathcal{T}$ denotes time-ordering which is important for GPDs and GTMDs that are defined at the amplitude level\footnote{\update{While we note that the explicit time-ordering can be omitted in the definition of GPDs and GTMDs to make the connection to the PDFs and TMDs more straightforward~\cite{Diehl:1998sm,Diehl:2003ny,Collins:2011zzd}, we keep it here to make perturbative calculations easier.}}; parton distributions defined at the level of the cross section, such as PDFs and TMDs, are discussed in Sec.~\ref{sec:TMDPDF}.
We will also implicitly set $r^+ =0$ for all parton distributions, and
the open indices $i,j$ of the field operators can be contracted with suitable projection operators. 
The Wilson line $\mathcal{W}$ connects the two field operators to make the quantity gauge invariant, and the exact path $C$ of the Wilson line depends on the process of interest.
Specifically, we define the Wilson line as
\begin{equation}
\label{eq:Wilson_line}
    \mathcal{W}^R_C\qty[x,y]
    = \mathcal{P} \exp( -i g_s \int_C \dd{s_\mu} A_a^\mu(s) t^a_R )
\end{equation}
where the path $C$ starts from $y$ and ends at $x$,
$\mathcal{P}$ denotes path-ordering along the path $C$ with the start point on the right and the end point on the left,
and $t_R$ correspond to the color matrices in the representation $R$ of the field operators.
In practice, we will write all of the quantities in this work in terms of the Wilson lines in the fundamental representation $F$ and denote $\mathcal{W}_C \equiv \mathcal{W}^F_C$.

We will be working in the symmetric frame with the definitions:
\begin{align}
    P &= \frac{1}{2}(p + p') ,
    &
    p &= P - \frac{1}{2} \Delta,
  \\
     \Delta &= p '- p,
    &
    p' &= P + \frac{1}{2} \Delta,
\end{align}
where $p$ and $p'$ are the momenta of the incoming and outgoing nucleon, respectively.
Specifically, the momenta are defined as
\begin{align}
    p &= \qty([1+\xi]P^+, \frac{1}{2 [1+\xi]P^+} \qty[M^2 + \frac{1}{4} \Deltat^2], -\frac{1}{2}\Deltat),
    \\
    p' &= \qty([1-\xi]P^+, \frac{1}{2 [1-\xi]P^+} \qty[M^2 + \frac{1}{4} \Deltat^2], \frac{1}{2}\Deltat),
    \\
    P &= \qty(P^+ ,\frac{1}{2[1-\xi^2]P^+} \qty[M^2 + \frac{1}{4} \Deltat^2],\boldsymbol{0}),
    \\
    \Delta &= \qty(-2\xi P^+,2\xi P^-, \Deltat),
\end{align}
where we use the convention $p^\pm = \frac{1}{\sqrt{2}} (p^0 \pm p^3)$, bolded symbols denote transverse components\footnote{For clarity we will denote $\Deltat$ instead of $\boldsymbol{\Delta}$, in order to distinguish it from the 4-vector $\Delta$.}, and we have defined  the skewness 
\begin{equation}
    \xi = \frac{p^+ - p'^+}{p^+ + p'^+} = - \frac{\Delta^+}{2 P^+}.
\end{equation}
We can also read the value of the Mandelstam $t$ variable from these momenta:
\begin{equation}
    t = (p' -p)^2 =  \Delta^2
    = -\frac{1}{1-\xi^2}\qty[ \Deltat^2+ 4 \xi^2 M^2].
\end{equation}

To employ the shockwave limit, it is easier to work in the position space for the target:
\begin{equation}
    \ket{p(\pt, p^+, \lambda)} = \sqrt{2p^+}\int  \dd{b^-} \dd[2]{\bt} e^{i \pt \vdot \bt - i p^+ b^- } \ket{p( \bt, b^-, \lambda )}
\end{equation}
where the factor $\sqrt{2p^+}$ comes from the normalization
\begin{equation}
    \bra{p'(\pt', p^{\prime +}, \lambda')}
    \ket{p(\pt, p^+, \lambda)}
    =  \delta_{\lambda  \lambda'} 2 p^+ (2\pi)^3 \delta(p^+ - p^{\prime +}) \delta^{(2)}(\pt - \pt').
\end{equation}
This allows for a very natural application of the shockwave limit where the target is localized at a specific light-cone time $b^-$.
Alternatively, it can be motivated by the eikonal approximation  that is valid in the high-energy limit, which states that an interaction with the target is instantaneous in the light-cone time.
Due to this instantaneous interaction the transverse coordinates are also conserved, allowing us to write the eikonal approximation as
\begin{equation}
\label{eq:eikonal_approximation}
    \mel{p'(\bt',b^{\prime -}, \lambda')}{\mathcal{O}}{p( \bt, b^-, \lambda )}
    = \delta^{(2)}(\bt' -\bt) \delta(b'^- -b^-)
    \expval{\mathcal{O}}_{b, \lambda  \lambda'},
\end{equation}
where the notation $\expval{\mathcal{O}}_{b, \lambda  \lambda'}$ means that the operator $\mathcal{O}$ is evaluated with a shockwave at $b^-$ and the center of the target at $\bt$.
We have also kept the helicities $\lambda, \lambda'$ of the incoming and outgoing target explicit.
Equation~\eqref{eq:eikonal_approximation} serves as our definition of the matrix element $  \expval{\mathcal{O}}_{b, \lambda  \lambda'}$.

Employing the eikonal approximation, we can rewrite matrix elements in the momentum space as
\begin{equation}
\begin{split}
    &\mel{p'(\pt', p^{\prime +}, \lambda')}{\mathcal{O}}{p(\pt, p^+, \lambda)} \\
    =&
2 \sqrt{p^+ p^{\prime +}}\int \dd{b^-} \dd[2]{\bt} \dd{b^{\prime -}} \dd[2]{\bt'}  e^{-i\pt' \vdot \bt' +i \pt \vdot \bt - i p^+ b^- + i p^{\prime +} b^{\prime -} }
\mel{p'(\bt',b^{\prime -}, \lambda')}{\mathcal{O}}{p( \bt, b^-, \lambda )} \\    
=&
2 \sqrt{p^+ p^{\prime +}}\int  \dd{b^-} \dd[2]{\bt}
e^{  i (\pt -\pt') \vdot \bt - i (p^+ -  p^{\prime +}) b^-  }
\expval{\mathcal{O}}_{b, \lambda  \lambda'} \\    
=&
2 P^+ \sqrt{1 - \xi^2 }
\int \dd{b^-} \dd[2]{\bt} 
e^{ -i \Deltat \vdot \bt - i 2  \xi P^+ b^-  }
\expval{\mathcal{O}}_{b, \lambda  \lambda'},
\end{split}
\end{equation}
and the parton distribution then becomes
\begin{equation}
\label{eq:parton_distribution2}
\begin{split}
   & F^{ij}(\kt, \Deltat,x, \xi) 
   \\
    =&
   \frac{(P^+)^n }{(2\pi)^3} 
2 P^+ \sqrt{1 - \xi^2 }
    \int \dd{r^-} \dd[2]{\rt}
 \dd{b^-}\dd[2]{\bt}
e^{ -i \Deltat \vdot \bt - i 2  \xi P^+ b^-  
+ixP^+ r^- - i \kt \vdot \rt} \\
&\times
\expval{
\mathcal{T}
\Phi^i \qty(-\frac{r}{2}) \mathcal{W}\qty[- \frac{r}{2},\frac{r}{2}] \Phi^j \qty(\frac{r}{2})}_{b, \lambda  \lambda'}
\\
    =&
   \frac{(P^+)^n }{(2\pi)^3} 
2 P^+ \sqrt{1 - \xi^2 }
    \int \dd{r^-} \dd[2]{\rt}
\dd{b^-} \dd[2]{\bt} 
e^{- i \Deltat \vdot \bt - i 2  \xi P^+ b^-  
+ixP^+ r^- - i \kt \vdot \rt} \\
&\times    \expval{ 
\mathcal{T}
\Phi^i \qty(-\frac{r}{2}-b) \mathcal{W}\qty[- \frac{r}{2}-b,\frac{r}{2}-b] \Phi^j \qty(\frac{r}{2} -b)}_{\lambda  \lambda'}\\
    =&
   \frac{(P^+)^n }{(2\pi)^3} 
2 P^+ \sqrt{1 - \xi^2 }
    \int \dd{x^-} \dd[2]{\xt}
 \dd{y^-}\dd[2]{\yt} \\
&\times e^{ i \Deltat \vdot \frac{1}{2}(\xt+\yt) + i   \xi P^+ (x^- + y^-)  
-ixP^+ (x^- - y^-) + i \kt \vdot (\xt-\yt)}
\expval{ 
\mathcal{T}
\Phi^i(x) \mathcal{W}[x,y] \Phi^j(y)}_{\lambda  \lambda'}.
\end{split}
\end{equation}
Here, we have shifted the coordinates using
\begin{align}
    \mathcal{O}(x) 
    &= e^{ i y \vdot \hat  P } \mathcal{O}(x-y) e^{ -i y \vdot \hat  P },  \\
    \ket{p(x)} 
    &= e^{ i y \vdot \hat  P }\ket{p(x-y)}  ,
\end{align}
such that the zero light-cone time corresponds to the interaction with the shockwave, and the center of the target is located at the zero transverse coordinate.
With this shift, it is natural to define the target average
\begin{equation}
    \expval{\mathcal{O}}_{\lambda \lambda'}
    = 
    \expval{\mathcal{O}}_{b,\lambda \lambda'}
    \Big\rvert_{b=0}
\end{equation}
which corresponds to the usual CGC average for $\lambda = \lambda'$.
Going to the last line, we have also redefined the integration variables:
\begin{align}
    r &= -x +y,
    &
    b &= -\frac{1}{2}(x +y).
\end{align}

Equation~\eqref{eq:parton_distribution2} is our main starting point for evaluating GTMDs and GPDs.
This definition needs to be modified slightly for
TMDs, PDFs, and diffractive distributions as will be  discussed in more detail in their corresponding sections~\ref{sec:TMDPDF} and \ref{sec:diffraction}.

\subsection{Feynman rules}

The evaluation of the parton distribution~\eqref{eq:parton_distribution2} can be done perturbatively, order by order in the strong coupling constant, when the relevant momentum scales are sufficiently large.
This gives rise to Feynman rules that can be used for a systematic way to organize different terms in the perturbation theory.
These Feynman rules are a combination of rules for the gauge links~\cite{Boussarie:2023izj}, corresponding to eikonal interactions of gluons, and rules for the scattering off the shockwave~\cite{Balitsky:1995ub,Balitsky:1998ya}.
For completeness, we provide a brief description of these rules below.

\begin{figure}
    \centering
\begin{align*}
    \begin{array}{l}
\begin{overpic}[width=0.3\textwidth]{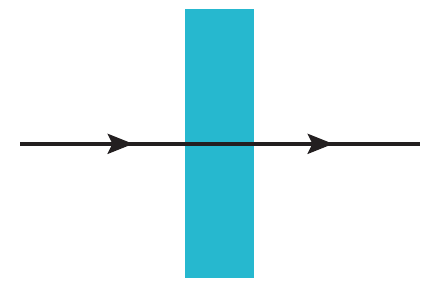}
    \put(25,45){$l$}
    \put(75,45){$l'$}
    \put(5,38){$i$}
    \put(93,38){$ j$}
    \put(48,40){$\xt$}
        \put(15,40){\vector(1,0){20}}
        \put(65,40){\vector(1,0){20}}
\end{overpic}
\end{array}
=&
   \gamma^- 2\pi\delta(l^{\prime -} - l^-)
   \int\dd[2]{\xt} 
     e^{ i b^- (l'^+ - l^+) -i \xt \vdot (\lt' -\lt)} 
     V(\xt )_{ji} 
     \\
     \begin{array}{l}
\begin{overpic}[width=0.3\textwidth]{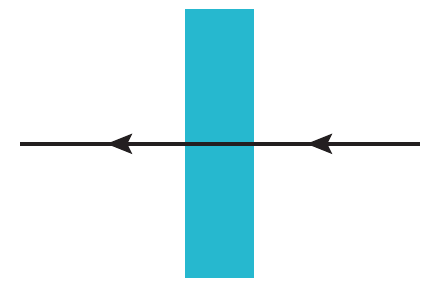}
    \put(25,45){$l$}
    \put(75,45){$l'$}
    \put(5,38){$i$}
    \put(93,38){$ j$}
    \put(48,40){$\xt$}
        \put(15,40){\vector(1,0){20}}
        \put(65,40){\vector(1,0){20}}
\end{overpic}
\end{array}
=&
  - \gamma^- 2\pi\delta(l^{\prime -} - l^-)
   \int\dd[2]{\xt} 
     e^{i b^- (l'^+ - l^+)-i \xt \vdot (\lt' -\lt)} 
     V^\dag(\xt)_{ij} 
     \\
     \begin{array}{l}
\begin{overpic}[width=0.3\textwidth]{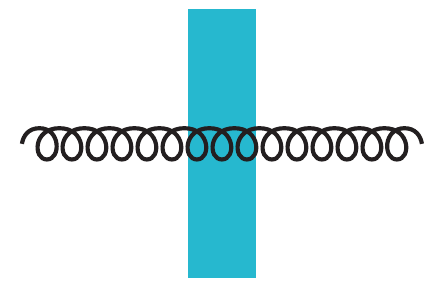}
    \put(48,40){$\xt$}
    \put(25,45){$l$}
    \put(75,45){$l'$}
    \put(-2,42){$a, \mu$}
    \put(90,42){$c, \nu$}
        \put(15,40){\vector(1,0){20}}
        \put(65,40){\vector(1,0){20}}
\end{overpic}
\end{array}
=&
-2 l^- g^{\mu \nu}   2\pi\delta( l^{\prime -} - l^-)
   \int\dd[2]{\xt} 
     e^{i b^- (l'^+ - l^+)-i \xt \vdot (\lt' -\lt)} 
     U(\xt)_{ca} 
\end{align*}
    \caption{Feynman rules for the shockwave with the target at the coordinate $b$.
    In practice, we can shift the coordinates such that in the actual computations we  take $b =0$.
    }
    \label{fig:shockwave_rules}
\end{figure}

\begin{figure}
    \centering
\begin{align*}
    \begin{array}{l}
\begin{overpic}[width=0.3\textwidth]{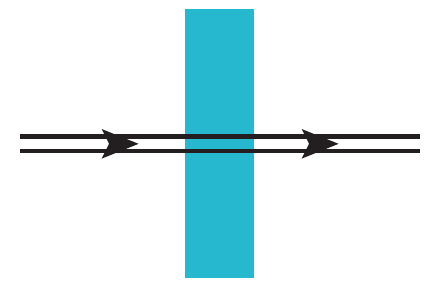}
    \put(5,40){$i$}
    \put(93,40){$ j$}
    \put(48,40){$\xt$}
\end{overpic}
\end{array}
=&
     V(\xt )_{ji} 
\end{align*}
    \caption{Feynman rule for an eikonal quark line crossing through the shockwave.
    For antiquarks we get \update{$V^\dag(\xt)_{ij}$} and for gluons $U(\xt)_{ji}$ instead.
    }
    \label{fig:eikonal_shockwave_rules}
\end{figure}

\begin{figure}
    \centering
    \begin{equation*}
           \begin{array}{l}
        \begin{overpic}[width=0.3\linewidth]{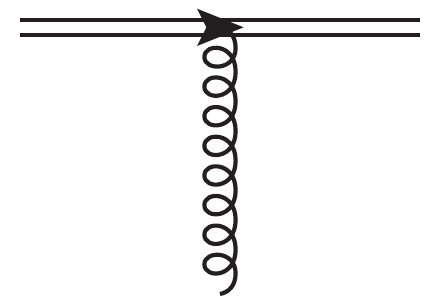}
            \put(65,35){$q$}
            \put(60,10){$a,\mu$}
            \put(50,72){$x$}
            \put(5,72){$x_i$}
            \put(90,72){$x_f$}
            \put(5,50){$i$}
            \put(90,50){$j$}
        \put(60,45){\vector(0,-1){20}}
        \end{overpic}
\end{array}
= - i g_s g^{\mu +}  t_{ji}^a
\int_{x_i^-}^{x_f^-} \dd{x^-}
 e^{i q \vdot x}
    \end{equation*}
    \caption{Feynman rule for a gluon interacting with an eikonal line in the fundamental representation.
        Here $x_i$ and $x_f$ denote the light-cone times for the start and end points of the eikonal line.
    }
    \label{fig:eikonal_rule}
\end{figure}

\begin{enumerate}
    
    \item \label{it:standard} Standard Feynman rules apply for regular propagators and vertices~\cite{Peskin:1995ev}, including minus signs that appear for e.g. quark loops.
    To make computations with the shockwave easier, we will be working in the light-cone gauge $A^- = 0$ for gluons.
    This corresponds to the following Lorentz structure of the gluon propagator:
    \begin{equation}
    \label{eq:gluon_prop}
        \Pi^{\mu \nu}(q)
        = - g^{\mu \nu} + \frac{q^\mu n^\nu + q^\nu n^\mu}{q \vdot n},
    \end{equation}
    where $n^\mu$ is a vector such that $n \vdot q = q^-$ for any vector $q^\mu$.
    Additionally, the sign of the quark--gluon vertex is chosen as $-i g_s$, consistent with the definition of the Wilson line~\eqref{eq:Wilson_line}.

    \item  \label{it:eikonal_setup}  \textbf{Eikonal Feynman rules}: Eikonal lines are treated in position space where they do not have  propagators associated with them.
        Instead, particle lines connecting to eikonal lines give additional exponential factors.
    \begin{enumerate}
    \item \label{it:eikonal_endpoint} For vertices corresponding to field operators, add a factor $e^{i q \vdot x}$ where 
    $x$ is the coordinate of the field operator and
    $q$ is the momentum flowing out of the vertex. 
    
    \item \label{it:eikonal_rule} Gluons can be connected to the eikonal lines.
    The corresponding Feynman rule for this vertex is given in Fig.~\ref{fig:eikonal_rule}.
    \end{enumerate}
    
    \item \label{it:shockwave} \textbf{Shockwave Feynman rules}: 
    Draw one vertical (instantaneous) shockwave for each amplitude.
    Shockwaves at different positions correspond to different diagrams, and these different diagrams have to be summed together to get the final result. 
    \begin{enumerate}
    
 \item \label{it:shockwave_interaction} For particles passing through the shockwave, we get a shockwave vertex shown in Fig.~\ref{fig:shockwave_rules}.

    \item \label{it:time_constraint} Because the shockwave is instantaneous in light-cone time, we must make sure that the field operators are restricted to before or after the shockwave depending on the Feynman diagram.
    For example, if we have a field operator at $x=(x^-,\xt)$ and a shockwave at $b^-$, we would need a step function $\theta(x^- - b^-)$ for a field operator after the shockwave or  $\theta(b^- - x^-)$ for a field operator before the shockwave.
    
    \item \label{it:eikonal_shockwave} For eikonal lines going through the shockwave, we get a Wilson line in the representation of the particle that has been eikonalized, as illustrated in Fig.~\ref{fig:eikonal_shockwave_rules} for an eikonal quark line.
    The form of the Wilson line depends on the representation of the eikonalized particle crossing the shockwave as follows:
    \begin{itemize}
        \item Eikonal quark: $V(\xt)$ (fundamental representation)
        \item Eikonal antiquark: \update{$V^\dag(\xt)$} (fundamental representation)
        \item Eikonal gluon: $U(\xt)$ (adjoint representation)
    \end{itemize}

    \item \label{it:subtract_rule} For each shockwave, subtract the noninteracting case corresponding to $V \to 1$ for the Wilson lines.
    This subtraction has to be done to exclude disconnected diagrams in the definition of the parton distribution.
    
    \end{enumerate}
     
    The Wilson lines $V$, $V^\dag$, and $U$ are understood to depend only on the gluon fields of the background shockwave, as opposed to the gauge link $\mathcal{W}$ which contains both perturbative gluons and the color field of the background shockwave.
    
\end{enumerate}

While these Feynman rules are standard in their respective communities, let us briefly give some motivation for the way they appear here.
The general exponential factors of the form $e^{i q \vdot x}$ can be traced to the relation between coordinate and momentum space propagators:
\begin{equation}
    D(q) 
    = 
    \int \dd[4]{x} e^{i q \vdot x } \widetilde D(x).
\end{equation}
This also explains the integrals over the transverse coordinates in the shockwave vertices in Fig.~\ref{fig:shockwave_rules}:
Because the shockwave is instantaneous in the light-cone time $x^-$, there is no associated integral.
Additionally, since the shockwave Wilson lines do not depend on the plus-coordinate $x^+$, this coordinate can be integrated over, leading to the minus-momentum-conserving delta functions.

The interaction with the eikonal line follows from the definition of the Wilson lines in the gauge link:
\begin{equation}
    \mathcal{W}\qty[(x_f^-, \xt),(x_i^-,\xt)]
    = \mathcal{P} \exp( -i g_s \int_{x_i^-}^{x_f^-} \dd{x^-} A^+(x) )
    = 1  - i g_s \int_{x_i^-}^{x_f^-} \dd{x^-} A^+(x) + \order{g_s^2}.
\end{equation}
Contracting with the gluon fields $A^+(x)$ and writing the corresponding propagator in momentum space leads to the term $g^{\mu + }  t^a e^{i q \vdot x}$.

The shockwave Wilson lines $V$, $V^\dag$, and $U$ correspond to classical color fields that carry the information about the shockwave.
The Wilson line $V$ is a color matrix in the fundamental representation, whereas we denote the adjoint representation Wilson line by $U$.
For a shockwave at the time $b^-$, these Wilson lines only have support in a narrow region $[b^- - L^-/2, b^- + L^-/2]$ in light-cone time:
\begin{equation}
    V(\xt) = \mathcal{P} \exp( -i g_s \int_{b^- -L^-/2}^{b^- + L^-/2} \dd{x^-} A_\cl^{+}(x) )
    = \mathcal{P} \exp( -i g_s \int_{-\infty}^{\infty} \dd{x^-} A_\cl^{+}(x) )
\end{equation}
where we used the fact that $A_\cl^+$, corresponding to the background color field, vanishes outside the shockwave.
The shockwave approximation then corresponds to taking the width of the shockwave to zero, $L^-  \to 0$, such that we can neglect the emission or absorption of perturbative gluons inside the shockwave.
The gluon fields $A_\cl$ are considered to be classical in the sense that they correspond to real numbers whose value is given by the target shockwave.
In general, the color-field configuration of the target is a fluctuating quantity, and for this reason the classical Wilson lines need to be evaluated as an expectation value such as:
\begin{equation}
    \expval{V}_{\lambda \lambda'}
    = 
    \int \mathcal{D} A_\cl \Psi_{\lambda \lambda'}[A_\cl] 
    V[A_\cl]
\end{equation}
where $\Psi$ is the wave function describing the weight of the color configuration.
In phenomenological applications, the wave function is often described using the McLerran--Venugopalan model~\cite{McLerran:1993ka,McLerran:1993ni,McLerran:1994vd}  which allows for an analytic evaluation of correlations of classical Wilson lines.

We note that Rule~\ref{it:subtract_rule}, corresponding to the subtraction of the identity, is often left out in small-$x$ computations.
This modification of no subtraction is usually valid, as the identity contribution corresponds to the non-interacting case where no momentum is transferred between the projectile and the target.
Such contributions generally vanish for cross sections of interest.
However, in this work we are doing computations at the operator level, and it is no longer clear whether the non-interacting case vanishes identically.
For this reason, we insist upon including this rule to ensure the consistency of our framework.

Finally, we note that Rule~\ref{it:eikonal_rule} is not needed at leading order, which is the focus of this work.
We include it here for completeness, as it would be needed for next-to-leading order calculations, and we do not expect any other modifications to the rules at higher orders apart from the choice of the regularization scheme required for rapidity divergences that start at next-to-leading order.

\section{Generalized transverse-momentum-dependent distributions}
\label{sec:GTMD}

As already alluded to before, GTMDs, despite their fundamental importance, have remained essentially unexplored in practical calculations. This is largely due to the formidable complexity of their operator structure and the absence of suitable nonperturbative tools for their evaluation. In this work, we overcome these obstacles by establishing a direct connection between GTMDs and the shockwave formalism. The latter provides a natural and efficient framework in which GTMDs can be computed systematically, placing them on the same footing as other parton distributions previously studied in the CGC.

\subsection{Gluon}

For gluons, we can define a general GTMD as (generalizing the GPD case in Ref.~\cite{Diehl:2003ny})
\begin{equation}
\label{eq:gluon_GTMD}
\begin{split}
       W^{ij}_{\lambda \lambda'}(P,x,\kt, \Delta;C,C') =&
   \frac{2}{P^+} \int \frac{\dd{r^-} \dd[2]{\rt}}{(2\pi)^3} 
    e^{ ixP^+ r^- - i \kt \vdot \rt}\\
    &\times
    \Tr
    \mel{p'}{
    \mathcal{T}
    G^{+ i} \qty(-\frac{r}{2})
    \mathcal{W}_C\qty[- \frac{r}{2},\frac{r}{2}] 
    G^{+j} \qty(\frac{r}{2})
    \mathcal{W}_{C'}\qty[\frac{r}{2},-\frac{r}{2}] 
    }{p} ,
\end{split}
\end{equation}
where we use the Latin alphabet to indicate summation over transverse indices.
Other possible gluon GTMDs vanish in the eikonal limit and will not be considered in this work.
The seemingly extra factor of 2 compared to Ref.~\cite{Diehl:2003ny} comes from the fact that we are writing the Wilson lines in the gauge link in the fundamental representation as opposed to the adjoint representation.
This allows for a more general gauge link structure as the paths $C$ and $C'$ can be completely independent.
In the shockwave limit, this becomes:
\begin{equation}
\label{eq:gluon_GTMD_shockwave}
\begin{split} 
W^{ij}_{\lambda \lambda'}
    =&
   \frac{4 }{(2\pi)^3} 
\sqrt{1 - \xi^2 }
    \int \dd{x^-} \dd[2]{\xt}
 \dd{y^-}\dd[2]{\yt}
e^{ i \Deltat \vdot \frac{1}{2}(\xt+\yt) + i   \xi P^+ (x^- + y^-)  
-ixP^+ (x^- - y^-) + i \kt \vdot (\xt-\yt)} \\
&\times
\Tr
\expval{
\mathcal{T}
G^{+ i} (x) 
\mathcal{W}_C\qty[x,y]
G^{+ j}(y)
\mathcal{W}_{C'}\qty[y,x]
}_{\lambda  \lambda'}.
\end{split}
\end{equation}

The leading order case for the gluon distribution comes completely from the classical gluon fields inside the shockwave.
This follows from the enhancement of the gluon field $A_\cl^+$ within the shockwave compared to the perturbative gluons, as generally we can have $g_s A_\cl^+ \sim 1$ in terms of power counting.
Such contributions then need to be resummed, which can be done by considering a narrow but finite shockwave.
As the contribution comes from gluons inside the shockwave, we can approximate $x^- = y^- = 0$ for the exponential factors, leading to e.g. $e^{i\xi P^+ x^-} \approx 1$.
The relevant contribution from the gluon field operators then reduces to 
\begin{equation}
\begin{split}
    &\int_{-L^-/2}^{L^-/2} \dd{x^-} \dd{y^-}
    \Tr
 G^{+ i}(x) 
 \mathcal{W}_C[x,y] 
 G^{+ j}(y)
 \mathcal{W}_{C'}[y,x] \\
    =&
    \int_{-L^-/2}^{L^-/2} \dd{x^-} \dd{y^-}
\Tr
 \partial^{i} A^{+}(x) 
 \mathcal{W}_C[x,y]  
 \partial^{j} A^{+}(y) 
 \mathcal{W}_{C'}[y,x]      
\end{split}
\end{equation}
where we have neglected other components of the gluon field operator that are not enhanced by the shockwave.
This can be simplified further by noting the identity
\begin{equation}
    \partial^i W_\xt[u^-, v^-]
    = -ig_s \int_{v^-}^{u^-} \dd{z^-} 
    W_\xt[u^-, z^-]
    \partial^i A^+(z^-, \xt)
    W_\xt[z^-, v^-],
\end{equation}
where we have denoted a Wilson line at a constant transverse coordinate $\xt$ by 
\begin{equation}
W_\xt[u^-, v^-] = \mathcal{P} \exp( - i g_s \int_{v^-}^{u^-} \dd{z^-} A^+(z^-,\xt) ) .
\end{equation}
A direct computation then shows that:
\begin{equation}
\begin{split}
    \Tr \Big\{ & \partial^i W_\xt[L^-/2, -L^-/2]
        \times
     \mathcal{W}_C[(-L^-/2,\xt),(-L^-/2,\yt)]  \\
     \times
     &
      \partial^j W_\yt[-L^-/2, L^-/2]
    \times
     \mathcal{W}_{C'}[(L^-/2,\yt),(L^-/2,\xt)]
     \Big\}
     \\
    =& g_s^2
    \int^{L^-/2}_{-L^-/2}\dd{x^-} \dd{y^-} 
    \Tr
   \partial^{i} A^{+}(x)   
   \mathcal{W}_C[x,y]  
   \partial^{j} A^{+}(y)   
   \mathcal{W}_{C'}[y,x].
\end{split}
\end{equation}
Since we can neglect perturbative gluons at this order, we can replace
\begin{align}
  W_\xt[L^-/2, -L^-/2] &= V(\xt),
  &
    \mathcal{W}_C[(-L^-/2,\xt),(-L^-/2,\yt)] 
    &= V_{C_{--}}[\xt,\yt],
    \\
  W_\yt[-L^-/2, L^-/2] &= V^\dag(\yt),
  &
    \mathcal{W}_{C'}[(L^-/2,\yt),(L^-/2,\xt)] 
    &= V_{C'_{++}}[\yt,\xt],
\end{align}
which allows us to write the gluon GTMD as
\begin{equation}
\label{eq:gluon_GTMD_shockwave2}
\begin{split} 
W^{ij}_{\lambda \lambda'}
    =&
   \frac{2 }{\as (2\pi)^4} 
\sqrt{1 - \xi^2 }
    \int  \dd[2]{\xt}\dd[2]{\yt} 
    e^{ i \Deltat \vdot \frac{1}{2}(\xt+\yt) + i \kt \vdot (\xt-\yt)}
  \\
&\times 
\Tr
    \expval{\partial^i V(\xt)
V_{C_{--}}\qty[\xt,\yt]
    \partial^j V^\dag(\yt)
V_{C_{++}'}\qty[\yt,\xt]
    }_{\lambda \lambda'}.
\end{split}
\end{equation}
Here $C_{--}$ indicates that the path $C$ is taken to begin and end at a negative (infinitesimal) light-cone time $x^-$, located before the shockwave.
Similarly, the path $C'_{++}$ begins and ends at a positive light-cone time, after the shockwave.

With this general notation, it is now easy to consider
different gauge link structures.
For the simplest gauge links, corresponding to  staples pointing in the past ($\sqsubset$) or future ($\sqsupset$) directions, we get:
\begin{align}
    V_{\sqsubset_{--}}\qty[\xt,\yt] &= 1
    &
    V_{\sqsubset_{++}}\qty[\yt,\xt] &= V(\yt) V^\dag(\xt)
    \\
    V_{\sqsupset_{--}}\qty[\xt,\yt] &= V^\dag(\xt) V(\yt)
    &
    V_{\sqsupset_{++}}\qty[\yt,\xt] &= 1.
\end{align}
These can be derived by noting that only Wilson lines crossing the shockwave yield a nonvanishing contribution.
With these gauge links, we find the following gluon GTMDs:
\begingroup
\allowdisplaybreaks
\begin{align}
\label{eq:gluon_dipole_GTMD1}
\begin{split} 
W^{ij}_{\lambda \lambda'}(C = \sqsubset,  C' = \sqsupset) 
    =&
   \frac{2 \sqrt{1 - \xi^2 } }{\as (2\pi)^4} 
    \int  \dd[2]{\xt}\dd[2]{\yt}
e^{ i \Deltat \vdot \frac{1}{2}(\xt+\yt) + i \kt \vdot (\xt-\yt)}
\\
&\times \Tr
    \expval{\partial^i V(\xt)
    \partial^j V^\dag(\yt)
    }_{\lambda \lambda'},
\end{split}
\\
\label{eq:gluon_dipole_GTMD2}
\begin{split} 
W^{ij}_{\lambda \lambda'}(C = \sqsupset,  C' = \sqsubset) 
    =&
   \frac{2 \sqrt{1 - \xi^2 } }{\as (2\pi)^4} 
    \int  \dd[2]{\xt}\dd[2]{\yt}
e^{ i \Deltat \vdot \frac{1}{2}(\xt+\yt) + i \kt \vdot (\xt-\yt)} \\
&\times  \Tr
    \expval{\partial^i V^\dag(\xt)
    \partial^j V(\yt)
    }_{\lambda \lambda'},
\end{split}
\\
\label{eq:gluon_WW_GTMD1}
\begin{split} 
W^{ij}_{\lambda \lambda'}(C = \sqsubset,  C' = \sqsubset) 
    =&
  - \frac{2 \sqrt{1 - \xi^2 } }{\as (2\pi)^4} 
    \int  \dd[2]{\xt}\dd[2]{\yt}
e^{ i \Deltat \vdot \frac{1}{2}(\xt+\yt) + i \kt \vdot (\xt-\yt)}\\
&\times  \Tr
    \expval{
    V^\dag(\xt)
    [\partial^i V(\xt)]
    V^\dag(\yt)
    [\partial^j V(\yt)]
    }_{\lambda \lambda'},
\end{split}
\\
\label{eq:gluon_WW_GTMD2}
\begin{split} 
W^{ij}_{\lambda \lambda'}(C = \sqsupset,  C' = \sqsupset) 
    =&
  - \frac{2 \sqrt{1 - \xi^2 } }{\as (2\pi)^4} 
    \int  \dd[2]{\xt}\dd[2]{\yt}
e^{ i \Deltat \vdot \frac{1}{2}(\xt+\yt) + i \kt \vdot (\xt-\yt)}
\\
&\times  \Tr
    \expval{[\partial^i V(\xt)]
    V^\dag(\xt)
    [\partial^j V(\yt)]
    V^\dag(\yt)
    }_{\lambda \lambda'},
\end{split}
\end{align}%
\endgroup
where the identity
\begin{equation}
\label{eq:Wilson_shuffling}
     \qty[\partial^i V(\xt)] V^\dag(\xt) = - V(\xt) \qty[\partial^i V^\dag(\xt)]
\end{equation}
has been used.
These results are consistent with Ref.~\cite{Hatta:2016dxp} after taking $\xi \to 0$.
The first two correspond to the so-called dipole gluon GTMD, and they are related to each other by the complex conjugation of the Wilson-line correlator.
Similarly, the last two are of the Weizsäcker--Williams type, and their Wilson-line structure is the complex conjugate of each other.
From this it is clear that even more complicated gauge link structures can be considered in a straightforward way.

For the dipole gluon GTMD, it is also convenient to consider this in terms of the dipole amplitude, defined as:
\begin{equation}
    N_{\lambda \lambda'} (\xt,\yt)= 1- \frac{1}{\nc} \Tr \expval{V(\xt )V^\dag(\yt)}_{\lambda \lambda'}.
\end{equation}
We will also define the momentum-space dipole amplitude as
\begin{equation}
\label{eq:dipole_amplitude_mom}
    \widetilde N_{\lambda \lambda'}(\pt, \qt)
    \equiv 
    \int  \frac{\dd[D-2]{\xt} \dd[D-2]{\yt}}{(2\pi)^{2(D-2)}}
    \frac{e^{i \pt \vdot \xt + i \qt \vdot \yt}}{\abs{\xt -\yt}^2}
     N_{\lambda \lambda'}(\xt, \yt),
\end{equation}
written in $D = 4 -2\varepsilon$ dimensions for generality.
We note that including the extra factor $1/\abs{\xt -\yt}^2$ here  is convenient, as it guarantees that the Fourier transform converges for any reasonable model of the dipole amplitude.
The dipole gluon GTMDs in $D=4$ dimensions can then be written as:
\begin{equation}
\label{eq:gluon_dipole_GTMD_momentum1}
\begin{split} 
&W^{ij}_{\lambda \lambda'}(C = \sqsubset,  C' = \sqsupset) \\
    =&
    -\frac{2 \nc }{\as } 
\sqrt{1 - \xi^2 }
\qty(\kt + \frac{1}{2}\Deltat)^i
\qty(-\kt + \frac{1}{2}\Deltat)^j
\nabla^2_\kt
    \widetilde N_{\lambda \lambda'}\qty(
\kt + \frac{1}{2}\Deltat
,-\kt + \frac{1}{2}\Deltat
    ),
\end{split}
\end{equation}
\begin{equation}
\label{eq:gluon_dipole_GTMD_momentum2}
\begin{split} 
&
W^{ij}_{\lambda \lambda'}(C = \sqsupset,  C' = \sqsubset) \\
    =&
    -\frac{2 \nc }{\as } 
\sqrt{1 - \xi^2 }
\qty(\kt + \frac{1}{2}\Deltat)^i
\qty(-\kt + \frac{1}{2}\Deltat)^j
\nabla^2_\kt
    \widetilde N_{\lambda \lambda'}\qty(
-\kt + \frac{1}{2}\Deltat
,\kt + \frac{1}{2}\Deltat
    ).
\end{split}
\end{equation}

To use these distributions in phenomenological studies, an explicit model for the dipole amplitude $N$ or its Fourier transform $\widetilde N$ is needed.
Here, we have also kept the helicities $\lambda$ and $\lambda'$ of the target general, although typically these are assumed to be the same: $\lambda = \lambda'$.
This is the case, for example, for the standard McLerran--Venugopalan model~\cite{McLerran:1993ka,McLerran:1993ni,McLerran:1994vd}, where one usually also neglects the impact-parameter dependence, although this can also be extended to a general transverse density of the target~\cite{Mantysaari:2024zxq,Penttala:2025tmp}.

\subsection{Quark}
\label{sec:quark_GTMD}

\begin{figure}
	\centering
    \begin{subfigure}{0.45\textwidth}
        \centering
        \begin{overpic}[width=\linewidth]{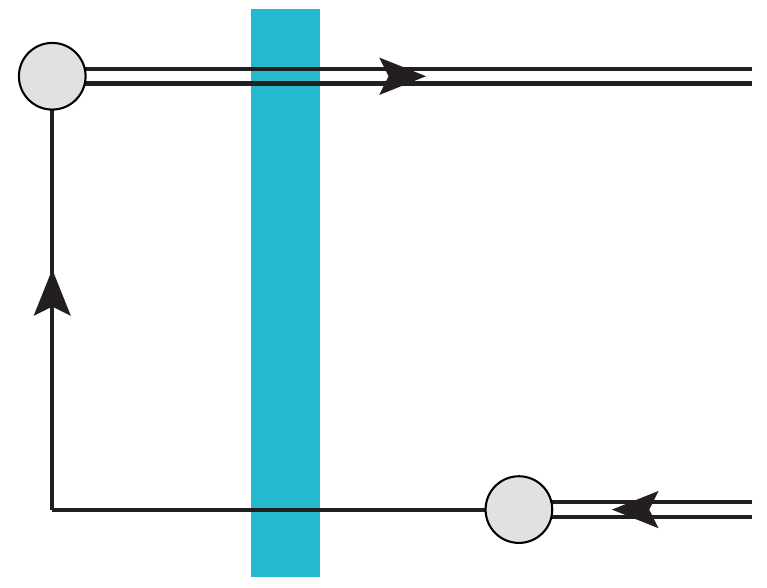}
            \put(17,40){$q$}
            \put(35,17){$\ut$}
            \put(5,75){$y$}
            \put(53,18){$q'$}
            \put(67,17){$x$}
        \put(14,30){\vector(0,1){20}}
        \put(63,14){\vector(-1,0){20}}
        \end{overpic}
        \caption{}
        \label{fig:quark_GTMD1}
    \end{subfigure}
    \begin{subfigure}{0.45\textwidth}
        \centering
        \begin{overpic}[width=\linewidth]{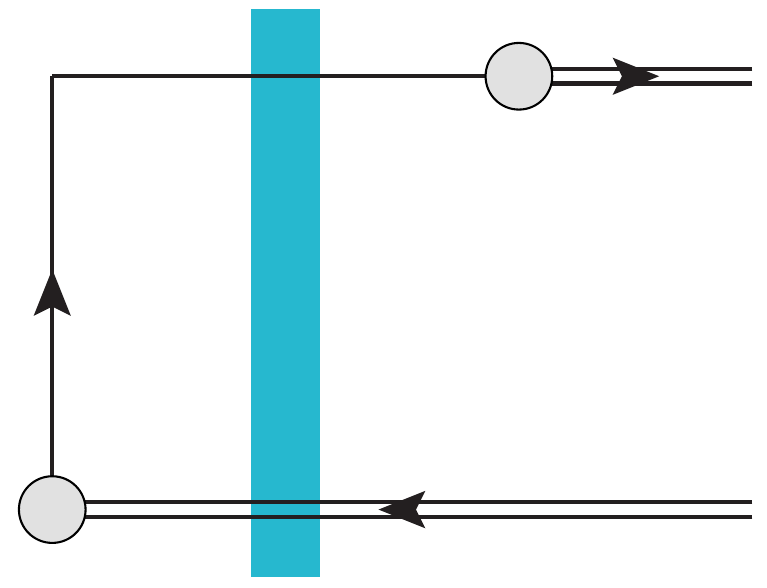}
            \put(17,40){$q'$}
            \put(35,70){$\ut$}
            \put(10,15){$x$}
            \put(53,57){$q$}
            \put(65,75){$y$}
        \put(14,30){\vector(0,1){20}}
        \put(43,62){\vector(1,0){20}}
        \end{overpic}
        \caption{}
        \label{fig:quark_GTMD2}
    \end{subfigure}
    \caption{Feynman diagrams for computing the quark GTMD.
    }
    \label{fig:quark_GTMD}
\end{figure}
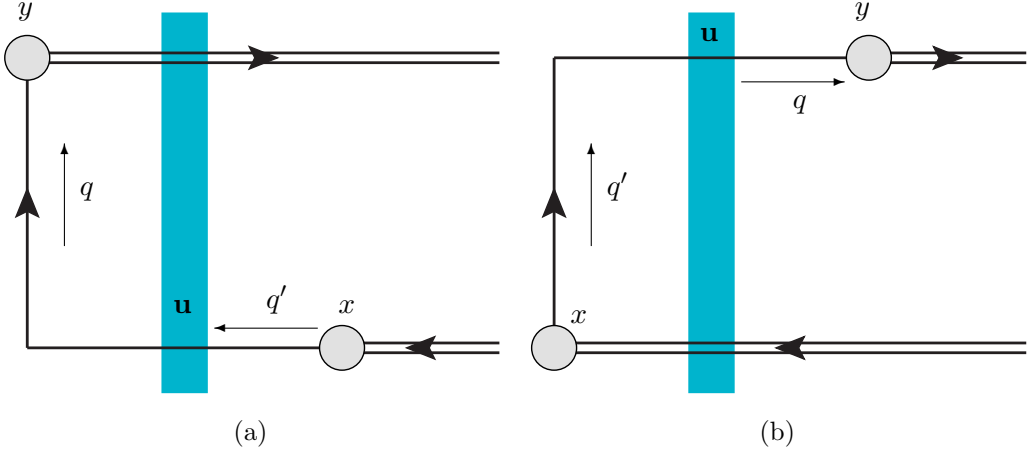

Following the notation from Refs.~\cite{Meissner:2009ww,Lorce:2025aqp}, we can write the quark GTMD as
\begin{equation}
\label{eq:quark_GTMD}
   W^{[\Gamma]}_{\lambda \lambda'}(P,x,\kt,2 \Delta;C) =
   \frac{1}{2} \int \frac{\dd{r^-} \dd[2]{\rt}}{(2\pi)^3} 
    e^{ ixP^+ r^- - i \kt \vdot \rt}
    \mel{p'}{
    \mathcal{T}
    \bar \psi \qty(-\frac{r}{2}) \Gamma \mathcal{W}_C\qty[- \frac{r}{2},\frac{r}{2}] \psi \qty(\frac{r}{2})}{p} ,
\end{equation}
where $\Gamma$ denotes a general gamma matrix and $C$ is the path for the gauge link.
Because of the eikonal approximation in the shockwave limit, structures with the $\gamma^5$ matrix such as $\Gamma = \gamma^+ \gamma^5$ and $\Gamma = i \sigma^{j+} \gamma^5$ vanish, and the only non-vanishing twist-2 GTMD is given by $\Gamma = \gamma^+$.
As outlined in Sec.~\ref{sec:distributions}, in the shockwave approximation we can write this GTMD as:
\begin{equation}
\label{eq:quark_GTMD_shockwave}
\begin{split}
    W_{\lambda \lambda'}^{[\gamma^+]}
    =&
   \frac{P^+ }{(2\pi)^3} 
\sqrt{1 - \xi^2 }
    \int \dd{x^-} \dd[2]{\xt}
 \dd{y^-}\dd[2]{\yt} \\
&\times e^{ i \Deltat \vdot \frac{1}{2}(\xt+\yt) + i   \xi P^+ (x^- + y^-)  
-ixP^+ (x^- - y^-) + i \kt \vdot (\xt-\yt)}
\expval{ 
\mathcal{T}
\bar \psi  (x) \gamma^+ \mathcal{W}_C\qty[x,y] \psi(y)}_{\lambda  \lambda'}.
\end{split}
\end{equation}
The relevant Feynman diagrams are then shown in Fig.~\ref{fig:quark_GTMD}, where for illustration purposes we show the future-pointing gauge link $C = \sqsupset$.
However, we note that the equations are computed for a completely general gauge link structure.
Diagram~\ref{fig:quark_GTMD1} yields:
\begin{equation}
\label{eq:quark_GTMD_fig1_1}
\begin{split}
    W_{\ref{fig:quark_GTMD1}}^{[\gamma^+]}
    &=
    \frac{P^+}{(2\pi)^3} \sqrt{1-\xi^2} \int \dd[2]{\xt} \dd{x^-} \dd[2]{\yt} \dd{y^-}
    e^{ i \Deltat \vdot \frac{1}{2}(\xt+\yt) + i   \xi P^+ (x^- + y^-)  
-ixP^+ (x^- - y^-) + i \kt \vdot (\xt-\yt)} \\
&\times
\theta(x^-) \theta( - y^-)
 \int \frac{\dd[4]{q}\dd[4]{q'}}{(2\pi)^{8}} 
(-1) \Tr[
\gamma^+ \frac{i\slashed{q}}{q^2 + i \varepsilon}  \gamma^- \frac{i \slashed{q'}}{q'^2 + i\varepsilon}
]
e^{-i q^+ y^- + i \qt \vdot \yt + i q^{\prime +}  x^- - i\qt' \vdot \xt}
\\
&
\times 2\pi \delta( q^- - q^{\prime -})
\int \dd[2]{\ut} e^{-i (\qt - \qt')\vdot \ut} 
\Tr \expval{  -V^\dag(\ut) V_{C_{+-}}[\xt,\yt]+1 }_{\lambda \lambda'},
\end{split}
\end{equation}
where the extra minus sign comes from having a quark loop.
We have also denoted by $V_{C_{+-}}[\xt,\yt]$ a gauge link with only classical gluon fields that starts at $(y^-,\yt)$ and ends at $(x^-,\xt)$ with $y^- < 0 <x^-$.
Such gauge links contribute only when there is a crossing through the shockwave.
For example, $V_{\sqsupset_{+-}}[\xt,\yt] = V(\yt)$ and $V_{\sqsubset_{+-}}[\xt,\yt] = V(\xt)$, but even more complicated gauge link structures are possible.

Evaluating the trace and integrating over $x^-$ and $y^-$, we have:
\begin{equation}
\label{eq:quark_GTMD_fig1_2}
\begin{split}
    W_{ \ref{fig:quark_GTMD1}}^{[\gamma^+]}
    =&
    \frac{P^+}{(2\pi)^3} \sqrt{1-\xi^2} \int \dd[2]{\xt} \dd[2]{\yt} 
    e^{ i \Deltat \vdot \frac{1}{2}(\xt+\yt) + i \kt \vdot (\xt-\yt)} 
 \int \frac{\dd[4]{q}\dd[4]{q'}}{(2\pi)^{8}} 
    \\
&\times
 \frac{1}{
 \qty[(\xi+x) P^+ - q^+ - i\varepsilon]
 \qty[ (\xi-x)P^+ + q'^+ + i\varepsilon ]
 }
 \frac{4 \qt \vdot \qt'}{\qty[q^2 + i \varepsilon]\qty[q'^2 + i \varepsilon]}
e^{ i \qt \vdot \yt  - i\qt' \vdot \xt}
\\
&
\times 2\pi \delta( q^- - q^{\prime -})
\int \dd[2]{\ut} e^{-i (\qt - \qt')\vdot \ut} 
\Tr \expval{  -V^\dag(\ut)  V_{C_{+-}}[\xt,\yt]+1 }_{\lambda \lambda'},
\end{split}
\end{equation}
where we have used the distributional identities
\begin{align}
    \int_{-\infty}^0 \dd{x^-} e^{ i x^- p^+} &= \frac{-i}{p^+ - i\varepsilon},
    \\
    \int_{0}^\infty \dd{x^-} e^{ i x^- p^+} &= \frac{i}{p^+ + i\varepsilon},
\end{align}
with an infinitesimal $\varepsilon > 0$.
We can then integrate over $q^+$, $q'^+$, and $q'^-$:
\begin{equation}
\label{eq:quark_GTMD_fig1_3}
\begin{split}
 W_{\ref{fig:quark_GTMD1}}^{[\gamma^+]}
    =&
    \frac{P^+}{(2\pi)^3} \sqrt{1-\xi^2} \int \dd[2]{\xt} \dd[2]{\yt} 
    e^{ i \Deltat \vdot \frac{1}{2}(\xt+\yt) + i \kt \vdot (\xt-\yt)} 
 \int \frac{\dd[2]{\qt}\dd[2]{\qt'} \dd{q^-}}{(2\pi)^{5}} 
    \\
&\times
 \frac{\theta(-q^-)}{
 \qty[-2q^-(\xi+x) P^+ + \qt^2 - i\varepsilon]
 \qty[ 2q^- (\xi-x)P^+ + \qt'^2 - i\varepsilon ]
 }
 (4 \qt \vdot \qt')
e^{ i \qt \vdot \yt  - i\qt' \vdot \xt}
\\
& \times
\int \dd[2]{\ut} e^{-i (\qt - \qt')\vdot \ut} 
\Tr \expval{  -V^\dag(\ut)  V_{C_{+-}}[\xt,\yt]+1 }_{\lambda \lambda'}.
\end{split}
\end{equation}
At this point, it is easiest to perform the Fourier transforms over $\qt$ and $\qt'$ using~\cite{Hanninen:2017ddy}:
\begin{align}
\label{eq:K1_FT}
   \int \frac{\dd[2]{\qt}}{(2\pi)^2} \frac{\qt^i}{\qt^2 + Q^2} e^{i \qt \vdot \rt}
   = \frac{i}{2\pi} \frac{\rt^i}{\abs{\rt}} \times Q K_1(\abs{\rt}Q).
\end{align}
This gives us
\begin{equation}
\label{eq:quark_GTMD_fig1_4}
\begin{split}
    W_{ \ref{fig:quark_GTMD1}}^{[\gamma^+]}
    =&
    \frac{2}{(2\pi)^6} \sqrt{1-\xi^2} \int \dd[2]{\xt} \dd[2]{\yt}  \dd[2]{\ut}
    e^{ i \Deltat \vdot \frac{1}{2}(\xt+\yt) + i \kt \vdot (\xt-\yt)} 
    \\
&\times
\frac{( \yt -\ut ) \vdot (\xt -\ut)}{\abs{\yt -\ut} \abs{\xt -\ut}}
\Tr \expval{  -V^\dag(\ut)  V_{C_{+-}}[\xt,\yt]+1 }_{\lambda \lambda'}
\sqrt{ (\xi +x - i\varepsilon) (-\xi +x - i\varepsilon) } \\
&\times
\int_0^\infty  \dd{\eta} \eta
K_1\qty(\abs{\yt -\ut} \sqrt{ \eta (\xi +x - i\varepsilon) } )
K_1\qty(\abs{\xt -\ut} \sqrt{ \eta (-\xi +x - i\varepsilon) } )
\end{split}
\end{equation}
where we have also defined $\eta = -2 q^- P^+$.
The integral over $\eta$ can be done analytically
(see Eq.~(6.521-3) of Ref.~\cite{MR2360010} in the limit $\nu \to 1$):
\begin{equation}
    \label{eq:bessel_integral}
    \int_0^\infty \dd{\eta} \eta K_1(\sqrt{a \eta} ) K_1(\sqrt{b \eta} )
    = \frac{4}{\sqrt{ab} (a - b)^3} \qty[ a^2 - b^2 + 2ab \log(\frac{b}{a}) ],
\end{equation}
where $a$ and $b$ are complex numbers that are not on the negative real axis, to avoid the branch cuts in the Bessel functions and the logarithm.
This allows us to write:
\begin{equation}
\label{eq:quark_GTMD_fig1_5}
\begin{split}
    W_{\ref{fig:quark_GTMD1}}^{[\gamma^+]}
    =&
    \frac{8}{(2\pi)^6} \sqrt{1-\xi^2} \int \dd[2]{\xt} \dd[2]{\yt}  \dd[2]{\ut}
    e^{ i \Deltat \vdot \frac{1}{2}(\xt+\yt) + i \kt \vdot (\xt-\yt)} 
    \\
&\times
\frac{( \yt -\ut ) \vdot (\xt -\ut)}{\abs{\yt -\ut}^2 \abs{\xt -\ut}^2}
\Tr \expval{  -V^\dag(\ut)   V_{C_{+-}}[\xt,\yt]+1 }_{\lambda \lambda'} \\
&\times
F\qty( \abs{\yt -\ut}^2 \qty[\xi +x - i\varepsilon] , \abs{\xt -\ut}^2\qty[-\xi +x - i\varepsilon]  )
\end{split}
\end{equation}
where we have denoted
\begin{equation}
    F(a,b) =  \frac{1}{(a - b)^3} \qty[ a^2 - b^2 + 2ab \log(\frac{b}{a}) ].
\end{equation}

For the full quark GTMD we also need Diagram~\ref{fig:quark_GTMD2}:
\begin{equation}
\label{eq:quark_GTMD_fig2_1}
\begin{split}
    W_{ \ref{fig:quark_GTMD2}}^{[\gamma^+]}
    &=
    \frac{P^+}{(2\pi)^3} \sqrt{1-\xi^2} \int \dd[2]{\xt} \dd{x^-} \dd[2]{\yt} \dd{y^-}
    e^{ i \Deltat \vdot \frac{1}{2}(\xt+\yt) + i   \xi P^+ (x^- + y^-)  
-ixP^+ (x^- - y^-) + i \kt \vdot (\xt-\yt)} \\
&\times
\theta(-x^-) \theta( y^-)
 \int \frac{\dd[4]{q}\dd[4]{q'}}{(2\pi)^{8}} 
(-1) \Tr[
\gamma^+ \frac{i\slashed{q}}{q^2 + i \varepsilon}  \gamma^- \frac{i \slashed{q'}}{q'^2 + i\varepsilon}
]
e^{-i q^+ y^- + i \qt \vdot \yt + i q^{\prime +}  x^- - i\qt' \vdot \xt}
\\
&
\times 2\pi \delta( q^- - q^{\prime -})
\int \dd[2]{\ut} e^{-i (\qt - \qt')\vdot \ut} 
\Tr \update{\expval{  V(\ut)  V_{C_{-+}}[\xt,\yt] - 1 }_{\lambda \lambda'}}.
\end{split}
\end{equation}
We can proceed similarly to Diagram~\ref{fig:quark_GTMD1}, leading to
\begin{equation}
\label{eq:quark_GTMD_fig2_2}
\begin{split}
    W_{ \ref{fig:quark_GTMD2}}^{[\gamma^+]}
    =&
    \frac{8}{(2\pi)^6} \sqrt{1-\xi^2} \int \dd[2]{\xt} \dd[2]{\yt}  \dd[2]{\ut}
    e^{ i \Deltat \vdot \frac{1}{2}(\xt+\yt) + i \kt \vdot (\xt-\yt)} 
    \\
&\times
\frac{( \yt -\ut ) \vdot (\xt -\ut)}{\abs{\yt -\ut}^2 \abs{\xt -\ut}^2}
\Tr \expval{  -V(\ut) V_{C_{-+}}[\xt,\yt]+1 }_{\lambda \lambda'} \\
&\times
F\qty( \abs{\yt -\ut}^2 \qty[\xi +x + i\varepsilon] , \abs{\xt -\ut}^2\qty[-\xi +x + i\varepsilon]  ).
\end{split}
\end{equation}

\begin{figure}
	\centering
        \begin{overpic}[width=0.5\linewidth]{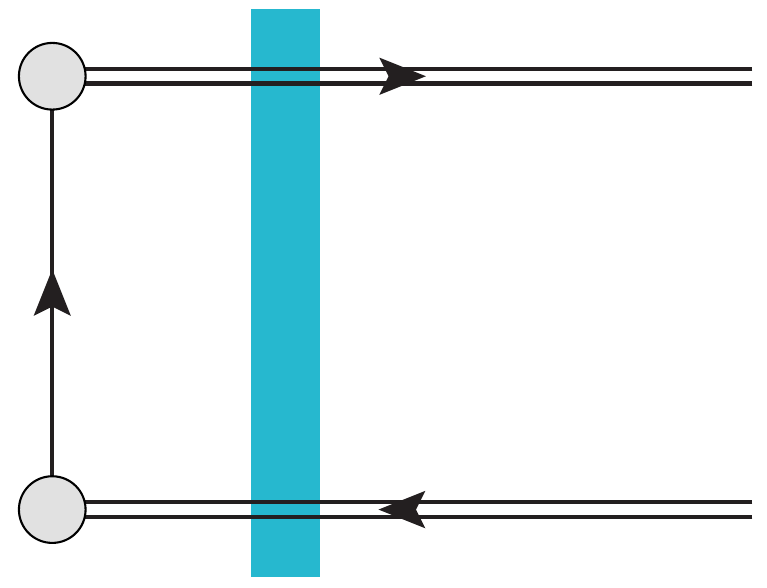}
            \put(-3,65){$y$}
            \put(-3,9){$x$}
        \end{overpic}
    \caption{An example contribution to the quark GTMD that evaluates to zero.
    }
        \label{fig:quark_GTMD3}
\end{figure}

By considering different possibilities for the location of the shockwave, we would seemingly also need to evaluate the diagram shown in Fig.~\ref{fig:quark_GTMD3}.
However, as can be checked by a direct calculation, the contribution from this diagram is identically zero.
\update{By following our Feynman rules, the contribution from this diagram can be written as:
\begin{equation}
\label{eq:quark_GTMD_fig3_1}
\begin{split}
    W_{\ref{fig:quark_GTMD3}}^{[\gamma^+]}
    &=
    \frac{P^+}{(2\pi)^3} \sqrt{1-\xi^2} \int \dd[2]{\xt} \dd{x^-} \dd[2]{\yt} \dd{y^-}
    e^{ i \Deltat \vdot \frac{1}{2}(\xt+\yt) + i   \xi P^+ (x^- + y^-)  
-ixP^+ (x^- - y^-) + i \kt \vdot (\xt-\yt)} \\
&\times
\theta(-x^-) \theta( - y^-)
 \int \frac{\dd[4]{q}}{(2\pi)^{4}} 
(-1) \Tr[
\gamma^+ \frac{i\slashed{q}}{q^2 + i \varepsilon}
]
e^{-i q^+ (y^-  - x^-) + i \qt \vdot (\yt-\xt) }
\\
&
\times
\Tr \expval{  V_{C_{--}}[\xt,\yt]-1 }_{\lambda \lambda'}.
\end{split}
\end{equation}
Here, we encounter the following integrals over the momentum $q$:
\begin{equation}
\begin{split}
 &   \int \frac{\dd[2]{\qt} \dd{q^-}}{(2\pi)^3}
    \Tr[
\gamma^+ \frac{i\slashed{q}}{q^2 + i \varepsilon}]
e^{ i \qt \vdot (\yt-\xt) }
=
    \int \frac{\dd[2]{\qt} \dd{q^-}}{(2\pi)^3}
\frac{4 i q^+}{q^2 + i \varepsilon}
e^{ i \qt \vdot (\yt-\xt) }
\\
=&
\text{sgn}( q^+)
    \int \frac{\dd[2]{\qt} }{(2\pi)^2}
e^{ i \qt \vdot (\yt-\xt) }
  =  
\text{sgn}( q^+)
\delta^{(2)}(\xt-\yt).
\end{split}
\end{equation}
The delta function $
\delta^{(2)}(\xt-\yt)$ then allows us to simplify the structure as
$\Tr( 1 - V_{C--}[\xt,\xt]  ) = \Tr( 1 - 1 ) = 0$ due to the unitarity of the Wilson lines, and the contribution from Fig.~\ref{fig:quark_GTMD3} vanishes.
}
This means that only the two diagrams shown in Fig.~\ref{fig:quark_GTMD} are needed for the quark GTMD.
The full expression is then given by:
\begin{equation}
\label{eq:quark_GTMD_final}
\begin{split}
    W_{}^{[\gamma^+]}
    =&
    \frac{8}{(2\pi)^6} \sqrt{1-\xi^2} \int \dd[2]{\xt} \dd[2]{\yt}  \dd[2]{\ut}
    e^{ i \Deltat \vdot \frac{1}{2}(\xt+\yt) + i \kt \vdot (\xt-\yt)} 
\frac{( \yt -\ut ) \vdot (\xt -\ut)}{\abs{\yt -\ut}^2 \abs{\xt -\ut}^2}
\\
\times
\Bigg\{&
\Tr \expval{  -V^\dag(\ut)   V_{C_{+-}}[\xt,\yt]+1 }_{\lambda \lambda'}
F\qty( \abs{\yt -\ut}^2 \qty[\xi +x - i\varepsilon] , \abs{\xt -\ut}^2\qty[-\xi +x - i\varepsilon]  )
\\
+&
\Tr \expval{  -V(\ut) V_{C_{-+}}[\xt,\yt]+1 }_{\lambda \lambda'} 
F\qty( \abs{\yt -\ut}^2 \qty[\xi +x + i\varepsilon] , \abs{\xt -\ut}^2\qty[-\xi +x + i\varepsilon]  )
\Bigg\}
.
\end{split}
\end{equation}
The special cases $C=\sqsupset$ and $C=\sqsubset$ can be read directly from this:
\begin{equation}
\label{eq:quark_GTMD_future}
\begin{split}
    W_{}^{[\gamma^+]}(\sqsupset)
    =&
    \frac{8 \nc}{(2\pi)^6} \sqrt{1-\xi^2} \int \dd[2]{\xt} \dd[2]{\yt}  \dd[2]{\ut}
    e^{ i \Deltat \vdot \frac{1}{2}(\xt+\yt) + i \kt \vdot (\xt-\yt)} 
\frac{( \yt -\ut ) \vdot (\xt -\ut)}{\abs{\yt -\ut}^2 \abs{\xt -\ut}^2}
\\
\times
\Bigg\{
&
N_{\lambda \lambda'}(\yt,\ut)
F\qty( \abs{\yt -\ut}^2 \qty[\xi +x - i\varepsilon] , \abs{\xt -\ut}^2\qty[-\xi +x - i\varepsilon]  )
\\
+&
N_{\lambda \lambda'}(\ut,\xt)
F\qty( \abs{\yt -\ut}^2 \qty[\xi +x + i\varepsilon] , \abs{\xt -\ut}^2\qty[-\xi +x + i\varepsilon]  )
\Bigg\}
,
\end{split}
\end{equation}
\begin{equation}
\label{eq:quark_GTMD_past}
\begin{split}
    W_{}^{[\gamma^+]}(\sqsubset)
    =&
    \frac{8 \nc}{(2\pi)^6} \sqrt{1-\xi^2} \int \dd[2]{\xt} \dd[2]{\yt}  \dd[2]{\ut}
    e^{ i \Deltat \vdot \frac{1}{2}(\xt+\yt) + i \kt \vdot (\xt-\yt)} 
\frac{( \yt -\ut ) \vdot (\xt -\ut)}{\abs{\yt -\ut}^2 \abs{\xt -\ut}^2}
\\
\times
\Bigg\{
&
N_{\lambda \lambda'}(\xt,\ut)
F\qty( \abs{\yt -\ut}^2 \qty[\xi +x - i\varepsilon] , \abs{\xt -\ut}^2\qty[-\xi +x - i\varepsilon]  )
\\
+&
N_{\lambda \lambda'}(\ut,\yt)
F\qty( \abs{\yt -\ut}^2 \qty[\xi +x + i\varepsilon] , \abs{\xt -\ut}^2\qty[-\xi +x + i\varepsilon]  )
\Bigg\}
.
\end{split}
\end{equation}
To our knowledge, these expressions for the quark GTMD are completely new and have not been considered before.

\section{Generalized parton distributions}
\label{sec:GPD}

Moving on to GPDs, we recall that experimental observables in deeply virtual exclusive processes are sensitive only to certain integrals of GPDs~\cite{Lorce:2025aqp}, rather than to the full distributions themselves. This limitation highlights the need for complementary theoretical approaches that can provide access to their complete structure~\cite{Qiu:2022pla,Qiu:2023mrm}. In this work, we take a step in this direction by computing GPDs within the shockwave formalism. This establishes a novel framework for investigating their detailed properties and yields results that complement ongoing efforts in both experiment and lattice QCD.

\subsection{Gluon}

We can get the relevant gluon GPD from the GTMD by integrating over $\kt$~\cite{Diehl:2003ny}:
\begin{equation}
    \begin{split}
    F^g =&
    \int \dd[D-2]{\kt} 
   W^{ii}_{\lambda \lambda'}(P,x,\kt, \Delta)  \\
   =& 
   \frac{2}{P^+} \int \frac{\dd{r^-} }{2\pi} 
    e^{ ixP^+ r^-}
    \Tr
    \mel{p'}{
    \mathcal{T}
    G^{+ i} \qty(-\frac{r}{2})
    \mathcal{W}\qty[- \frac{r}{2},\frac{r}{2}] 
    G^{+ i} \qty(\frac{r}{2})
    \mathcal{W}\qty[ \frac{r}{2},-\frac{r}{2}] 
    }{p} 
    \bigg\lvert_{\rt=\mathbf{0}}\\
    =&
   \frac{2 }{\pi} 
\sqrt{1 - \xi^2 }
    \int \dd{x^-} \dd[D-2]{\xt}
 \dd{y^-}
 e^{ i \Deltat \vdot \frac{1}{2}(\xt+\yt) + i   \xi P^+ (x^- + y^-)  
-ixP^+ (x^- - y^-)} \\
&\times
\Tr
\expval{
\mathcal{T}
G^{+ i} (x) 
\mathcal{W}\qty[x,y]
G^{+ i}(y)
\mathcal{W}\qty[y,x]
}_{\lambda  \lambda'}
\bigg\lvert_{\yt = \xt}
    \end{split}
\end{equation}
where we have written this in $D=4-2\varepsilon$ dimensions in anticipation of the quark GPD calculation in Sec.~\ref{sec:quark_GPD}.
Because the transverse coordinates of the external fields are the same ($\xt = \yt$), the path of the gauge link does not matter.
In practice, we will take it to be a straight line along the minus direction connecting $x^-$ and $y^-$.
We can simply read the result from the dipole gluon GTMD:
\begin{equation}
\label{eq:gluon_GPD}
    \begin{split}
    F^g =&
   \frac{2 }{\as (2\pi)^2} 
    \qty(\frac{4\pi e^{-\gamma_E}}{\mu^2})^\varepsilon
\sqrt{1 - \xi^2 }
    \int  \dd[D-2]{\xt}
e^{ i \Deltat \vdot \xt}
  \Tr
    \expval{\partial^i V(\xt)
    \partial^i V^\dag(\xt)
    }_{\lambda \lambda'} \\
    =&
    -\frac{2 \nc }{\as } 
    \qty(\frac{ e^{-\gamma_E}}{\pi \mu^2})^\varepsilon
\sqrt{1 - \xi^2 }
\int \dd[D-2]{\kt}
\qty(\frac{1}{4}\Deltat^2 - \kt^2)
\nabla^2_\kt
    \widetilde N_{\lambda \lambda'}\qty(
\kt + \frac{1}{2}\Deltat
,-\kt + \frac{1}{2}\Deltat
    )    \\
    =&
    \frac{4 \nc (D-2) }{\as} 
    \qty(\frac{e^{-\gamma_E}}{\pi \mu^2})^\varepsilon
\sqrt{1 - \xi^2 }
\int \dd[D-2]{\kt}
    \widetilde N_{\lambda \lambda'}\qty(
\kt + \frac{1}{2}\Deltat
,-\kt + \frac{1}{2}\Deltat
    ) 
    \end{split}
\end{equation}
where going to the last line we have used the fact that the dipole amplitude vanishes for zero dipole sizes, $N_{\lambda \lambda'}(\xt,\xt)=0$.
We have also included the $\msbar$ factor by
substituting $g_s^2\to g_s^2 \times (\mu^2/[4\pi  e^{-\gamma_E}])^{\varepsilon} $.

\subsection{Quark}
\label{sec:quark_GPD}

\begin{figure}
	\centering
    \begin{subfigure}{0.3\textwidth}
        \centering
        \begin{overpic}[width=\linewidth]{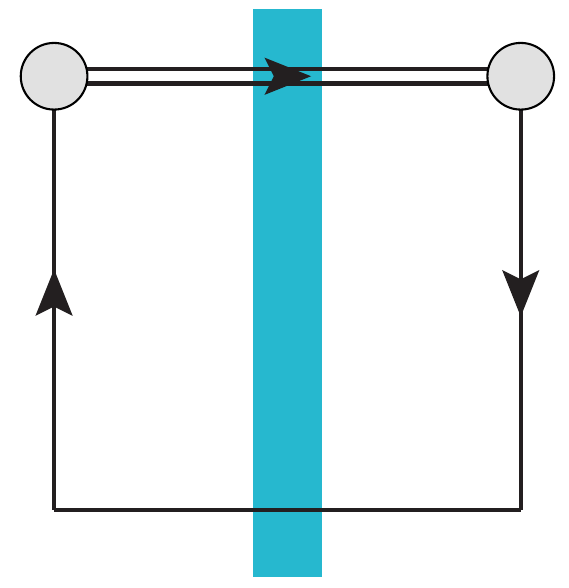}
            \put(17,50){$q$}
            \put(46,18){$\ut$}
            \put(3,97){$(y^-,\xt)$}
            \put(76,50){$q'$}
            \put(70,97){$(x^-,\xt)$}
        \put(14,40){\vector(0,1){20}}
        \put(83,60){\vector(0,-1){20}}
        \end{overpic}
        \caption{ }
        \label{fig:quark_GPD1}
    \end{subfigure}
    \begin{subfigure}{0.3\textwidth}
        \centering
        \begin{overpic}[width=\linewidth]{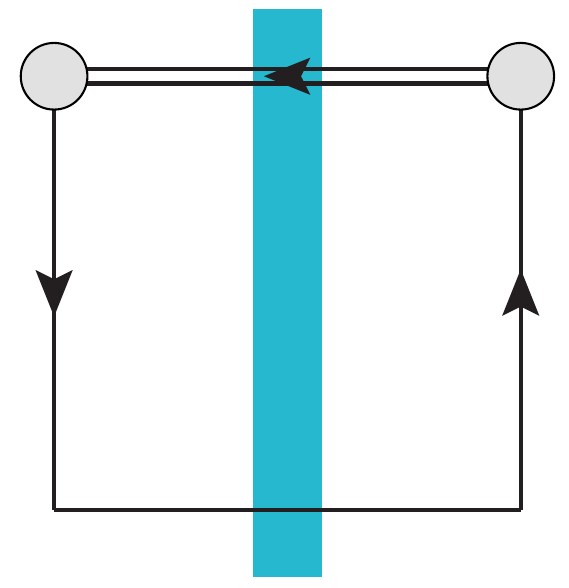}
            \put(17,50){$q'$}
            \put(46,18){$\ut$}
            \put(7,97){$(x^-,\xt)$}
            \put(76,50){$q$}
            \put(77,97){$(y^-,\xt)$}
        \put(14,60){\vector(0,-1){20}}
        \put(83,40){\vector(0,1){20}}
        \end{overpic}
        \caption{}
        \label{fig:quark_GPD2}
    \end{subfigure}
    \caption{Feynman diagrams for computing the quark GPD.
    }
    \label{fig:quark_GPD}
\end{figure}

Similarly to the gluon case, 
the relevant quark GPD can be obtained from the GTMD by integrating over $\kt$~\cite{Diehl:2003ny}:
\begin{equation}
    \begin{split}
    F^q =&
    \int \dd[2]{\kt} 
   W^{[\gamma^+]}_{\lambda \lambda'}(P,x,\kt, \Delta;C)  \\
   =& 
   \frac{1}{2} \int \frac{\dd{r^-}}{2\pi} 
    e^{ ixP^+ r^- }
    \mel{p'}{
    \mathcal{T}
    \bar \psi \qty(-\frac{r}{2}) \gamma^+ \mathcal{W}\qty[- \frac{r}{2},\frac{r}{2}] \psi \qty(\frac{r}{2})}{p} 
    \bigg\lvert_{\rt=\mathbf{0}}\\
    =&
   \frac{P^+ }{2\pi} 
\sqrt{1 - \xi^2 }
    \int \dd{x^-} \dd[2]{\xt}
 \dd{y^-} \\
&\times e^{ i \Deltat \vdot \xt + i   \xi P^+ (x^- + y^-)  
-ixP^+ (x^- - y^-) }
\expval{
\mathcal{T}
\bar \psi  (x) \gamma^+ \mathcal{W}\qty[x,y] \psi(y)}_{\lambda  \lambda'}
\bigg\lvert_{\yt = \xt}.
    \end{split}
\end{equation}
As in the case of  gluon GPD, the path of the gauge link does not matter, and we can choose it to be a straight line along the minus direction.

It turns out that in this case the GPD will develop a UV divergence due to the same transverse coordinate $\xt =\yt$ in the quark field operators.
This divergence will be related to the GPD evolution as we will discuss later.
Because of this divergence, we need to work in $D$ dimensions, and we cannot directly read the result from the quark GTMD.
To evaluate the quark GPD, we need  to consider the diagrams shown in Fig.~\ref{fig:quark_GPD}.
The first one of these gives the expression:
\begin{equation}
\label{eq:quark_GPD0}
\begin{split}
    F_{\ref{fig:quark_GPD1}}^q =& 
    \sqrt{1-\xi^2}
    {\frac{P^+}{2\pi} \int 
    \dd{x^-} \dd{y^-}\dd[D-2]{\xt}
    e^{-ixP^+ (x^- -y^-)  +i \Deltat \vdot \xt +i \xi P^+ (x^- + y^-)}}
  {  \theta(- y^-) \theta(x^- )}
    \\
 &\times
 {
 \int \frac{\dd[D]{q} \dd[D]{q'}}{(2\pi)^{2D}}}
 {(-1)}
 {\Tr[
{ 
 \frac{i \slashed{q}}{q^2 + i\varepsilon}}
 {\gamma^-} 
 {\frac{i \slashed{q'}}{q'^2 + i\varepsilon}}
{ \gamma^+}]}
{ e^{i \xt \vdot (\qt -\qt') - i y^- q^+ + i x^- q'^+ }}
 \\
&\times 
{
\int \dd[D-2]{\ut}
 (2\pi)  \delta(q^- - q^{\prime -}) e^{-i (\qt -\qt') \vdot \ut}}
 \Tr \expval{{-} {V(\xt)} {V^\dag(\ut)} {+ 1} }_{\lambda \lambda'}.
\end{split}
\end{equation}
Note that there is again a minus sign from the quark loop.

Evaluating the Dirac trace and integrating over
 $x^-$ and $y^-$, we get:
\begin{equation}
\label{eq:quark_GPD1}
    \begin{split}
          F_{\ref{fig:quark_GPD1}}^q =&
         \sqrt{1-\xi^2}  \frac{P^+}{\pi}
           \int \dd[D-2]{\xt} \dd[D-2]{\ut}
           \int \frac{\dd[D]{q}\dd[D]{q'}}{(2\pi)^{2D}}
           e^{i \Deltat \vdot \xt + i(\xt-\ut) \vdot(\qt-\qt') }\\
           &\times 
           (2\pi) \delta(q^- - q^{\prime -})
 \Tr \expval{{-} {V(\xt)} {V^\dag(\ut)} {+ 1} }_{\lambda \lambda'} \\
           &\times
           \frac{2\qt \vdot \qt'}{(q^2 + i \varepsilon)(q^{\prime 2} + i\varepsilon)}
           \frac{1}{
           (-xP^+ + \xi P^+ + q^{\prime +} + i\varepsilon )
           (xP^+ + \xi P^+ - q^{+} - i\varepsilon )
           }.
    \end{split}
\end{equation}
We can also evalute the integrals over $q^{\prime -}$, $q^{\prime +}$, and $q^{+}$ directly:
\begin{equation}
\label{eq:quark_GPD2}
    \begin{split}
          F_{\ref{fig:quark_GPD1}}^q =&
         \sqrt{1-\xi^2}  
           \frac{P^+}{\pi}
           \int \dd[D-2]{\xt} \dd[D-2]{\ut}
           \int \frac{\dd[D-2]{\qt}\dd[D-2]{\qt'} \dd{q^-}}{(2\pi)^{2D - 3}}
           \\
           &\times
           e^{i \Deltat \vdot \xt + i(\xt-\ut) \vdot(\qt-\qt') }
 \Tr \expval{{-} {V(\xt)} {V^\dag(\ut)} {+ 1} }_{\lambda \lambda'}\\
           &\times
           \frac{2\qt \vdot \qt' \theta(- q^-)}{
           \qty[2q^- ( xP^+ -\xi P^+ -i\varepsilon ) -\qt^{\prime 2} + i \varepsilon]
           \qty[2 q^- ( xP^+ + \xi P^+ -i\varepsilon  ) -\qt^2 + i\varepsilon]}.
    \end{split}
\end{equation}
The rest of the integrals are more difficult.
We will use the Schwinger parametrization to massage them into a nicer form:
\begin{equation}
\begin{split}
&\frac{1}{
\qty[2q^- ( xP^+ -\xi P^+ -i\varepsilon ) -\qt^{\prime 2} + i \varepsilon]
\qty[2 q^- ( xP^+ + \xi P^+ -i\varepsilon  ) -\qt^2 + i\varepsilon]}\\
=&
-\int_0^\infty \dd{t} \dd{u}
e^{
it  \qty[2q^- ( xP^+ -\xi P^+ -i\varepsilon ) -\qt^{\prime 2} + i \varepsilon])
+iu  \qty[2 q^- ( xP^+ + \xi P^+ -i\varepsilon  ) -\qt^2 + i\varepsilon]
}    ,
\end{split}
\end{equation}
which allows us to do the integrals over $q^-$, $\qt$ and $\qt'$:
\begin{equation}
    \int_{-\infty}^0 \dd{q^-}
    e^{ 2i q^- P^+ [ t (x -\xi -i\varepsilon) + u (x+\xi -i\varepsilon) ] }
    = \frac{-i}{2 P^+ [ t (x -\xi -i\varepsilon) + u (x+\xi -i\varepsilon) ] },
\end{equation}
\begin{equation}
   \int \frac{\dd[D-2]{\qt}}{(2\pi)^{D-2}}
    \qt^i
    e^{i (\xt -\ut) \vdot \qt - i u \qt^2  }
    = \frac{1}{(4\pi i u)^{(D-2)/2}}
    \frac{i (\xt -\ut)^i}{2 i u} \exp[ - (\xt -\ut)^2 /( 4iu ) ],
\end{equation}
\begin{equation}
    \int\frac{\dd[D-2]{\qt'}}{(2\pi)^{D-2}}
    \qt^{\prime i}
    e^{- i (\xt -\ut) \vdot \qt' - i t \qt^{\prime 2}  }
    = \frac{1}{(4\pi i t)^{(D-2)/2}}
    \frac{- i (\xt -\ut)^i}{2 i t} \exp[ - (\xt -\ut)^2 /( 4it ) ].
\end{equation}
Using these equations, the result can be written as
\begin{equation}
\label{eq:quark_GPD3}
    \begin{split}
          F_{\ref{fig:quark_GPD1}}^q =&
         \sqrt{1-\xi^2}  
           \frac{2}{(2\pi)^2}
           \int \dd[D-2]{\xt} \dd[D-2]{\ut}
           e^{i \Deltat \vdot \xt  }
 \Tr \expval{{-} {V(\xt)} {V^\dag(\ut)} {+ 1} }_{\lambda \lambda'}\\
           &\times
           \int_0^\infty \dd{t} \dd{u}
           \frac{i}{ t (x -\xi -i\varepsilon) + u (x+\xi -i\varepsilon)}\\
           &\times 
           \frac{1}{\qty[(4\pi)^2 i^2 tu]^{(D-2)/2}  }
           \frac{(\xt-\ut)^2}{4i^2 tu}
           \exp( - \frac{(\xt-\ut)^2}{4i} \qty[\frac{1}{t} + \frac{1}{u}] )\\
           =& 
            \sqrt{1-\xi^2}  
           \frac{2}{(2\pi)^2}
           \int \dd[D-2]{\xt} \dd[D-2]{\ut}
           e^{i \Deltat \vdot \xt  }
           \Tr[ -V(\xt) V^\dag(\ut) + 1] \\
           &\times
           \frac{1}{\qty[ \pi(\xt-\ut)^2]^{D-2}} 
           \int_0^\infty \frac{\dd{t} \dd{u}}{\qty[tu]^{D/2}}
           \frac{1}{ t (x -\xi -i\varepsilon) + u (x+\xi -i\varepsilon)}
           \exp( - \qty[\frac{1}{t} + \frac{1}{u}] ).
    \end{split}
\end{equation}
The final integrals over $t$ and $u$ are done using the following identity, which is valid for general complex numbers $a$ and $b$ that are not on the negative real axis: 
\begin{equation}
\begin{split}
    &\int_0^\infty \frac{\dd{t}\dd{u}}{\qty[tu]^\nu} \frac{1}{at +bu} 
           \exp( - \qty[\frac{1}{t} + \frac{1}{u}] )
    =
    \int_0^\infty \frac{\dd{t}\dd{w}}{t^{2\nu}\qty[w]^\nu} \frac{1}{a +bw} 
    \exp( -\frac{1}{t}  \frac{w+1}{w} ) \\
    =&
    \Gamma(2\nu -1)
    \int_0^\infty \dd{w} \frac{w^{\nu-1}}{\qty(a +bw) \qty(w+1)^{2\nu -1}} 
    =
    \frac{\Gamma(\nu)^2}{a (2\nu -1)} \times {}_2 F_1\qty(1 ,\nu; 2\nu; 1- \frac{b}{a}),
\end{split}
\end{equation}
where in the last line we used Eq.~(3.197-1) of Ref.~\cite{MR2360010}.
Finally, we get:
\begin{equation}
\label{eq:quark_GPD4}
    \begin{split}
            F_{\ref{fig:quark_GPD1}}^q =&
         \sqrt{1-\xi^2}  
           \frac{2}{(2\pi)^2}
           \frac{\Gamma\qty( D/2 )^2}{D-1}
           \frac{1}{x-\xi-i\varepsilon}
           {}_2 F_1\qty(1 ,D/2; D; 1- \frac{x+\xi -i \varepsilon}{x-\xi-i\varepsilon})\\
           &\times
           \int \dd[D-2]{\xt} \dd[D-2]{\ut}
           e^{i \Deltat \vdot \xt  }
 \Tr \expval{{-} {V(\xt)} {V^\dag(\ut)} {+ 1} }_{\lambda \lambda'}
           \frac{1}{\qty[ \pi(\xt-\ut)^2]^{D-2}}\\
           =&   \sqrt{1-\xi^2}  
           \frac{2^{D-1} \nc}{(2\pi)^D}
           \frac{\Gamma\qty( D/2 )^2}{D-1}
         \frac{1}{x-i \varepsilon}
     {_2F_1}\qty(\frac{1}{2},1;\frac{D+1}{2}; \frac{\xi^2}{(x-i \varepsilon)^2})\\
           &\times
           \int \dd[D-2]{\xt} \dd[D-2]{\ut}
           e^{i \Deltat \vdot \xt  }
           \frac{N_{\lambda \lambda'}(\xt,\ut)}{ \abs{\xt-\ut}^{2(D-2)} },
    \end{split}
\end{equation}
where 
we used Eq.~(9.134-1) of Ref.~\cite{MR2360010}
to rewrite  the hypergeometric function as
\begin{equation}
    \frac{1}{x - \xi - i\varepsilon}  {_2F_1}\qty(1,D/2;D; 1- \frac{x+\xi - i\varepsilon}{x-\xi -i \varepsilon})
    = \frac{1}{x-i \varepsilon}
     {_2F_1}\qty(\frac{1}{2},1;\frac{D+1}{2}; \frac{\xi^2}{(x-i \varepsilon)^2}).
\end{equation}
To make the UV divergence explicit, it is instructive to write Eq.~\eqref{eq:quark_GPD4} in terms of the momentum-space dipole amplitude~\eqref{eq:dipole_amplitude_mom}:
\begin{equation}
\label{eq:GPD3}
\begin{split}
            F_{\ref{fig:quark_GPD1}}^q =&
         \sqrt{1-\xi^2}   
 \frac{2^{D-1} N_c}{(2\pi)^{4-D}} \frac{\Gamma(D/2)^2}{D-1}  \frac{1}{x-i \varepsilon}
     {_2F_1}\qty(\frac{1}{2},1;\frac{D+1}{2}; \frac{\xi^2}{(x-i \varepsilon)^2})\\
&\times  \frac{1}{(4\pi)^{(D-2)/2}}\frac{\Gamma \qty( -\frac{D-4}{2} )}{ \Gamma(D-3)}
 \times
 \int \dd[D-2]{\qt}  \widetilde N_{\lambda \lambda'}(\Deltat - \qt,\qt) 
  \qty(\frac{4}{\qt^2})^{-(D-4)/2} .
\end{split}
\end{equation}
Here we can see the divergence $\Gamma \qty(- \frac{D-4}{2}) = \Gamma(\varepsilon) = \frac{1}{\varepsilon} + \ldots$ that needs to be renormalized.

For the second diagram in Fig.~\ref{fig:quark_GPD}, we have:
\begin{equation}
\begin{split}
    F_{\ref{fig:quark_GPD2}}^q =&
    \sqrt{1-\xi^2}
    {\frac{P^+}{2\pi} \int 
    \dd{x^-} \dd{y^-}\dd[D-2]{\xt}
    e^{-ixP^+ (x^- -y^-)  +i \Deltat \vdot \xt +i \xi P^+ (x^- + y^-)}}
  {  \theta(y^-) \theta(-x^- )}
    \\
 &\times
 {
 \int \frac{\dd[D]{q} \dd[D]{q'}}{(2\pi)^{2D}}}
 {(-1)}
 {\Tr[
{ 
 \frac{i \slashed{q}}{q^2 + i\varepsilon}}
 {\gamma^-} 
 {\frac{i \slashed{q'}}{q'^2 + i\varepsilon}}
{ \gamma^+}]}
{ e^{i \xt \vdot (\qt -\qt') - i y^- q^{\prime +} + i x^- q^{+} }}
 \\
&\times 
{
\int \dd[D-2]{\ut}
 (2\pi)  \delta(q^- - q^{\prime -}) e^{-i (\qt -\qt') \vdot \ut}}
 \Tr \expval{{-} {V^\dag(\xt)} {V(\ut)} {+ 1} }_{\lambda \lambda'}.
\end{split}
\end{equation}
Comparing to Eq.~\eqref{eq:quark_GPD0}, we see that this is just the complex conjugate of Diagram~\ref{fig:quark_GPD1}.
Summing the two contributions together, the quark GPD is given by
\begin{equation}
\label{eq:quark_GPD_bare}
\begin{split}
    F^q =&
    \sqrt{1-\xi^2}
 \frac{2 N_c}{ \pi} \frac{\Gamma(D/2)^2}{D-1} 
\frac{\Gamma \qty( -\frac{D-4}{2} )}{ \Gamma(D-3)}
 \int \dd[D-2]{\qt}
  \qty(\frac{1}{\pi \qt^2})^{-(D-4)/2}
 \\
 \times
 \Biggl\{ 
 &
 \frac{\widetilde N_{\lambda \lambda'}(\Deltat + \qt,-\qt)}{x - i\varepsilon}
   {_2F_1}\qty(\frac{1}{2},1;\frac{D+1}{2}; \frac{\xi^2}{(x-i \varepsilon)^2})\\
 +&
 \frac{\widetilde N_{\lambda \lambda'}( \qt,\Deltat -\qt)}{x + i\varepsilon}
   {_2F_1}\qty(\frac{1}{2},1;\frac{D+1}{2}; \frac{\xi^2}{(x+i \varepsilon)^2})
   \Bigg\}
\\
  =& \frac{\nc}{ 2\pi} \frac{1}{\varepsilon}
  \times \frac{2\update{\sqrt{1-\xi^2}}}{\xi^3} \qty[ 2 x \xi + (\xi^2-x^2) \log( \abs{ \frac{x+\xi}{x - \xi } } ) ]
   \int \dd[2]{\qt}  \widetilde N_{\lambda \lambda'}(\Deltat + \qt,-\qt) + ...
\end{split}
\end{equation}
where we have expanded in $\varepsilon$ in the final line to show explicitly the UV pole.
The pole has to be renormalized, which can be achieved by introducing a counterterm~\cite{Ji:1997nk,Belitsky:1997rh,Mankiewicz:1997bk,Boussarie:2023xun,Moffat:2023svr}:
\begin{equation}
\label{eq:quark_GPD_renormalization}
    F^q( \Deltat,x, \xi, \mu_R)
    =  F^q( \Deltat,x, \xi) 
    - \frac{\as}{2\pi} \frac{1}{\varepsilon}  \qty(\frac{\mu^2}{\mu_R^2})^\varepsilon \int_{-\infty}^{\infty} \frac{\dd{y}}{y^2} K^{q g}(x,y; \xi) \times F^g( \Deltat,y, \xi),
\end{equation}
where $\mu_R^2$ is the renormalization scale and $K^{qg}$ corresponds to the GPD evolution kernel for the $g \to q$ channel. 
This is the only channel we need at this accuracy, which follows from the $1/\as$ enhancement of the gluon GPD in Eq.~\eqref{eq:gluon_GPD}.

We note that typically the $y$-integral in Eq.~\eqref{eq:quark_GPD_renormalization} is restricted to the interval $[-1,1]$ due to momentum conservation.
However, in the strict shockwave approximation we lose this information when writing the target in the position space, and for this reason momentum conservation should be considered a subeikonal effect that we do not want to impose here.
Another way to motivate this choice is to note that here $x$ will be integrated over at the cross-section level, as opposed to the skewness $\xi$  which has a fixed value that depends on the kinematics of the process.
For the shockwave approximation to be valid, we will in practice need $\xi$ to be small as discussed in Sec.~\ref{sec:theory}.
The variable $x$ appears as $x/\xi$ in our expressions, and as such $\xi$ also sets a characteristic scale for $x$.
We are thus only sensitive to values $x \sim \xi \ll 1$, meaning that the contribution from the non-physical values $\abs{x} > 1$ is expected to be small.
Any sensitivity related to imposing the momentum conservation constraint $-1 \leq x \leq 1$ would then be an indication of a breakdown of the shockwave approximation.

Let us now evaluate the counterterm in Eq.~\eqref{eq:quark_GPD_renormalization}.
First of all, the kernel $K^{qg}$ is given by~\cite{Ji:1997nk,Belitsky:1997rh,Mankiewicz:1997bk,Moffat:2023svr}
\begin{equation}
\begin{split}
    K^{qg}(x,y; \xi)
    =& P_{qg}\qty( \frac{x}{y}, \frac{\xi}{y} )  \qty[ \theta(\xi < x< y) - \theta(  y < x< -\xi) ]\\
    &
    +
    \frac{1}{2}
    \theta( -\xi < x < \xi )
    \qty[
    P'_{qg}\qty( \frac{x}{y}, \frac{\xi}{y} ) \theta(x<y)
     -P'_{qg}\qty( \frac{x}{y}, -\frac{\xi}{y} ) \theta(y<x)
     ],
\end{split}
\end{equation}
where the splitting functions are:
\begin{align}
    P_{qg}\qty( \hat x, \hat \xi ) &=
    T_F \frac{\hat x^2 +  (1-\hat x)^2 - \hat \xi^2}{ (1-\hat \xi^2)^2},
    \\
    P'_{qg}\qty( \hat x, \hat \xi ) &=
    T_F \frac{(\hat x + \hat \xi) ( 1 -2 \hat x + \hat \xi)}{\hat \xi (1 + \hat \xi) (1- \hat \xi^2)},
\end{align}
with $T_F=\frac{1}{2}$
and we use the generalized notation
\begin{equation}
    \theta(X) =
    \begin{cases}
        &1 \text{ if $X$ is true,}\\
        &0 \text{ if $X$ is false,}
    \end{cases}
\end{equation}
for the theta functions.
These splitting functions differ from the usual definition by a factor $2 n_f$ as we only consider a single flavor and we do not include antiquarks yet.
As the gluon GPD $F^g( \Deltat,y, \xi, \mu_R)$ does not depend on $y$ at leading order, the $y$-integral is simply given by
\begin{equation}
    \int_{-\infty}^\infty \frac{\dd{y}}{\updatetwo{y^2}}  K^{qg}(x,y; \xi)
    =\frac{T_F}{2\xi^3} \qty[ 2x \xi + (\xi^2 -x^2) \log( \abs{\frac{\xi+x}{\xi-x}} ) ].
\end{equation}
We note that this cancels exactly the $\varepsilon$-pole in the quark GPD~\eqref{eq:quark_GPD_bare}, rendering the quark distribution finite.

The full renormalized quark GPD is then given by:
\update{
\begin{equation}
\label{eq:quark_GPD_final}
\begin{split}
   F^q(\Deltat , x ,\xi,\mu_R) 
&= \frac{  \sqrt{1-\xi^2}}{\xi}
 \frac{N_c}{2 \pi} 
 \\
 \times
 \int \dd[2]{\qt} 
 & \qty[\widetilde N_{\lambda \lambda'}(\Deltat + \qt,-\qt)
 \hat F^q\qty(\hat \xi_-,
\frac{\mu_R^2}{\qt^2}) 
 +
 \widetilde N_{\lambda \lambda'}( \qt,\Deltat-\qt)
 \hat F^q\qty(\hat \xi_+,
\frac{\mu_R^2}{\qt^2})]
\end{split}
\end{equation}
}
% \begin{equation}
% \label{eq:quark_GPD_final}
% \begin{split}
%   &  F^q(\Deltat , x ,\xi,\mu_R) 
% =\frac{  \sqrt{1-\xi^2}}{\xi}
%  \frac{N_c}{ \pi} 
%  \Re
%  \int \dd[2]{\qt} 
%  \frac{\widetilde N_{\lambda \lambda'}(\Deltat + \qt,-\qt)}{
%  \hat \xi_-^2
%  }
%  \\
%  &\times
%  \Biggl\{ 
%  \qty[
% \log(
% \frac{\mu_R^2}{\qt^2}
% )
% -\frac{1}{3}
%  ]
%   \qty[
%      2 \hat \xi_-
%      -(1-\hat \xi_-^2) \log(
%      \frac{1+\hat \xi_-}{1-\hat\xi_-}
%      )
%      ]
% +
% h(\hat \xi_-)
%    \Bigg\},
% \end{split}
% \end{equation}
\update{
where $\hat \xi_\pm = \xi/(x \pm i\varepsilon) $ and we have defined
\begin{equation}
    \hat F^q\qty(\hat \xi,
\frac{\mu_R^2}{\qt^2})
=
\frac{1}{\hat \xi^2}
\qty{
     \qty[
\log(
\frac{\mu_R^2}{\qt^2}
)
-\frac{1}{3}
 ]
  \qty[
     2 \hat \xi
     -(1-\hat \xi^2) \log(
     \frac{1+\hat \xi}{1-\hat\xi}
     )
     ]
+
h(\hat \xi)
}.
\end{equation}
}%
% and we have used the following expansion for the hypergeometric function evaluated with the \texttt{HypExp Mathematica }
We have used the following expansion for the hypergeometric function evaluated with the \texttt{HypExp Mathematica }package~\cite{Huber:2005yg,Huber:2007dx}:
\begin{equation}
\begin{split}
  &{_2F_1}\qty(\frac{1}{2},1;\frac{5}{2} - \varepsilon; \hat \xi^2)
 =
     \frac{3}{4 \hat \xi^3}
     \qty[
     2 \hat \xi
     -(1-\hat \xi^2) \log(
     \frac{1+\hat \xi}{1-\hat\xi}
     )
     + \varepsilon h(\hat \xi)
     \update{+\order{\varepsilon^2}}
     ],
\end{split}
\end{equation}
where
\begin{equation}
\begin{split}
 h(\hat \xi)
  =&
    \frac{2}{3} \hat \xi
     - (1 -\hat \xi^2 )
     \Biggl[
     2 \qty(
     \li(\hat \xi) - \li(-\hat \xi)
     )
        -\frac{5}{3} \log(\frac{1+\hat \xi}{1-\hat \xi})
     \\
     &
     +
     \li\qty(\frac{1-\hat \xi}{2})
     -\li\qty(\frac{1+\hat \xi}{2})
     +\frac{1}{2}\log^2 \qty(\frac{1-\hat \xi}{2})
     -\frac{1}{2}\log^2 \qty(\frac{1+\hat \xi}{2})
     \Biggr].
\end{split}
\end{equation}

To our knowledge this is the first time the full quark GPD, including the finite terms, has been derived in the shockwave limit.
Apart from the overall factor $\sqrt{1-\xi^2}/\xi$, its dependence on the $\xi$ and $x$ variables is fully given in terms of the combination $\xi/x \equiv  \hat \xi$.
Writing the quark GPD in terms of $\xi$ and $\hat \xi$, the dependence on the skewness $\xi$ is then similar to that of the gluon GPD~\eqref{eq:gluon_GPD}, noting that often the gluon GPD is defined with an additional factor $1/x = \hat \xi/\xi$~\cite{Ji:1998pc}.

The quark GPD also depends on the renormalization scale $\mu_R^2$, as expected after the renormalization.
The dependence on the renormalization scale is dictated by the GPD evolution, such that Eq.~\eqref{eq:quark_GPD_final} can be used as an initial condition to the evolution equation at some initial scale $\mu_R^2 = \mu_i^2$.
To minimize the logarithmic term in Eq.~\eqref{eq:quark_GPD_final},
the initial scale should be chosen as $\mu_i^2 \sim Q_s^2$, where $Q_s^2$ is the saturation scale corresponding to the characteristic momentum scale in the dipole amplitude.

\section{Tranverse-momentum-dependent and collinear parton distributions}
\label{sec:TMDPDF}

While the GPDs and GTMDs are defined at the amplitude level, the TMDs and PDFs are defined at the \textit{cross section} level.
Because of this, the relation between these different parton distributions is not immediately clear in the shockwave approximation.
As a cross-check, we will compute the TMDs and PDFs at the cross section level starting from their definition, and verify that the expected relations to the GTMDs and GPDs are valid.

Because we are now working at the cross section level, we will have final-state particles that need to be integrated over.
This can be done by summing over the helicities of the final-state particles and introducing the phase-space integral
\begin{equation}
    \int \frac{\dd[4]{p_i}}{(2\pi)^4} 2\pi \delta(p_i^2) \theta( p_i^- )
\end{equation}
for each particle $i$ in the final state.

\subsection{Gluon}

The gluon TMD is defined as~\cite{Boussarie:2023izj}:
\begin{equation}
\label{eq:gluon_TMD}
\begin{split}
  f^{ij}_{g,\lambda}(x,\kt;C) =&
   \frac{1}{x p^+} \int \frac{\dd{r^-} \dd[2]{\rt}}{(2\pi)^3} 
    e^{ ix p^+ r^- - i \kt \vdot \rt}
    \mel{p}{G_a^{+ i} \qty(-\frac{r}{2})\mathcal{W}^{ab}_C\qty[- \frac{r}{2},\frac{r}{2}] G_b^{+j} \qty(\frac{r}{2})}{p}^c
    \\
    =& 
    \frac{2 }{(2\pi)^3 x } 
    \int \dd{x^-} \dd[2]{\xt}
 \dd{y^-}\dd[2]{\yt} \\
&\times e^{   
-ix p^+ (x^- - y^-) + i \kt \vdot (\xt -\yt)}
\expval{G_a^{+ i} (x) \mathcal{W}^{ab}_C\qty[x,y]G_b^{+ j}(y)}^c_{\lambda  \lambda},
\end{split}
\end{equation}
where $c$ denotes that we have a ``cut'' in the final state, corresponding to evaluating this at the cross section level.
Similarly, the gluon PDF is defined as~\cite{Collins:2011zzd}:
\begin{equation}
    \begin{split}
    f^g(x) =&
   \frac{1}{x p^+} \int \frac{\dd{r^-}}{2\pi} 
    e^{ ixp^+ r^-}
    \frac{1}{2}
    \sum_\lambda
    \mel{p}{G_a^{+ i} \qty(-\frac{r}{2})  \mathcal{W}^{ab}_C\qty[- \frac{r}{2},\frac{r}{2}]G_b^{+ i} \qty(\frac{r}{2})}{p}^c
    \bigg\lvert_{\rt=\mathbf{0}}\\
    =&
   \frac{1}{x \pi} 
    \int \dd{x^-} \dd[2]{\xt}
 \dd{y^-} 
 e^{ 
-ix p^+ (x^- - y^-) }
    \frac{1}{2}
    \sum_\lambda
\expval{G_a^{+ i}  (x) \mathcal{W}^{ab}_C\qty[x,y] G_b^{+ i} (y)}^c_{\lambda  \lambda}
\bigg\lvert_{\yt = \xt}.
    \end{split}
\end{equation}
As in the GPD case, the path of the gauge link does not matter.

In the shockwave approximation, the leading-order gluon distribution comes from 
gluon-field operators inside the shockwave, and there are no final-state particles to integrate over.
Therefore, the derivation of the gluon distributions follows that of the GTMD and GPD, and we can directly read the results for the TMD and the PDF:
\begin{equation}
\label{eq:gluon_TMD_PDF}
\begin{split}
  f^{ij}_{g,\lambda }(x,\kt) =&
    \frac{1}{x} W^{ij}_{\lambda \lambda}(\Delta \to 0) \\
    =& 
   \frac{2 }{\as (2\pi)^4} 
   \frac{1}{x}
    \int  \dd[2]{\xt}\dd[2]{\yt} 
    e^{  i \kt \vdot (\xt-\yt)}
\Tr
    \expval{\partial^i V(\xt)
V_{C_{--}}\qty[x,y]
    \partial^j V^\dag(\yt)
V_{C_{++}'}\qty[y,x]
    }_{\lambda \lambda}
    ,
\end{split}
\end{equation}

    \begin{equation}
    \label{eq:gluon_PDF}
  f^g(x) =
    \frac{1}{2}
    \sum_\lambda
    \frac{1}{x} F^g(\Delta \to 0, \lambda' = \lambda)
    =
    \frac{4 \nc (D-2) }{\as}
    \frac{1}{x}
    \qty(\frac{e^{-\gamma_E}}{\pi \mu^2})^\varepsilon
\int \dd[D-2]{\kt}
    \frac{1}{2}
    \sum_\lambda
    \widetilde N_{\lambda \lambda}\qty(
\kt 
,-\kt 
    ) 
    .
    \end{equation}
These results for the gluon TMD and PDF are consistent with those found in the literature~\cite{Dominguez:2011wm,Kovchegov:2012mbw}.

\subsection{Quark TMD}

\begin{figure}
	\centering
\begin{equation*}
\begin{split}
&\qty(
    \begin{array}{l}    
\begin{overpic}[width=0.3\textwidth]{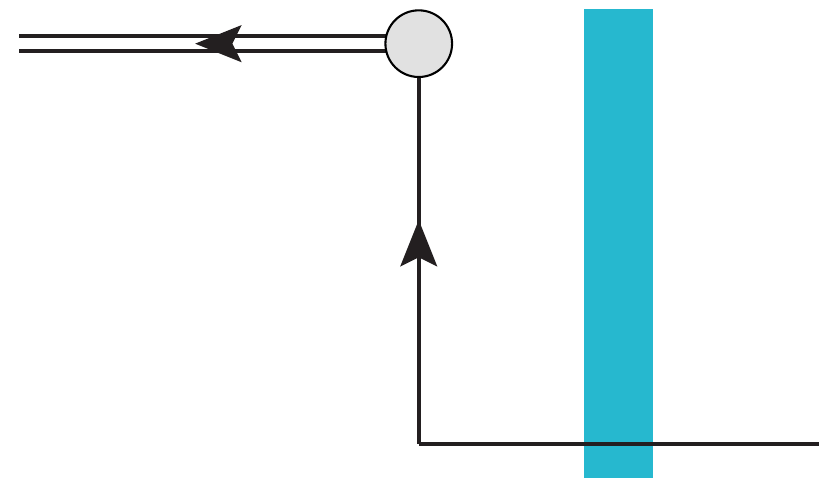}
    \put(48,62){$y$}
    \put(60,29){$q$}
    \put(88,15){$l$}
    \put(72,9){$\vt$}
        \put(57,21){\vector(0,1){16}}
        \put(80,10){\vector(1,0){16}}
\end{overpic}
\end{array}
+
    \begin{array}{l}    
\begin{overpic}[width=0.3\textwidth]{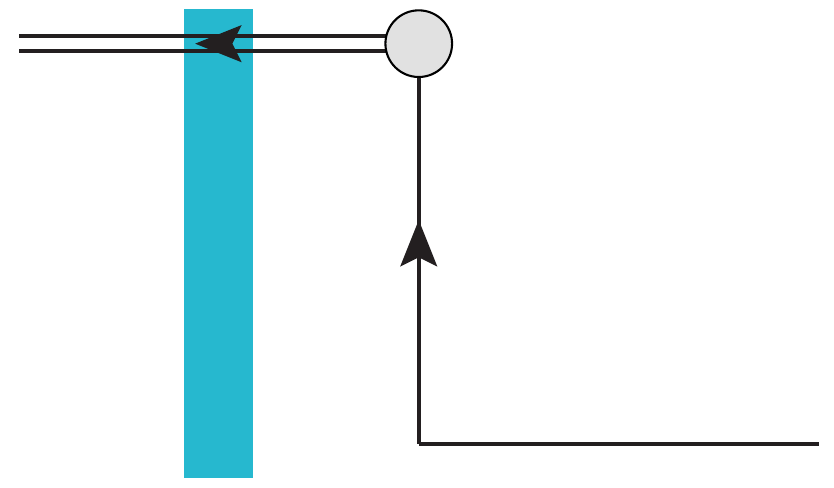}
    \put(48,62){$y$}
    \put(88,15){$l$}
        \put(80,10){\vector(1,0){16}}
\end{overpic}
\end{array}
)
\\
\times
&
\qty(
    \begin{array}{l}    
\begin{overpic}[width=0.3\textwidth]{figures/quark_TMD_past2.pdf}
    \put(48,62){$x$}
    \put(60,29){$q'$}
    \put(88,15){$l$}
    \put(72,9){$\ut$}
        \put(57,21){\vector(0,1){16}}
        \put(80,10){\vector(1,0){16}}
\end{overpic}
\end{array} 
+
    \begin{array}{l}    
\begin{overpic}[width=0.3\textwidth]{figures/quark_TMD_past1.pdf}
    \put(48,62){$x$}
    \put(88,15){$l$}
        \put(80,10){\vector(1,0){16}}
\end{overpic}
\end{array} 
)^*
\end{split}
\end{equation*}
    \caption{Feynman diagrams for computing the quark TMD.
    }
    \label{fig:quark_TMD}
\end{figure}

The relevant quark TMD in the shockwave limit is defined as~\cite{Boussarie:2023izj}:
\begin{equation}
\label{eq:quark_TMD}
\begin{split}
  f^{[\gamma^+]}_{q,\lambda}(x,\kt) =&
    \frac{1}{2} \int \frac{\dd{r^-} \dd[2]{\rt}}{(2\pi)^3} 
    e^{ ix p^+ r^- - i \kt \vdot \rt}
    \mel{p}{\bar \psi \qty(-\frac{r}{2})
    \gamma^+
    \mathcal{W}_C\qty[- \frac{r}{2},\frac{r}{2}]  \psi \qty(\frac{r}{2})}{p}^c\\
    =& 
    \frac{p^+ }{(2\pi)^3 } 
    \int \dd{x^-} \dd[2]{\xt}
 \dd{y^-}\dd[2]{\yt} 
 e^{   
-ix p^+ (x^- - y^-) + i \kt \vdot (\xt -\yt)}
\expval{\bar \psi(x) \gamma^+ \mathcal{W}_C\qty[x,y] \psi(y)}^c_{\lambda  \lambda}.
\end{split}
\end{equation}
Let us first consider a general gauge link.
We then need to evaluate the diagrams shown in Fig.~\ref{fig:quark_TMD}, where for illustration purposes we show the case $C=\sqsubset$.
The two diagrams for the amplitude (and the complex conjugate) can be \update{combined to the following expression}:
\begin{equation}
\label{eq:quark_TMD_1}
\begin{split}
  f^{[\gamma^+]}_{q,\lambda}(x,\kt) =& 
\frac{p^+}{(2\pi)^3} \int \dd[2]{\xt} \dd{x^-} \dd[2]{\yt} \dd{y^-}
e^{-i (x^- - y^-)x p^+ + i (\xt - \yt) \vdot \kt}
\times
\theta(- x^-) \theta( - y^-)
\\
&\times
 \int \frac{\dd[4]{q}\dd[4]{q'} \dd[4]{l}}{(2\pi)^{12}}
\Tr[
\gamma^+ \frac{i\slashed{q}}{q^2 + i \varepsilon} \gamma^- \slashed{l} \gamma^- \frac{-i \slashed{q'}}{q'^2 - i\varepsilon}
]
e^{-i q^+ y^- + i \qt \vdot \yt + i q^{\prime +}  x^- - i\qt' \vdot \xt}
\\
&
\times
 2\pi  \delta(l^2) \theta(l^-) \times
(2\pi)^2 \delta(l^- + q^-) \delta(l^- + q'^-)
\int \dd[2]{\ut} \dd[2]{\vt} e^{-i (\lt +\qt)\vdot \vt + i (\lt + \qt')\vdot \ut} \\
&\times
\Tr \expval{ \qty(-V(\ut) V^\dag_{C^x_-}[\xt] + V^\dag_{C^x_+}[\xt]) 
V_{C^{xy}}[\xt,\yt]
\qty( -V_{C^y_-}[\yt]V^\dag(\vt) +V_{C^y_+}[\yt]) }_{\lambda \lambda}.
\end{split}
\end{equation}
\update{Here, we have rewritten the ``quark field after shockwave'' contributions, corresponding to the diagrams on the right in Fig.~\ref{fig:quark_TMD}, using the following identity:
\begin{equation}
\begin{split}
    &
\int \dd{y^-}
\theta(y^-)
\gamma^+ \slashed{l}
e^{iy^- x p^+ +i l^+ y^- - i\lt \vdot \yt}
\delta(l^2) \theta(l^-)
\\
=&
\int \dd{y^-}
\theta(-y^-)
\int \frac{\dd[4]{q}}{(2\pi)^4}
\gamma^+\frac{i \slashed{q}}{q^2 + i\varepsilon} \gamma^- \slashed{l}
\times
e^{iy^- x p^+ -i q^+ y^- + i\qt \vdot \yt}
\\
&
\times
\delta(l^2) \theta(l^-)
\times (2\pi) \delta(l^- + q^-)
\int \dd[2]{\vt} e^{-i (\lt +\qt) \vdot \vt}
    ,
\end{split}
\end{equation}
which can be checked by a direct calculation.
With this trick, the only difference to the ``quark field before shockwave'' contributions is in the color structure.
}

Additionally, we have divided the gauge link structure as 
\begin{equation}
 \mathcal{W}_C[x,y] = \mathcal{W}^\dag_{C^x}[x]\mathcal{W}_{C^{xy}}[x,y]\mathcal{W}_{C^y}[y]   ,
\end{equation}
where $C^y$ is a straight line from $y$ to either plus or minus infinity, and similarly for $C^x$.
The path $C^{xy}$ contains the rest of the gauge link.
We use the notation $C^{y}_+$ for the case $y^- >0$, and correspondingly $y^- < 0$ for $C^{y}_-$, to clarify whether we start the Wilson line before or after the shockwave.
For example, for the simplest gauge links we have:
\begin{align}
    \mathcal{W}_\sqsupset[x,y] &= \mathcal{W}^\dag_{\neg}[x]
    \mathcal{W}_{\neg}[y] ,
    \\
    \mathcal{W}_\sqsubset[x,y] &= \mathcal{W}^\dag_{\ineg}[x]
    \mathcal{W}_{\ineg}[y],
\end{align}
where we have denoted the gauge links to the future and past by:
\begin{align}
    \mathcal{W}_\neg[y] &= 
    \mathcal{P} \exp( -i g_s \int_{y-}^\infty \dd{z^-}  A^+(z^-,\yt)),
    \\
    \mathcal{W}_\ineg[y] &= 
    \mathcal{P} \exp( -i g_s \int_{y-}^{-\infty} \dd{z^-}  A^+(z^-,\yt)).
\end{align}
We can then identify:
\begin{align}
    {V}_{\neg_-}\qty[\xt] &= V(\xt),
    &
    {V}_{\neg_+}\qty[\xt] &=1,
    \\
    {V}_{\ineg_-}\qty[\xt] &= 1,
    &
    {V}_{\ineg_+}\qty[\xt] &= V^\dag(\xt),
\end{align}
where again only Wilson lines crossing the shockwave give a nontrivial contribution.

To compute the quark TMD~\eqref{eq:quark_TMD_1}, we can calculate the Dirac trace and 
integrate over $l^+$, $\lt$, $x^-$, and $y^-$:
\begin{equation}
\label{eq:quark_TMD_2}
\begin{split}
  f^{[\gamma^+]}_{q,\lambda}(x,\kt) =& 
\frac{p^+}{(2\pi)^3} \int \dd[2]{\xt}\dd[2]{\yt} 
e^{ i (\xt - \yt) \vdot \kt}
\int \frac{\dd[4]{q}\dd[4]{q'} \dd{l^-}}{(2\pi)^{9}}  \theta(l^-)
\\
&\times
 \frac{4 \qt \vdot \qt'}{\qty[q^2 + i \varepsilon]\qty[q'^2 - i\varepsilon]}
 \frac{-1}{\qty[-xp^+ +q'^+ - i\varepsilon] \qty[xp^+ - q^+ - i \varepsilon]}
e^{ i \qt \vdot \yt  - i\qt' \vdot \xt}
\\
&
\times (2\pi)^2 \delta(l^- + q^-) \delta(l^- + q'^-)
\int \dd[2]{\ut} \dd[2]{\vt} e^{-i \qt\vdot \vt + i \qt'\vdot \ut} \delta^{(2)}(\ut -\vt) \\
&\times
\Tr \expval{ \qty(-V(\ut) V^\dag_{C^x_-}[\xt] + V^\dag_{C^x_+}(\xt)) 
V_{C^{xy}}[\xt,\yt]
\qty( -V_{C^y_-}[\yt]V^\dag(\vt) +V_{C^y_+}[\yt]) }_{\lambda \lambda}.
\end{split}
\end{equation}
We can now evaluate the integrals over $q^+$ and $q'^+$ using the residue theorem, and use the delta functions to integrate over $\vt$, $q^-$, and $q'^-$:
\begin{equation}
\label{eq:quark_TMD_3}
\begin{split}
  f^{[\gamma^+]}_{q,\lambda}(x,\kt) =& 
\frac{p^+}{(2\pi)^3} \int \dd[2]{\xt}\dd[2]{\yt} \dd[2]{\ut}
\int \frac{\dd[2]{\qt}\dd[2]{\qt'} \dd{l^-}}{(2\pi)^{5}}
  \theta(l^-)
e^{ i (\xt - \yt) \vdot \kt + i \qt \vdot(\yt -\ut) + i \qt' \vdot (\ut -\xt)}
\\
&\times
 \frac{4 \qt \vdot \qt'}{\qty[2 xp^+ l^- +\qt^2]\qty[2 xp^+ l^- +\qt'^2]}\\
 &\times 
\Tr \expval{ \qty(-V(\ut) V^\dag_{C^x_-}[\xt] + V^\dag_{C^x_+}(\xt)) 
V_{C^{xy}}[\xt,\yt]
\qty( -V_{C^y_-}[\yt]V^\dag(\ut) +V_{C^y_+}[\yt]) }_{\lambda \lambda}.
\end{split}
\end{equation}
Integrals over $\qt$ and $\qt'$ can be done using Eq.~\eqref{eq:K1_FT}:
\begin{equation}
\label{eq:quark_TMD_4}
\begin{split}
  f^{[\gamma^+]}_{q,\lambda}(x,\kt) =& 
\frac{2}{x(2\pi)^6} \int \dd[2]{\xt}\dd[2]{\yt} \dd[2]{\ut}
e^{ i (\xt - \yt) \vdot \kt}
\frac{(\yt - \ut) \vdot (\xt -\ut)}{\abs{\yt -\ut}\abs{\xt -\ut}}
\\
&\times
\Tr \expval{ \qty(-V(\ut) V^\dag_{C^x_-}[\xt] + V^\dag_{C^x_+}(\xt)) 
V_{C^{xy}}[\xt,\yt]
\qty( -V_{C^y_-}[\yt]V^\dag(\ut) +V_{C^y_+}[\yt]) }_{\lambda \lambda}
\\
&\times \int_0^\infty \dd{\eta} \eta
K_1( \abs{\yt -\ut} \sqrt{\eta} ) K_1( \abs{\xt -\ut} \sqrt{\eta} ),
\end{split}
\end{equation}
where we have also denoted $\eta = 2x p^+ l^-$.
Using now Eq.~\eqref{eq:bessel_integral}, this becomes:
\begin{equation}
\label{eq:quark_TMD_5}
\begin{split}
  f^{[\gamma^+]}_{q,\lambda}(x,\kt) =& 
\frac{8 }{x(2\pi)^6} \int \dd[2]{\xt}\dd[2]{\yt} \dd[2]{\ut}
e^{ i (\xt - \yt) \vdot \kt}
\frac{(\yt - \ut) \vdot (\xt -\ut)}{\abs{\yt -\ut}^2\abs{\xt -\ut}^2}
F\qty(\abs{\yt -\ut}^2 , \abs{\xt -\ut}^2 )
\\
&\times
\Tr \expval{ \qty(-V(\ut) V^\dag_{C^x_-}[\xt] + V^\dag_{C^x_+}(\xt)) 
V_{C^{xy}}[\xt,\yt]
\qty( -V_{C^y_-}[\yt]V^\dag(\ut) +V_{C^y_+}[\yt]) }_{\lambda \lambda}.
\end{split}
\end{equation}
Finally, we can simplify the Wilson line notation somewhat by noting that:
\begin{align}
     V^\dag_{C^x_\pm}[\xt] 
V_{C^{xy}}[\xt,\yt]V_{C^y_\mp}[\yt] &= V_{C_{\pm \mp}}[\xt,\yt] 
\end{align}
using the notation we introduced for the quark GTMD in Sec.~\ref{sec:quark_GTMD}.
This allows us to write:
\begin{equation}
\label{eq:quark_TMD_6}
\begin{split}
  f^{[\gamma^+]}_{q,\lambda}(x,\kt) =& 
\frac{8 }{x(2\pi)^6} \int \dd[2]{\xt}\dd[2]{\yt} \dd[2]{\ut}
e^{ i (\xt - \yt) \vdot \kt}
\frac{(\yt - \ut) \vdot (\xt -\ut)}{\abs{\yt -\ut}^2\abs{\xt -\ut}^2}
F\qty(\abs{\yt -\ut}^2 , \abs{\xt -\ut}^2 )
\\
&\times
\Tr \expval{ 
V_{C_{--}}[\xt,\yt]
+
V_{C_{++}}[\xt,\yt]
-
V(\ut) V_{C_{-+}}[\xt,\yt]
-
V^\dag(\ut) V_{C_{+-}}[\xt,\yt]
}_{\lambda \lambda}.
\end{split}
\end{equation}

We can now compare to the quark GTMD~\eqref{eq:quark_GTMD_final}, expecting:
\begin{equation}
      f^{[\gamma^+]}_{q,\lambda} =    W_{\lambda \lambda}^{[\gamma^+]}(\Delta \to 0)
\end{equation}
when $x>0$.
Noting that $F(a x, bx) = \frac{1}{x} F(a,b)$, we see that this is true if we can show that
we
can replace
$V_{C_{--}}[\xt,\yt] \to 1$ and
$V_{C_{++}}[\xt,\yt] \to 1$ in Eq.~\eqref{eq:quark_TMD_6}.
This is easiest to see starting from Eq.~\eqref{eq:quark_TMD_3}, as for these terms there is no dependence on $\ut$ in the Wilson lines.
We can then integrate over $\ut$, setting $\qt=\qt'$, which
gives us the following integral:
\begin{equation}
\begin{split}
     &\int \dd[2]{\xt}\dd[2]{\yt} 
\int \frac{\dd[2]{\qt}}{(2\pi)^{2}}
\int_0^\infty \dd{l^-}
e^{ i (\xt - \yt) \vdot \kt + i \qt \vdot(\yt -\xt)}
 \frac{ \qt^2 }{\qty[2 xp^+ l^- +\qt^2]^2}
\Tr \expval{V_{C}[\xt,\yt] }_{\lambda \lambda} \\
=& 
 \frac{1}{2 xp^+ }\int \dd[2]{\xt}\dd[2]{\yt} 
\int \frac{\dd[2]{\qt}}{(2\pi)^{2}}
e^{ i (\xt - \yt) \vdot \kt + i \qt \vdot(\yt -\xt)}
\Tr \expval{V_{C}[\xt,\yt] }_{\lambda \lambda}\\
=& 
 \frac{1}{2 xp^+ }\int \dd[2]{\xt}\dd[2]{\yt} 
\delta^{(2)}(\xt -\yt)
e^{ i (\xt - \yt) \vdot \kt }
\Tr \expval{V_{C}[\xt,\yt] }_{\lambda \lambda}.
\end{split}
\end{equation}
Because of the delta function, this expression only depends on $V_C[\xt,\xt] =1$.
In the end, we can then rewrite the quark TMD as
\begin{equation}
\label{eq:quark_TMD_final}
\begin{split}
  f^{[\gamma^+]}_{q,\lambda}(x,\kt) =& 
\frac{8 }{x(2\pi)^6} \int \dd[2]{\xt}\dd[2]{\yt} \dd[2]{\ut}
e^{ i (\xt - \yt) \vdot \kt}
\frac{(\yt - \ut) \vdot (\xt -\ut)}{\abs{\yt -\ut}^2\abs{\xt -\ut}^2}
F\qty(\abs{\yt -\ut}^2 , \abs{\xt -\ut}^2 )
\\
&\times
\Tr \expval{ 
2
-
V(\ut) V_{C_{-+}}[\xt,\yt]
-
V^\dag(\ut) V_{C_{+-}}[\xt,\yt]
}_{\lambda \lambda}
\end{split}
\end{equation}
which verifies the connection to the quark GTMD.

For specific gauge links, we get
\begin{equation}
\label{eq:quark_TMD_SIDIS}
\begin{split}
  f^{[\gamma^+]}_{q,\lambda}(x,\kt;\sqsupset) =& 
\frac{8 \nc}{x(2\pi)^6} \int \dd[2]{\xt}\dd[2]{\yt} \dd[2]{\ut}
e^{ i (\xt - \yt) \vdot \kt}
\frac{(\yt - \ut) \vdot (\xt -\ut)}{\abs{\yt -\ut}^2\abs{\xt -\ut}^2}
F\qty(\abs{\yt -\ut}^2 , \abs{\xt -\ut}^2 )
\\
&\times
\qty[ N_{\lambda \lambda}(\yt,\ut) + N_{\lambda \lambda}(\ut,\xt)  ].
\end{split}
\end{equation}
\begin{equation}
\label{eq:quark_TMD_DY}
\begin{split}
  f^{[\gamma^+]}_{q,\lambda}(x,\kt;\sqsubset) =& 
\frac{8 \nc}{x(2\pi)^6} \int \dd[2]{\xt}\dd[2]{\yt} \dd[2]{\ut}
e^{ i (\xt - \yt) \vdot \kt}
\frac{(\yt - \ut) \vdot (\xt -\ut)}{\abs{\yt -\ut}^2\abs{\xt -\ut}^2}
F\qty(\abs{\yt -\ut}^2 , \abs{\xt -\ut}^2 )
\\
&\times
\qty[ N_{\lambda \lambda}(\xt,\ut) + N_{\lambda \lambda}(\ut,\yt)  ].
\end{split}
\end{equation}
The result for the future-pointing gauge link agrees with the quark TMD computed in Ref.~\cite{Marquet:2009ca}, \update{along with Ref.~\cite{Mueller:1999wm} once the large-$Q^2$ limit is taken (see Ref.~\cite{Marquet:2009ca} for the detailed derivation of this limit)}.
We see that these expressions only differ in the Wilson-line structures that are related to each other by complex conjugation.
For this reason, the imaginary parts of the corresponding distributions in position space differ by a minus sign, which is a direct verification of the Sivers asymmetry between the TMDs that appear in Drell--Yan and SIDIS ~\cite{Collins:2002kn,Brodsky:2002cx,Boussarie:2023izj}.

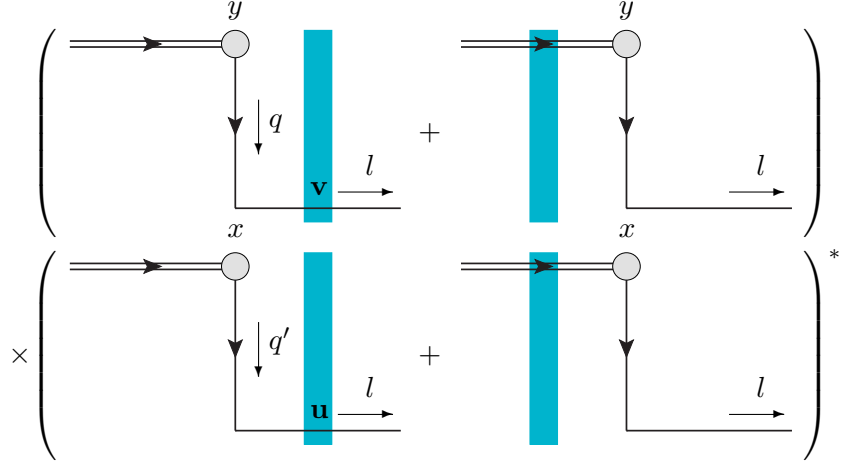
\begin{figure}
	\centering
\begin{equation*}
\begin{split}
&\qty(
    \begin{array}{l}    
\begin{overpic}[width=0.3\textwidth]{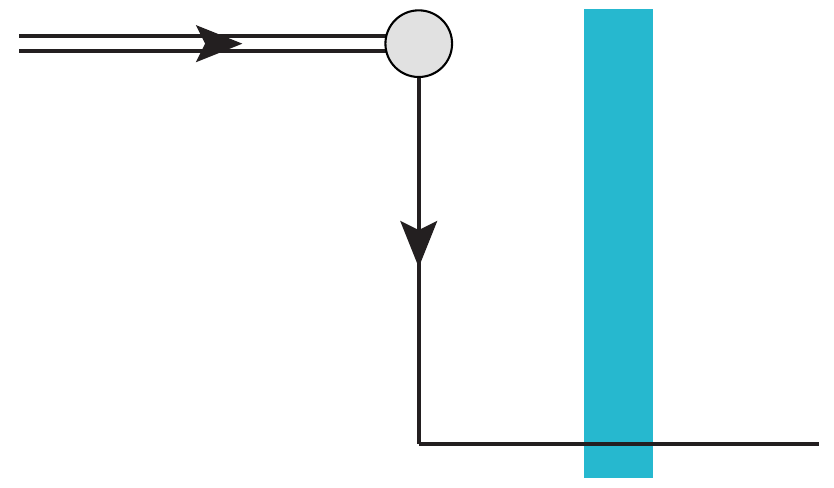}
    \put(48,62){$y$}
    \put(60,29){$q$}
    \put(88,15){$l$}
    \put(72,9){$\vt$}
        \put(57,37){\vector(0,-1){16}}
        \put(80,10){\vector(1,0){16}}
\end{overpic}
\end{array}
+
    \begin{array}{l}    
\begin{overpic}[width=0.3\textwidth]{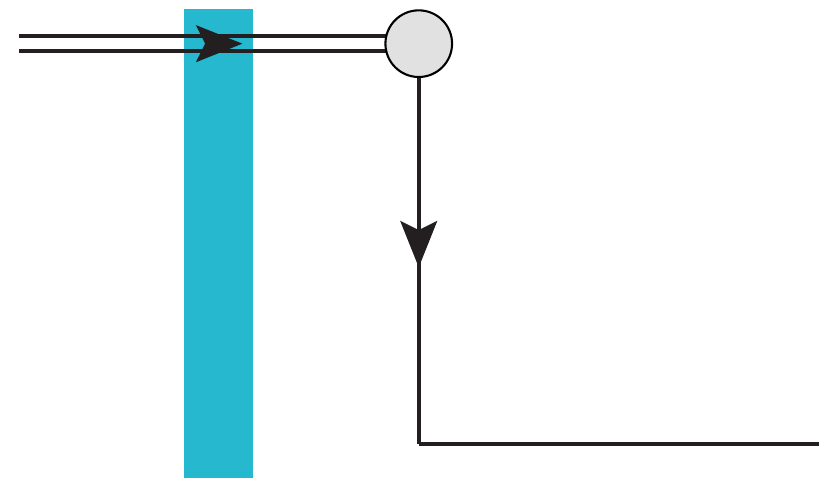}
    \put(48,62){$y$}
    \put(88,15){$l$}
        \put(80,10){\vector(1,0){16}}
\end{overpic}
\end{array}
)
\\
\times
&
\qty(
    \begin{array}{l}    
\begin{overpic}[width=0.3\textwidth]{figures/antiquark_TMD_past2.pdf}
    \put(48,62){$x$}
    \put(60,29){$q'$}
    \put(88,15){$l$}
    \put(72,9){$\ut$}
        \put(57,37){\vector(0,-1){16}}
        \put(80,10){\vector(1,0){16}}
\end{overpic}
\end{array} 
+
    \begin{array}{l}    
\begin{overpic}[width=0.3\textwidth]{figures/antiquark_TMD_past1.pdf}
    \put(48,62){$x$}
    \put(88,15){$l$}
        \put(80,10){\vector(1,0){16}}
\end{overpic}
\end{array} 
)^*
\end{split}
\end{equation*}
    \caption{Feynman diagrams for computing the antiquark TMD.
    }
    \label{fig:antiquark_TMD}
\end{figure}

For antiquarks, the TMD is defined as~\cite{Collins:2011zzd}:
\begin{equation}
\label{eq:antiquark_TMD}
\begin{split}
  f^{[\gamma^+]}_{\bar q,\lambda}(x,\kt) =&
    \frac{1}{2} \int \frac{\dd{r^-} \dd[2]{\rt}}{(2\pi)^3} 
    e^{ ix p^+ r^- - i \kt \vdot \rt}
  \Tr   \mel{p}{  \gamma^+      \psi \qty(-\frac{r}{2})
   \mathcal{W}^\dag_C\qty[-\frac{r}{2},\frac{r}{2}] 
\bar \psi \qty(\frac{r}{2})}{p}^c
   \\
    =& 
    \frac{p^+ }{(2\pi)^3 } 
    \int \dd{x^-} \dd[2]{\xt}
 \dd{y^-}\dd[2]{\yt} 
 e^{   
-ix p^+ (x^- - y^-) + i \kt \vdot (\xt -\yt)} \\
&\times
 \Tr\expval{ \gamma^+   \psi(x) \mathcal{W}^\dag_C\qty[x,y] \bar \psi(y)}^c_{\lambda  \lambda}.
\end{split}
\end{equation}
The relevant diagrams are shown in Fig.~\ref{fig:antiquark_TMD} , giving us:
\begin{equation}
\label{eq:antiquark_TMD_1}
\begin{split}
  f^{[\gamma^+]}_{\bar q,\lambda}(x,\kt) =& 
  \frac{p^+}{(2\pi)^3} \int \dd[2]{\xt} \dd{x^-} \dd[2]{\yt} \dd{y^-}
e^{-i (x^- - y^-)x p^+ + i (\xt - \yt) \vdot \kt}
\times
\theta(- x^-) \theta( - y^-)
\\
&\times
 \int \frac{\dd[4]{q}\dd[4]{q'} \dd[4]{l}}{(2\pi)^{12}} 
\Tr[
\gamma^+ \frac{-i\slashed{q'}}{q'^2 - i \varepsilon} \gamma^- \slashed{l} \gamma^- \frac{i \slashed{q}}{q^2 + i\varepsilon}
]
e^{i q^+ y^- - i \qt \vdot \yt - i q^{\prime +}  x^- + i\qt' \vdot \xt}
\\
&
\times 2\pi  \delta(l^2) \theta(l^-) \times  (2\pi)^2 \delta(p^- - q^-) \delta(p^- - q'^-)
\int \dd[2]{\ut} \dd[2]{\vt} e^{-i (\lt - \qt)\vdot \vt + i (\lt - \qt')\vdot \ut} \\
&\times
\Tr \expval{ V^\dag_{C^{xy}}[\xt,\yt] \qty(- V_{C^x_-}[\xt] V^\dag(\ut) + V_{C^x_+}(\xt)) 
\qty( -V(\vt) V^\dag_{C^y_-}[\yt] +V^\dag_{C^y_+}[\yt]) }_{\lambda \lambda}.
\end{split}
\end{equation}
Switching the integration variables $q \to -q$ and $q' \to - q'$, we can compare this to the quark case~\eqref{eq:quark_TMD_SIDIS} and note that only the Wilson-line structure is changed to its complex conjugate:
\begin{equation}
\label{eq:antiquark_TMD_final}
\begin{split}
  f^{[\gamma^+]}_{\bar q,\lambda}(x,\kt) =& 
\frac{8 }{x(2\pi)^6} \int \dd[2]{\xt}\dd[2]{\yt} \dd[2]{\ut}
e^{ i (\xt - \yt) \vdot \kt}
\frac{(\yt - \ut) \vdot (\xt -\ut)}{\abs{\yt -\ut}^2\abs{\xt -\ut}^2}
F\qty(\abs{\yt -\ut}^2 , \abs{\xt -\ut}^2 )
\\
&\times
\Tr \expval{ 
2
-
V^\dag(\ut) V^\dag_{C_{-+}}[\xt,\yt]
-
V(\ut) V^\dag_{C_{+-}}[\xt,\yt]
}_{\lambda \lambda}.
\end{split}
\end{equation}
This allows us to verify the connection to the quark TMD:
\begin{equation}
     f^{[\gamma^+]}_{\bar q,\lambda}(x,\kt) 
     =
     -f^{[\gamma^+]}_{ q,\lambda}(-x,-\kt),
\end{equation}
which is expected from the operator-level definition~\cite{Collins:2011zzd}.

\subsection{Quark PDF}

\begin{figure}
	\centering
\begin{equation*}
\qty(
    \begin{array}{l}    
\begin{overpic}[width=0.3\textwidth]{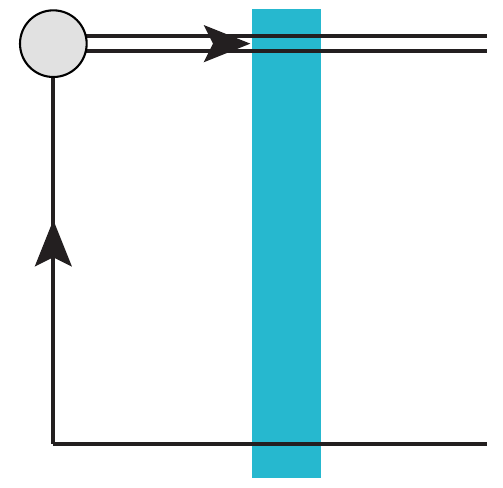}
    \put(9,100){$y$}
    \put(20,45){$q$}
    \put(85,20){$l$}
    \put(55,15){$\vt$}
        \put(17,35){\vector(0,1){20}}
        \put(75,15){\vector(1,0){20}}
\end{overpic}
\end{array}
)
\times
\qty(
    \begin{array}{l}    
\begin{overpic}[width=0.3\textwidth]{figures/quark_TMD_LO.pdf}
    \put(9,100){$x$}
    \put(20,45){$q'$}
    \put(85,20){$l$}
    \put(55,15){$\ut$}
        \put(17,35){\vector(0,1){20}}
        \put(75,15){\vector(1,0){20}}
\end{overpic}
\end{array} 
)^*
\end{equation*}
    \caption{Feynman diagrams for computing the quark PDF.
    }
    \label{fig:quark_PDF}
\end{figure}

The quark PDF is defined as~\cite{Collins:2011zzd}:
\begin{equation}
    \begin{split}
    f^q(x) =&
   \frac{1}{2} \int \frac{\dd{r^-}}{2\pi} 
    e^{ ixp^+ r^-}
    \frac{1}{2}\sum_\lambda
    \mel{p}{\bar \psi \qty(-\frac{r}{2}) \gamma^+ \mathcal{W}\qty[- \frac{r}{2},\frac{r}{2}] \psi \qty(\frac{r}{2})}{p}^c
    \bigg\lvert_{\rt=\mathbf{0}}\\
    =&
   \frac{p^+ }{2\pi} 
    \int \dd{x^-} \dd[2]{\xt}
 \dd{y^-} 
 e^{ 
-ix p^+ (x^- - y^-) }
    \frac{1}{2}\sum_\lambda
\expval{ \bar \psi  (x) \gamma^+ \mathcal{W}\qty[x,y] \psi(y)}^c_{\lambda  \lambda}
\bigg\lvert_{\yt = \xt}.
    \end{split}
\end{equation}
As in the GPD case, this needs to be computed in $D$ dimensions to regulate an infrared singularity that is related to the evolution of the distribution.
We again note that the PDF does not depend on the choice of the path for the gauge link, and therefore for the computation we will choose the future-pointing staple as shown in Fig.~\ref{fig:quark_PDF}.
This choice is slightly simpler than the past-pointing staple in Fig.~\ref{fig:quark_TMD} as now we only have one Feynman diagram at the amplitude level instead of two.
Computing this diagram, along with its complex conjugate, yields:
\begin{equation}
    \begin{split}
    f^q(x) =& 
\frac{p^+}{2\pi} \int \dd[D-2]{\xt} \dd{x^-}  \dd{y^-}
e^{-i (x^- - y^-)x p^+ }
\times
\theta(- x^-) \theta( - y^-)
\\
&\times
 \int \frac{\dd[D]{q}\dd[D]{q'} \dd[D]{l}}{(2\pi)^{3D}} 
\Tr[
\gamma^+ \frac{i\slashed{q}}{q^2 + i \varepsilon} \gamma^- \slashed{l} \gamma^- \frac{-i \slashed{q'}}{q'^2 - i\varepsilon}
]
e^{-i q^+ y^- + i (\qt - \qt') \vdot \xt + i q^{\prime +}  x^- }
\\
&
\times 
2\pi  \delta(l^2) \theta(l^-)
\times
(2\pi)^2 \delta(l^- + q^-) \delta(l^- + q'^-)
\int \dd[D-2]{\ut} \dd[D-2]{\vt} e^{-i (\lt +\qt)\vdot \vt + i (\lt + \qt')\vdot \ut} \\
&\times
    \frac{1}{2}\sum_\lambda
\Tr \expval{ \qty(-V^\dag (\xt)V(\ut) + 1) \qty( -V^\dag(\vt) V(\xt)+1) }_{\lambda \lambda}.
\end{split}
\end{equation}

We can follow similar steps to those in the TMD calculation. This leads to:
\begin{equation}
    \begin{split}
    f^q(x) =& 
\frac{p^+}{2\pi} \int \dd[D-2]{\xt} \dd[D-2]{\ut}
\int \frac{\dd[D-2]{\qt}\dd[D-2]{\qt'} \dd{l^-}}{(2\pi)^{2D-3}}
  \theta(l^-)
e^{  i (\qt -\qt') \vdot(\xt -\ut) }
\\
&\times
 \frac{4 \qt \vdot \qt'}{\qty[2 xp^+ l^- +\qt^2]\qty[2 xp^+ l^- +\qt'^2]}
 \times
    \frac{1}{2}\sum_\lambda
\Tr \expval{2 -V^\dag (\xt)V(\ut)  -V^\dag(\ut) V(\xt) }_{\lambda \lambda}.
\end{split}
\end{equation}
Noting that the terms $V^\dag (\xt)V(\ut)$ and  $V^\dag(\ut) V(\xt) $ give the same contribution, we can compare to the quark GPD calculation, Eq.~\eqref{eq:quark_GPD2}, and note that the quark PDF is proportional to the GPD in the limit $\Delta=0$, $\lambda= \lambda'$.
Using
\begin{equation}
    {}_2 F_1\qty(\alpha, \beta ; \gamma ;0) = 1,
\end{equation}
we can read the final expression from the GPD case~\eqref{eq:quark_GPD4}:
\begin{equation}
\label{eq:quark_PDF}
    \begin{split}
    f^q(x) 
           =&   
           \frac{2^{D} \nc}{x( 2\pi)^D}
           \frac{\Gamma\qty( D/2 )^2}{D-1}
           \int \dd[D-2]{\xt} \dd[D-2]{\ut}
    \frac{1}{2}\sum_\lambda
           \frac{N_{\lambda \lambda}(\xt,\ut)}{ \abs{\xt-\ut}^{2(D-2)} } \\
        =& 
 \frac{4 N_c}{x \pi} \frac{\Gamma(D/2)^2}{D-1}
 \frac{\Gamma \qty( -\frac{D-4}{2} )}{ \Gamma(D-3)}
 \times
 \int \dd[D-2]{\qt} 
    \frac{1}{2}\sum_\lambda
 \widetilde N_{\lambda \lambda}( - \qt,\qt) 
  \qty(\frac{1}{ \pi \qt^2})^{-(D-4)/2}.
\end{split}
\end{equation}
As in the GPD case, we can see the divergence $\Gamma\qty( -\frac{D-4}{2}) = \frac{1}{\varepsilon} + \ldots$ that needs to be renormalized.
This is achieved with the following counterterm:
\begin{equation}
\label{eq:quark_PDF_renormalization}
    f^{q}(x,\mu_R) 
    = f^q(x) - \frac{1}{\varepsilon} \qty(\frac{\mu^2}{\mu_R^2})^\varepsilon
     \frac{\as}{2\pi}
    \int_x^\infty \frac{\dd{y}}{y} P_{qg}\qty(\frac{x}{y}) f^g\qty(y,\mu_R)
\end{equation}
which corresponds to the case $\xi=0$ of the quark GPD renormalization~\eqref{eq:quark_GPD_renormalization}.
We have only included the splitting function for the gluon-to-quark case:
\begin{equation}
    P_{qg}(z)
    = P_{qg}\qty( z,  0 ) 
    = T_F \qty[z^2 + (1-z)^2],
\end{equation}
which is relevant at the leading order due to the $1/\as$ enhancement of the gluon PDF~\eqref{eq:gluon_PDF}.
We also note that we extend the $y$-integral in Eq.~\eqref{eq:quark_PDF_renormalization} to the unphysical region $y>1$ due to the shockwave approximation, similarly to the GPD case~\eqref{eq:quark_GPD_renormalization}.

Noting the $x$-dependence of the gluon PDF is given by $f^g(x) \sim 1/x$, the integral over splitting function evaluates to
\begin{equation}
\int_x^\infty \frac{\dd{y}}{y^2} P_{qg}\qty(\frac{x}{y})
=
    \int_0^1 \frac{\dd{z}}{z} P_{qg}(z) \times \frac{z}{x}
    = \frac{1}{3x}.
\end{equation}
The counterterm then cancels exactly the $\varepsilon$-pole in Eq.~\eqref{eq:quark_PDF}, and we can write the renormalized quark PDF as
\begin{equation}
\label{eq:quark_PDF_renormalized}
    \begin{split}
    f^q(x, \mu_R) 
        =& 
 \frac{4 N_c}{3x \pi} 
 \times
 \int \dd[2]{\qt} 
    \frac{1}{2}\sum_\lambda
 \widetilde N_{\lambda \lambda}( - \qt,\qt) 
 \qty[
 \log(
 \frac{\mu_R^2}{\qt^2}) - \frac{1}{3}
 ].
\end{split}
\end{equation}
As in  the quark GPD case, this is the first time the quark PDF has been computed including the finite piece.

\begin{figure}
	\centering
\begin{equation*}
\qty(
    \begin{array}{l}    
\begin{overpic}[width=0.3\textwidth]{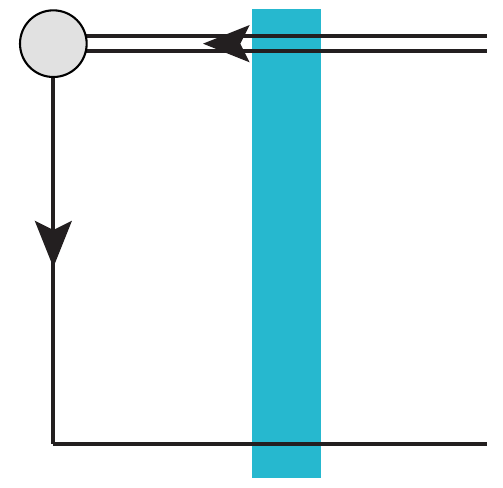}
    \put(9,100){$y$}
    \put(20,45){$q$}
    \put(85,20){$l$}
    \put(55,15){$\vt$}
        \put(17,55){\vector(0,-1){20}}
        \put(75,15){\vector(1,0){20}}
\end{overpic}
\end{array}
)
\times
\qty(
    \begin{array}{l}    
\begin{overpic}[width=0.3\textwidth]{figures/antiquark_TMD_future.pdf}
    \put(9,100){$x$}
    \put(20,45){$q'$}
    \put(85,20){$l$}
    \put(55,15){$\ut$}
        \put(17,55){\vector(0,-1){20}}
        \put(75,15){\vector(1,0){20}}
\end{overpic}
\end{array} 
)^*
\end{equation*}
    \caption{Feynman diagrams for computing the antiquark PDF.
    }
    \label{fig:antiquark_PDF}
\end{figure}

Similarly, the antiquark PDF is given by:
\begin{equation}
    \begin{split}
    f^{\bar q}(x) =&
   \frac{1}{2} \int \frac{\dd{r^-}}{2\pi} 
    e^{ ixp^+ r^-}
    \frac{1}{2}\sum_\lambda
    \Tr \gamma^+ \mel{p}{ \psi \qty(-\frac{r}{2})  \mathcal{W}^\dag_C\qty[- \frac{r}{2},\frac{r}{2}] \bar \psi \qty(\frac{r}{2})}{p}^c
    \bigg\lvert_{\rt=\mathbf{0}}\\
    =&
   \frac{p^+ }{2\pi} 
    \int \dd{x^-} \dd[2]{\xt}
 \dd{y^-} 
 e^{ 
-ix p^+ (x^- - y^-) }
    \frac{1}{2}\sum_\lambda
    \Tr
\expval{ \gamma^+ \psi  (x) \mathcal{W}^\dag_C\qty[x,y] \bar \psi(y)}^c_{\lambda  \lambda}
\bigg\lvert_{\yt = \xt}.
    \end{split}
\end{equation}
This can be computed from the diagrams in Fig.~\ref{fig:antiquark_PDF}:
\begin{equation}
    \begin{split}
    f^{\bar q}(x) =& 
\frac{p^+}{2\pi} \int \dd[D-2]{\xt} \dd{x^-}  \dd{y^-}
e^{-i (x^- - y^-)x p^+ }
\times
\theta(- x^-) \theta( - y^-)
\\
&\times
 \int \frac{\dd[D]{q}\dd[D]{q'} \dd[D]{l}}{(2\pi)^{3D}}
\Tr[
\gamma^+ \frac{i\slashed{q'}}{q'^2 - i \varepsilon} \gamma^- \slashed{l} \gamma^- \frac{i \slashed{q}}{q^2 + i\varepsilon}
]
e^{i q^+ y^- - i (\qt - \qt') \vdot \xt - i q^{\prime +}  x^- }
\\
&
\times  2\pi  \delta(l^2) \theta(l^-)
\times (2\pi)^2 \delta(l^- - q^-) \delta(l^- - q'^-)
\int \dd[D-2]{\ut} \dd[D-2]{\vt} e^{-i (\lt -\qt)\vdot \vt + i (\lt - \qt')\vdot \ut} \\
&\times
    \frac{1}{2}\sum_\lambda
\Tr \expval{ \qty(-V(\xt)V^\dag(\ut) + 1) \qty( -V(\vt) V^\dag(\xt)+1) }_{\lambda \lambda}.
\end{split}
\end{equation}
Changing $q \to -q$ and $q' \to -q'$, we note that the only difference to the quark case is in the Wilson-line structure, which is now the complex conjugate of the quark case.
However, it turns out that the PDF only depends on the real part of the dipole amplitude, as can be verified explicitly from Eq.~\eqref{eq:quark_PDF}.
We therefore arrive at
\begin{equation}
    f^{\bar q}(x) = f^q(x)
\end{equation}
for both bare and renormalized PDFs.
They also satisfy the expected relation $f^{\bar q}(x) = -f^q(-x)$ between the quark and antiquark PDFs~\cite{Collins:2011zzd}.
We can also explicitly verify the relation to the GPDs:
\begin{equation}
    \frac{1}{2}\sum_\lambda   F^q(\Delta \to 0)  
    = 
    \begin{cases}
        f^q(x) &\text{ if } x>0, \\
        - f^{\bar q}(-x) &\text{ if } x<0,
    \end{cases}
\end{equation}
which is as expected \cite{Diehl:2003ny}.

\section{Diffractive distributions}
\label{sec:diffraction}

Finally, we wish to consider parton distributions that are relevant for describing diffractive processes.
Diffractive processes are characterized by a rapidity gap between the target remnants and the produced particles, which indicates a color-neutral interaction between the projectile and the target.
This includes both exclusive and inclusive processes, with exclusive processes being described in terms of GPDs and inclusive processes by diffractive parton distributions.
We can further divide diffractive processes into coherent and incoherent processes, where in the coherent case the target stays intact while in the incoherent production it dissociates further.
Diffractive parton distributions are then also divided into two classes: the total diffractive distributions summing over all diffractive states, which includes both coherent and incoherent production, and coherent distributions that only include the coherent production.

For inclusive diffraction, it is more convenient to work in the asymmetric frame where the incoming target has no transverse momentum.
This means that we define the momenta for the incoming target and the outgoing target remnants $Y$ as:
\begin{align}
    p &= \qty(p^+, \frac{M^2}{2 p^+}, \boldsymbol{0}) ,
    \\
    p_Y &= \qty([1-\xpom] p^+, \frac{M_Y^2 + \Deltat^2}{2 [1-\xpom] p^+}, \Deltat),
\end{align}
where $M_Y$ is the invariant mass of the target remnants.
For coherent production, the state $Y$ is the same as the incoming target, and we will denote $p' = p_Y$.
We have also defined the momentum fraction of the target carried by the ``pomeron'' $\mathbb{P}$ as
\begin{equation}
    \xpom = \frac{p^+ - p_Y^+}{p^+}.
\end{equation}
This terminology comes from the Regge theory, where diffraction can be interpreted as an exchange of a color-neutral quasi-particle called pomeron.
Additionally, we define the variable $\beta$ to represent the plus-momentum fraction of the pomeron carried by the parton, which is related to the standard momentum fraction $x$ by
\begin{equation}
    x = \xpom \beta.
\end{equation}

There are several different definitions for diffractive distributions that differ by constant factors.
We will choose to follow the definition of Refs.~\cite{Berera:1995fj,Buchmuller:1998jv,Golec-Biernat:2001gyl,Hatta:2022lzj,Hatta:2024vzv};
this differs from Ref.~\cite{Collins:2001ga} by a factor of $1/(16 \pi^2 \xpom)$, from Ref.~\cite{Hauksson:2024bvv} by a factor of $1/\xpom^2$, and from Ref.~\cite{Lee:2025fml} by a factor of $4/\pi$.
Then, the coherent diffractive TMD distribution is defined as
\begin{equation}
\begin{split}
f_{ij}^{\coh}( \kt, \Deltat, \xpom, \beta; \lambda)
=& 
\frac{(xp^+)^n}{16\pi^2}
 \sumint[X]
 \sum_{\lambda'}
\int \frac{\dd{r^-} \dd[2]{\rt}}{(2\pi)^3} e^{i x r^- p^+ - i \rt \vdot \kt} \\
&\times
    \mel{p}{\Phi^i \qty(-\frac{r}{2})\mathcal{W}_\neg^\dag\qty[- \frac{r}{2}]}{p' X}\mel{p'X}{\mathcal{W}_\neg\qty[\frac{r}{2}] \Phi^j \qty(\frac{r}{2})}{p}.
\end{split}
\end{equation}
Here, we have an extra factor of $1/(16 \pi^2)$ compared to the TMDs and PDFs, which comes from the outgoing target nucleon phase space that we do not integrate over (see e.g. Ref.~\cite{Berera:1995fj}).
The final-state particles $X$ are separated from the outgoing target $p'$ by a rapidity gap and they have to be integrated over in the same way as for TMDs and PDFs.

Due to the large rapidity gap between $p'$ and $X$, we can ignore the final state $X$ when using the eikonal approximation~\eqref{eq:eikonal_approximation}. 
This allows us to write the diffractive distribution as
\begin{equation}
\begin{split}
&f_{ij}^{(D)}( \kt, \Deltat, \xpom, \beta; \lambda)\\
=& 
\frac{(xp^+)^n}{16\pi^2}
\frac{1 - \xpom}{(2\pi)^3}
(2 p^+)^2 
 \sumint[X] \sum_{\lambda'}
\int \dd{r^-} \dd[2]{\rt} 
\int \dd{b^-} \dd[2]{\bt} 
\int \dd{b^{\prime -}} \dd[2]{\bt'} \\
& \times e^{i x r^- p^+ - i \rt \vdot \kt
- i \xpom (b^- - b^{\prime -}) p^+ - i (\bt-\bt') \vdot \Deltat} 
\\
&\times
    \expval{\Phi^i \qty(-\frac{r}{2}) \mathcal{W}_\neg^\dag\qty[- \frac{r}{2}]}_{b',\lambda',\lambda}
   \ketbra{X}
    \expval{\mathcal{W}_\neg\qty[\frac{r}{2}] \Phi^j \qty(\frac{r}{2})}_{b,\lambda \lambda'}\\
=& 
\frac{(xp^+)^n}{16\pi^2}
\frac{1 - \xpom}{(2\pi)^3}
(2 p^+)^2 
\sumint[X] \sum_{\lambda'}
\int \dd{r^-} \dd[2]{\rt} 
\int \dd{b^-} \dd[2]{\bt} 
\int \dd{b^{\prime -}} \dd[2]{\bt'} \\
& \times e^{i x r^- p^+ - i \rt \vdot \kt
+ i (b^- - b^{\prime -}) (-\xpom p^+ + p_X^+) - i (\bt-\bt') \vdot (\Deltat + \pt_X )
} 
\\
&\times
    \expval{\Phi^i \qty(-\frac{r}{2}-b') \mathcal{W}_\neg^\dag\qty[- \frac{r}{2}-b']}_{\lambda'\lambda}
    \ketbra{X}
    \expval{\mathcal{W}_\neg\qty[\frac{r}{2}-b] \Phi^j \qty(\frac{r}{2}-b)}_{\lambda \lambda'},
\end{split}
\end{equation}
where we have again shifted the target to the origin.
By changing the integration variables $b$ and $b'$ to
\begin{align}
    x &= -\frac{r}{2}  -b' ,
    \\
    y &= \frac{r}{2} - b,
\end{align}
we can integrate over $r^-$ and $\rt$:
\begin{equation}
\begin{split}
&f_{ij}^{\coh}( \kt, \Deltat, \xpom, \beta; \lambda)\\
=& 
\frac{(xp^+)^n}{16\pi^2}
\frac{1 - \xpom}{(2\pi)^3}
(2 p^+)^2 
  \sumint[X] \sum_{\lambda'}
\int \dd{r^-} \dd[2]{\rt} 
\int \dd{x^-} \dd[2]{\xt} 
\int \dd{y^{-}} \dd[2]{\yt} \\
& \times e^{i x r^- p^+ - i \rt \vdot \kt
+i(x^- -y^- + r^-)  (-\xpom p^+ + p_X^+) - i (\xt-\yt+\rt) \vdot (\Deltat + \pt_X )
} 
\\
&\times
    \expval{\Phi^i \qty(x)\mathcal{W}_\neg^\dag\qty[x]}_{\lambda' \lambda}
   \ketbra{X}
    \expval{\mathcal{W}_\neg\qty[y]  \Phi^j \qty(y)}_{\lambda \lambda'}\\
=& 
\frac{(xp^+)^n}{16\pi^2}
(1 - \xpom)
(2 p^+)^2 
 \sumint[X] \sum_{\lambda'}
\int \dd{x^-} \dd[2]{\xt} 
\int \dd{y^{-}} \dd[2]{\yt}\\
& \times e^{
- i x p^+(x^- -y^- ) + i (\xt-\yt) \vdot \kt
} 
\delta(  p_X^+ - \xpom \qty[1-\beta] p^+)  \delta^{(2)}(\Deltat + \pt_X + \kt)
\\
&\times
    \expval{\Phi^i \qty(x) \mathcal{W}_\neg^\dag\qty[x]}_{\lambda' \lambda}
  \ketbra{X}
    \expval{\mathcal{W}_\neg\qty[y] \Phi^j \qty(y)}_{\lambda \lambda'}.
\end{split}
\end{equation}
This is the result that we will use to compute the coherent diffractive distributions.

While most studies focus on coherent production, we can also consider the total diffractive production where the target does not necessarily stay intact. 
Instead, we have a general outgoing state $Y$ that is separated from $X$ by a rapidity gap.
The corresponding distribution is defined as
\begin{equation}
\begin{split}
&f_{ij}^{(D)}( \kt, \Deltat, \xpom, \beta; \lambda)\\=& 
\frac{(xp^+)^n}{16\pi^2}
 \sumint[X, Y]
\int \frac{\dd{r^-} \dd[2]{\rt}}{(2\pi)^3} e^{i x r^- p^+ - i \rt \vdot \kt} 
\times 2 p_Y^+ (2\pi)^3 \delta( p_Y^+ - [1-\xpom]p^+ )
\delta^{(2)}( \pt_Y - \Deltat )
\\
&\times
    \mel{p}{\Phi^i \qty(-\frac{r}{2})\mathcal{W}_\neg^\dag\qty[- \frac{r}{2}]}{Y X}\mel{Y X}{\mathcal{W}_\neg\qty[\frac{r}{2}] \Phi^j \qty(\frac{r}{2})}{p}.
\end{split}
\end{equation}
To write this in the shockwave approximation, we wish to get rid of the final state $Y$.
This is achieved by first rewriting the momenta $p_Y$ in terms of a general momentum operator $\hat P$ acting on the state $\ket{Y}$:
\begin{equation}
\begin{split}
    &2 p_Y^+ (2\pi)^3 \delta( p_Y^+ - [1-\xpom]p^+ ) 
    \delta^{(2)}( \pt_Y - \Deltat )\ketbra{Y} \\
    =&
    2 [1-\xpom] p^+
    \int \dd{b^-} \dd[2]{\bt} 
    e^{ i b^- ( \hat P^+ - [1-\xpom]p^+ ) - i \bt \vdot ( \hat \Pt - \Deltat) }
    \ketbra{Y}    .
\end{split}
\end{equation}
Noting that
\begin{equation}
    e^{i b \vdot \hat P} \ket{YX} = \qty[ e^{i b \vdot \hat P} \ket{Y} ] \otimes e^{i b \vdot p_X} \ket{X},
\end{equation} 
we then get
\begin{equation}
\begin{split}
&f_{ij}^{(D)}( \kt, \Deltat, \xpom, \beta; \lambda)\\=& 
\frac{(xp^+)^n}{16\pi^2}
 [1-\xpom ] 2p^+
 \sumint[X, Y]
\int \frac{\dd{r^-} \dd[2]{\rt}}{(2\pi)^3} e^{i x r^- p^+ - i \rt \vdot \kt} 
    \int \dd{b^-} \dd[2]{\bt} 
    e^{ - i b^- ( p_X^+ + [1-\xpom]p^+ ) + i \bt \vdot ( \pt_X + \Deltat) }
\\
&\times
    \mel{p}{\Phi^i \qty(-\frac{r}{2})\mathcal{W}_\neg^\dag\qty[- \frac{r}{2}]
    e^{i b^- \hat P^+ - i \bt \vdot \hat \Pt}
      }{ YX}\mel{YX}{ 
    \mathcal{W}_\neg\qty[\frac{r}{2}] \Phi^j \qty(\frac{r}{2})}{p}
    \\
    =& 
\frac{(xp^+)^n}{16\pi^2}
 [1-\xpom ] 2p^+
 \sumint[X,Y]
\int \frac{\dd{r^-} \dd[2]{\rt}}{(2\pi)^3} e^{i x r^- p^+ - i \rt \vdot \kt} 
    \int \dd{b^-} \dd[2]{\bt} 
    e^{ - i b^- ( p_X^+ -\xpom p^+ ) + i \bt \vdot ( \pt_X + \Deltat) }
\\
&\times
    \mel{p}{\Phi^i \qty(-\frac{r}{2} -b)\mathcal{W}_\neg^\dag\qty[- \frac{r}{2}-b]
   }{ YX}\mel{YX}{
    \mathcal{W}_\neg\qty[\frac{r}{2}] \Phi^j \qty(\frac{r}{2})}{p}.
\end{split}
\end{equation}

We would then like to use the completeness relation
\begin{equation}
\label{eq:completeness}
    \sumint[Y] \ketbra{Y} = 1
\end{equation}
to cancel the dependence on $Y$. However, when defining the diffractive distribution, we require the final states $X$ and $Y$ to be separated by a rapidity gap, which introduces a nontrivial dependence between the two states.
This can be remedied by noting that in the high-energy limit the requirement of the rapidity gap is equivalent to a color-neutral exchange with the target, and hence the rapidity gap can be achieved by projecting the state $X$ into a color singlet.
We can thus write
\begin{equation}
    \ketbra{YX} = \hat C_S \ketbra{X} \hat C_S^\dag \otimes \ketbra{Y}
\end{equation}
where $\hat C_S$ is the color-singlet projection operator.
We can then apply the completeness relation, yielding:
\begin{equation}
\begin{split}
&f_{ij}^{(D)}( \kt, \Deltat, \xpom, \beta; \lambda)\\
    =& 
\frac{(xp^+)^n}{16\pi^2}
 [1-\xpom ] 2p^+
 \sumint[X]
\int \frac{\dd{r^-} \dd[2]{\rt}}{(2\pi)^3} e^{i x r^- p^+ - i \rt \vdot \kt} 
    \int \dd{b^-} \dd[2]{\bt} 
    e^{ - i b^- ( p_X^+ -\xpom p^+ ) + i \bt \vdot ( \pt_X + \Deltat) }
\\
&\times
    \mel{p}{\Phi^i \qty(-\frac{r}{2} -b)\mathcal{W}_\neg^\dag\qty[- \frac{r}{2}-b]
   \hat C_S}{ X}\mel{X}{ \hat C_S^\dag 
    \mathcal{W}_\neg\qty[\frac{r}{2}] \Phi^j \qty(\frac{r}{2})}{p}.
\end{split}
\end{equation}
Writing the proton states in position space and using the eikonal approximation, this becomes
\begin{equation}
\begin{split}
&f_{ij}^{(D)}( \kt, \Deltat, \xpom, \beta; \lambda)\\
    =& 
\frac{(xp^+)^n}{16\pi^2}
 [1-\xpom ] (2p^+)^2
 \sumint[X]
\int \frac{\dd{r^-} \dd[2]{\rt}}{(2\pi)^3} e^{i x r^- p^+ - i \rt \vdot \kt} 
    \int \dd{b^-} \dd[2]{\bt} 
    e^{ - i b^- ( p_X^+ -\xpom p^+ ) + i \bt \vdot ( \pt_X + \Deltat) }
\\
&\times
\int \dd{b'^-} \dd[2]{\bt'}
\expval{\Phi^i \qty(-\frac{r}{2} -b)\mathcal{W}_\neg^\dag\qty[- \frac{r}{2}-b]
   \hat C_S \ketbra{X} \hat C_S^\dag 
    \mathcal{W}_\neg\qty[\frac{r}{2}] \Phi^j \qty(\frac{r}{2})}_{b',\lambda \lambda}\\
    =& 
\frac{(xp^+)^n}{16\pi^2}
 [1-\xpom ] (2p^+)^2
 \sumint[X]
\int \frac{\dd{r^-} \dd[2]{\rt}}{(2\pi)^3} 
    \int \dd{b^-} \dd[2]{\bt} 
\int \dd{b'^-} \dd[2]{\bt'}
\\
&\times 
    e^{ i x r^- p^+ - i \rt \vdot \kt- i b^- ( p_X^+ -\xpom p^+ ) + i \bt \vdot ( \pt_X + \Deltat) }
\\
&\times
\expval{\Phi^i \qty(-\frac{r}{2} -b-b')\mathcal{W}_\neg^\dag\qty[- \frac{r}{2}-b-b']
   \hat C_S \ketbra{X} \hat C_S^\dag 
    \mathcal{W}_\neg\qty[\frac{r}{2}-b'] \Phi^j \qty(\frac{r}{2}-b')}_{\lambda \lambda}.
\end{split}
\end{equation}
Finally, we will define the variables
\begin{align}
    x &= -\frac{r}{2} -b  -b' ,
    \\
    y &= \frac{r}{2} - b',
\end{align}
so that we can integrate over $r$:
\begin{equation}
\begin{split}
&f_{ij}^{(D)}( \kt, \Deltat, \xpom, \beta; \lambda)\\
    =& 
\frac{(xp^+)^n}{16\pi^2}
 [1-\xpom ] (2p^+)^2
 \sumint[X]
\int \frac{\dd{r^-} \dd[2]{\rt}}{(2\pi)^3}
\int \dd{x^- }\dd[2]{\xt}\dd{y^- }\dd[2]{\yt}
e^{i x r^- p^+ - i \rt \vdot \kt} 
\\
&\times
    e^{i (x^- -y^- + r^-) ( p_X^+ -\xpom p^+ ) - i (\xt-\yt+\rt) \vdot ( \pt_X + \Deltat) }
\expval{\Phi^i \qty(x)\mathcal{W}_\neg^\dag\qty[x]
   \hat C_S \ketbra{X} \hat C_S^\dag 
    \mathcal{W}_\neg\qty[y] \Phi^j \qty(y)}_{\lambda \lambda}\\
    =& 
\frac{(xp^+)^n}{16\pi^2}
 [1-\xpom ] (2p^+)^2
 \sumint[X]
\int \dd{x^- }\dd[2]{\xt}\dd{y^- }\dd[2]{\yt}
\delta(p_X^+ - [1-\xpom]p^+)
\delta^{(2)}(\pt_X + \kt + \Deltat)
\\
&\times
    e^{-i (x^- -y^- ) xp^+ + i (\xt-\yt) \vdot \kt }
\expval{\Phi^i \qty(x)\mathcal{W}_\neg^\dag\qty[x]
   \hat C_S \ketbra{X} \hat C_S^\dag 
    \mathcal{W}_\neg\qty[y] \Phi^j \qty(y)}_{\lambda \lambda}.
\end{split}
\end{equation}
This result can be used for computing the total diffractive distribution which also includes incoherent production.
In practice, this only affects the target-averaging procedure, meaning that we can read off the result from coherent distributions by simply taking the target average $\expval{\ldots}$ at the level of the cross section instead of the amplitude.

Diffractive PDFs 
can be obtained from diffractive TMDs by integrating over the transverse momentum $\kt$ and averaging over the helicity of the incoming target nucleon.
For example, the coherent diffractive PDF is given as
\begin{equation}
\label{eq:DPDF}
\begin{split}
    f_{ij}^{(D)}( \Deltat, \xpom, \beta)
    =& \frac{1}{2}\sum_\lambda \int \dd[2]{\kt} f_{ij}^{(D)}(\kt, \Deltat, \xpom, \beta; C,\lambda) \\
    =&
\frac{(xp^+)^n}{32\pi^2}
\sumint[X] \sum_{\lambda \lambda'}
\int \frac{\dd{r^-} \dd[2]{\rt}}{(2\pi)^3} e^{i x r^- p^+ } \\
&\times  
    \mel{p}{\Phi^i \qty(-\frac{r}{2})\mathcal{W}\qty[- \frac{r}{2},\infty^-]}{p' X}\mel{p'X}{\mathcal{W}\qty[\infty^-,\frac{r}{2}] \Phi^j \qty(\frac{r}{2})}{p}
    \bigg\lvert_{\rt=\mathbf{0}}\\
    =&
\frac{(xp^+)^n}{32\pi^2}
(1 - \xpom)
(2 p^+)^2 
  \sumint[X] \sum_{\lambda \lambda'}
\int \dd{x^-} \dd[2]{\xt} 
\int \dd{y^{-}} \dd[2]{\yt}\\
& \times e^{
- i x p^+(x^- -y^- ) - i (\xt-\yt) \vdot (\Deltat + \pt_X)
} 
\delta(  p_X^+ - \xpom \qty[1-\beta] p^+)  
\\
&\times
    \expval{\Phi^i \qty(x) \mathcal{W}^\dag_{\neg}\qty[x]}_{\lambda' \lambda}
  \ketbra{X}
    \expval{\mathcal{W}_\neg\qty[y] \Phi^j \qty(y)}_{\lambda \lambda'},
\end{split}
    \end{equation}
and the total diffractive PDF can be computed similarly.

\subsection{Gluon}

The gluon diffractive TMD is defined as
\begin{equation}
\begin{split}
&f_g^{\coh}( \kt, \Deltat, \xpom, \beta; \lambda)\\
=&
\frac{2}{x p^+} \frac{1}{16\pi^2} 
\sumint[X] \sum_{\lambda'}
\int \frac{\dd{r^-} \dd[2]{\rt}}{(2\pi)^3} e^{i x r^- p^+ - i \rt \vdot \kt} \\
&\times
\Tr
\mel{p}{ \mathcal{W}_\neg \qty[-\frac{r}{2}] G^{+i}\qty(-\frac{r}{2}) \mathcal{W}^\dag_\neg \qty[-\frac{r}{2}]}{p' X}  \mel{p' X}{ \mathcal{W}_\neg \qty[\frac{r}{2}] G^{+i}\qty(\frac{r}{2})
 \mathcal{W}^\dag_\neg \qty[\frac{r}{2}]
}{p} \\
=& 
\frac{p^+}{2\pi^2}
\frac{1-\xpom}{x}
 \sumint[X] \sum_{\lambda'}
\int \dd{x^-} \dd[2]{\xt} 
\int \dd{y^{-}} \dd[2]{\yt}\\
& \times e^{
- i x p^+(x^- -y^- ) + i (\xt-\yt) \vdot \kt
} 
\delta(  p_X^+ - \xpom \qty[1-\beta] p^+)  \delta^{(2)}(\Deltat + \pt_X + \kt)
\\
&\times
\Tr
    \expval{ \mathcal{W}_\neg\qty[x] G^{+i} \qty(x) \mathcal{W}_\neg^\dag\qty[x]}_{\lambda' \lambda}
  \ketbra{X}
    \expval{\mathcal{W}_\neg\qty[y] G^{+i}(y) \mathcal{W}^\dag_\neg\qty[y]}_{\lambda \lambda'}.
\end{split}
\end{equation}
For gluons, the diffractive case is quite different from the other distributions we have considered so far.
This is because now contributions from ``gluons inside the shockwave'' vanish: 
following as in the TMD case for a future-pointing gauge link,
these would give a contribution like
\begin{equation}
\label{eq:diff_gluon_shockwave}
\Tr
    \expval{[\partial^k V(\xt)] V^\dag(\xt)}
    \expval{V(\yt) [\partial^k V^\dag(\yt)]}.
\end{equation}
Due to the color neutrality of the target, the expectation values project out the color-singlet contribution from the operators:
\begin{equation}
\label{eq:diffractive_trace}
    \expval{\mathcal{O}_{ij}} = \frac{\delta^{ij} }{D_R} \Tr \expval{\mathcal{O}}
\end{equation}
where $D_R$ is the dimension of the representation: $D_R = \nc$ for fundamental and $D_R = \nc^2 - 1$ for adjoint.
For Eq.~\eqref{eq:diff_gluon_shockwave}, this means that we get 
\begin{equation}
     \expval{
     \qty{
     [\partial^k V(\xt)] V^\dag(\xt)
     }_{ij}
     }
     =
     \frac{\delta^{ij} }{\nc} 
     \Tr \expval{[\partial^k V(\xt)] V^\dag(\xt)}
     = 0
\end{equation}
which vanishes due to the cyclicity of the trace and Eq.~\ref{eq:Wilson_shuffling}.

\begin{figure}
	\centering
\begin{equation*}
\qty(
    \begin{array}{l}    
\begin{overpic}[width=0.3\textwidth]{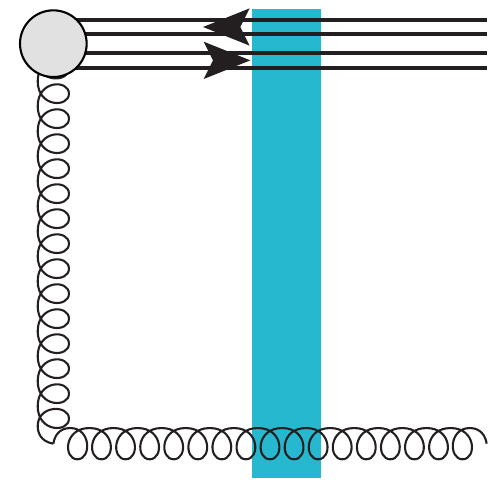}
    \put(9,100){$y$}
    \put(20,45){$q$}
    \put(85,20){$l$}
    \put(55,15){$\vt$}
        \put(17,55){\vector(0,-1){20}}
        \put(75,15){\vector(1,0){20}}
\end{overpic}
\end{array}
)
\times
\qty(
    \begin{array}{l}    
\begin{overpic}[width=0.3\textwidth]{figures/gluon_DTMD.pdf}
    \put(9,100){$x$}
    \put(20,45){$q'$}
    \put(85,20){$l$}
    \put(55,15){$\ut$}
        \put(17,55){\vector(0,-1){20}}
        \put(75,15){\vector(1,0){20}}
\end{overpic}
\end{array} 
)^*
\end{equation*}
    \caption{Feynman diagrams for computing the diffractive gluon distributions.
    }
    \label{fig:gluon_DTMD}
\end{figure}
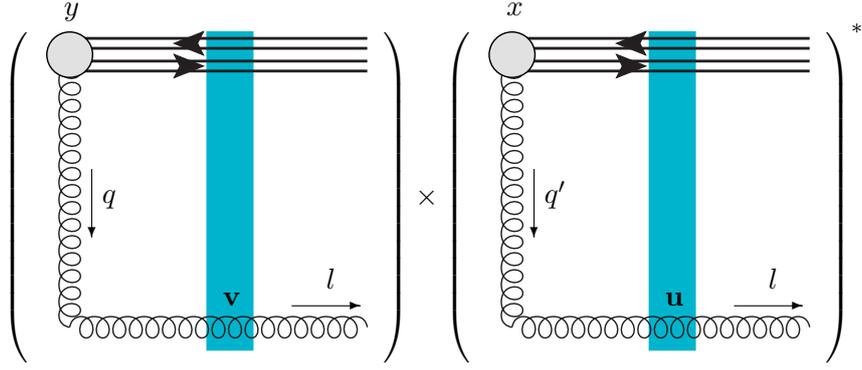

For this reason, our treatment of the gluon case is now similar to what we have done for quarks, in the sense that we need to compute the Feynman diagrams in Fig.~\ref{fig:gluon_DTMD}.
We also note that the gluon field $G_a^{\mu \nu} = \partial^\mu A_a^\nu - \partial^\nu A_a^\mu - g f_{abc} A^\mu_b A^\nu_c $
contains derivatives that will need to be taken into account.
We then get:
\begin{equation}
\label{eq:gluon_DTMD_1}
\begin{split}
  f_g^{\coh} =& 
\frac{p^+}{2\pi^2}
\frac{1-\xpom}{x}
\int \dd[2]{\xt} \dd{x^-} \dd[2]{\yt} \dd{y^-}
e^{-i (x^- - y^-)x p^+ + i (\xt - \yt) \vdot \kt}
\\
&\times
 \int \frac{\dd[4]{q}\dd[4]{q'} \dd[4]{l}}{(2\pi)^{12}}
 \frac{(-2 l^-)^2 \Pi_{\mu \nu}(l) }{\qty[q^2 + i \varepsilon]\qty[q'^2 - i \varepsilon]}
\delta(  l^+ - \xpom \qty[1-\beta] p^+)  \delta^{(2)}(\Deltat + \lt + \kt) 
\\
&\times 
\qty[
\Pi^{\mu i}(q) \partial_-^y
+
\Pi^{\mu +}(q) \partial_i^y
]
\qty(
\theta( - y^-)
e^{i q^+ y^- - i \qt \vdot \yt}
) \\
&\times 
\qty[
\Pi^{\nu i}(q') \partial_-^x
+
\Pi^{\nu +}(q') \partial_i^x
]
\qty(
\theta( - x^-)
e^{-i q^{\prime +}  x^- + i\qt' \vdot \xt}
)
\\
&
\times
 2\pi  \delta(l^2) \theta(l^-) \times
(2\pi)^2 \delta(l^- - q^-) \delta(l^- - q'^-)
\int \dd[2]{\ut} \dd[2]{\vt} e^{-i (\lt -\qt)\vdot \vt + i (\lt - \qt')\vdot \ut} \\
&\times
\sum_{\lambda'}
\Tr \expval{-U_{ab}(\ut) V(\xt) t^b V^\dag(\xt) + t^a}_{\lambda' \lambda}
\expval{
 -V(\yt)t^c V^\dag(\yt) U_{ac}(\vt) + t^a }_{\lambda \lambda'},
\end{split}
\end{equation}
where $\Pi_{\mu \nu}$ is the numerator of the gluon propagator~\eqref{eq:gluon_prop}.
The theta functions $\theta(-y^-)$ and $\theta(-x^-)$ should be understood to come from the gluon propagator, and for this reason the derivatives from the gluon field $G^{\mu \nu}$ also act on them.
Taking the derivatives of these theta functions then gives
us additional delta function contributions that set $y^- =0$ or $x^- =0$.
With this in mind, we can integrate over $x^-$, $y^-$, $l$, $q^-$ and $q'^-$:
\begin{equation}
\label{eq:gluon_DTMD_2}
\begin{split}
  f_g^{\coh} =& 
\frac{p^+}{\pi^2}
\frac{1-\xpom}{x}
\frac{(l^-)^2}{l^+ (2\pi)^3}
\int \dd[2]{\xt} \dd[2]{\yt}
\int \dd[2]{\ut} \dd[2]{\vt}
 \int \frac{\dd[2]{\qt}  \dd[2]{\qt'} \dd{q^+} \dd{q'^+} }{(2\pi)^{6}}
e^{ i (\xt - \yt) \vdot \kt}
\\
&\times
 \frac{-1}{\qty[2l^- q^+ -\qt^2 + i \varepsilon]\qty[2 l^- q'^+ -\qt'^2 - i \varepsilon]
 \qty[xp^+ + q^+ - i\varepsilon]
 \qty[-xp^+ - q'^+ - i\varepsilon]
 }
\\
&\times 
\qty[
\Pi^{j i}(q) x p^+
+
\Pi^{j +}(q) \qt^i
]
\qty[
\Pi^{j i}(q') x p^+
+
\Pi^{j +}(q') \qt'^i
]
 e^{-i \qt \vdot \yt + i \qt' \vdot \xt}
\\
&
\times e^{-i (\lt -\qt)\vdot \vt + i (\lt - \qt')\vdot \ut} 
\sum_{\lambda'}
\frac{1}{2} \expval{-U_{ab}(\ut) U_{db}(\xt) + \boldsymbol{1}_{ad}}_{\lambda' \lambda}
\expval{
 -U_{cd}(\yt)U_{ca}(\vt) + \boldsymbol{1}_{ad} }_{\lambda \lambda'},
\end{split}
\end{equation}
where  $l^+ = \xpom \qty[1-\beta] p^+$, $\lt = - \kt - \Deltat$, and $l^- =  \lt^2/(2l^+)$.
We have also used the light-cone gauge condition $\Pi^{- \mu} = \Pi^{\mu -} = 0$ to write 
\begin{equation}
\Pi^{\mu \rho}(q) \Pi^{\nu \sigma}(q') \Pi_{\mu \nu}(l) =
\Pi^{j \rho}(q) \Pi^{k \sigma}(q') \Pi^{jk}(l)
=\Pi^{j \rho}(q) \Pi^{j \sigma}(q'),   
\end{equation}
and the identity
\begin{equation}
    V(\xt)t^a V^\dag(\xt) = U_{ba}(\xt) t^b
\end{equation}
connecting the fundamental and adjoint Wilson lines.
Noting that the dependence on $q^+$ and $q'^+$ is only in the denominators and not in the functions $\Pi^{\mu \nu}(q)$ and $\Pi^{\mu \nu}(q')$, we can integrate over these variables:
\begin{equation}
\label{eq:gluon_DTMD_3}
\begin{split}
  f_g^{\coh} =& 
\frac{p^+}{2\pi^2}
\frac{1-\xpom}{x}
\frac{(l^-)^2}{l^+ (2\pi)^3}
\int \dd[2]{\xt}  \dd[2]{\yt}
 \dd[2]{\ut} \dd[2]{\vt}
 \int \frac{\dd[2]{\qt}  \dd[2]{\qt'}  }{(2\pi)^{4}}
e^{ i (\xt - \yt) \vdot \kt
+ i \lt \vdot (\ut -\vt)}
\\
&\times 
\qty[
 \delta^{ij} x p^+
+
\frac{\qt^i \qt^j}{l^-}
]
\qty[
 \delta^{ij} x p^+
+
\frac{\qt'^i \qt'^j}{l^-}
]
\frac{e^{
-i \qt \vdot (\yt -\vt) + i \qt' \vdot (\xt-\ut)}}{
\qty[ 2 l^- x p^+ + \qt^2 ]
\qty[ 2 l^- x p^+ + \qt'^2 ]
}
\\
&
\times
\sum_{\lambda'}
\Tr \expval{-U(\ut) U^\dag(\xt) + 1}_{\lambda' \lambda}
\expval{
 -U(\yt)U^\dag(\vt) + 1 }_{\lambda \lambda'}.
\end{split}
\end{equation}
The Fourier transforms over $\qt$ and $\qt'$ can be done analytically:
\begin{equation}
\begin{split}
    &\int \frac{\dd[2]{\qt}}{(2\pi)^2}
    \frac{e^{i \rt \vdot \qt}}{
    \qt^2 + m^2
    }
    \qty[
    - \delta^{ij}
    + 2 \frac{\qt^i \qt^j}{m^2}
    ]
    =
    \frac{1}{m^2}
    \int \frac{\dd[2]{\qt}}{(2\pi)^2}
    e^{i \rt \vdot \qt}
    \qty[
 \delta^{ij}
 +
 \frac{- \delta^{ij} \qt^2
    + 2 \qt^i \qt^j}{\qt^2 +m^2}
    ]
    \\
    =&
    \frac{\delta^{ij}}{m^2}
    \delta^{(2)}(\rt)
    +
    \frac{1}{2\pi m^2}
    \qty[
    \delta^{ij}
    \nabla^2_\rt
    - 2 \partial^i_\rt \partial^j_\rt
    ]
    K_0(m \abs{\rt})
    =
    \frac{\delta^{ij}}{m^2}
    \delta^{(2)}(\rt)
    +
    \frac{1}{2\pi}
    \qty[
    \delta^{ij}
    - 2 \frac{\rt^i \rt^j}{\rt^2}
    ]
    K_2(m \abs{\rt}),
\end{split}
\end{equation}
where we have used~\cite{Hanninen:2017ddy}
\begin{equation}
    \int \frac{\dd[2]{\qt}}{(2\pi)^2}
    \frac{e^{i \rt \vdot \qt}}{
    \qt^2 + m^2
    }
    = \frac{1}{2\pi} K_0(m \abs{\rt}).
\end{equation}
The delta function contribution can be neglected when combined with the Wilson lines, allowing us to write:
\begin{equation}
\label{eq:gluon_DTMD_4}
\begin{split}
  f_g^{\coh} =& 
\frac{p^+}{2\pi^2}
\frac{1-\xpom}{x}
\frac{(l^-)^2}{l^+ (2\pi)^5}
(xp^+)^2
\int \dd[2]{\xt}  \dd[2]{\yt}
 \dd[2]{\ut} \dd[2]{\vt}
e^{ i (\xt - \yt) \vdot \kt
+ i \lt \vdot (\ut -\vt)}
\\
&\times 
    \qty[
    \delta^{ij}
    - 2 \frac{(\yt -\vt)^i (\yt -\vt)^j}{(\yt -\vt)^2}
    ]
    \qty[
     \delta^{ij}
    - 2 \frac{(\xt -\ut)^i (\xt -\ut)^j}{(\xt -\ut)^2}
    ] \\
    &\times
    K_2\qty(
 \abs{\yt -\vt}
\abs{\lt }
 \sqrt{\frac{\beta}{1-\beta}}
 )
 K_2\qty(
 \abs{\xt -\ut}
 \abs{\lt }
 \sqrt{\frac{\beta}{1-\beta}}
 )
\\
&
\times
\sum_{\lambda'}
\Tr \expval{-U(\ut) U^\dag(\xt) + 1}_{\lambda' \lambda}
\expval{
 -U(\yt)U^\dag(\vt) + 1 }_{\lambda \lambda'}.
\end{split}
\end{equation}
We can now use the color neutrality~\eqref{eq:diffractive_trace} to write this in terms of the adjoint dipole amplitude
\begin{equation}
    \mathcal{N}_{\lambda \lambda'}(\xt ,\yt)
    = 1 - \frac{1}{\nc^2 - 1}
    \Tr \expval{U(\xt) U^\dag(\yt)}_{\lambda \lambda'},
\end{equation}
giving us:
\begin{equation}
\label{eq:gluon_DTMD_5}
\begin{split}
  f_g^{\coh} =& 
  \frac{\nc^2-1
  }{(2\pi)^7}
\frac{(1-\xpom)\beta}{\xpom^2(1-\beta)^3}
(\kt + \Deltat)^4
\int \dd[2]{\xt}  \dd[2]{\yt}
 \dd[2]{\ut} \dd[2]{\vt}
e^{ -i (\xt - \yt) \vdot \Deltat
- i (\ut -\xt + \yt -\vt) \vdot (\kt + \Deltat)}
\\
    &\times
    K_2\qty(
 \abs{\yt -\vt}
\abs{\kt + \Deltat }
 \sqrt{\frac{\beta}{1-\beta}}
 )
 K_2\qty(
 \abs{\xt -\ut}
 \abs{\kt + \Deltat }
 \sqrt{\frac{\beta}{1-\beta}}
 )
\\
&
\times
 \cos(2 \qty[\varphi_{\yt-\vt} - \varphi_{\xt-\ut}   ])
 \times
\sum_{\lambda'}
\mathcal{N}_{\lambda \lambda'}(\yt,\vt)
 \mathcal{N}_{\lambda' \lambda}(\ut,\xt),
\end{split}
\end{equation}
where we have substituted the value of the momentum $l$, and $\varphi_\rt$ denotes the azimuthal angle of the transverse vector $\rt$.

This expression can be compared to that found in Ref.~\cite{Hatta:2022lzj}.
We find exact matching when $\Deltat =0$, but for a general transverse momentum exchange $\Deltat$ we find that the results in Ref.~\cite{Hatta:2022lzj} need to be slightly modified\footnote{We thank Feng Yuan for discussions about the validity of the results in Ref.~\cite{Hatta:2022lzj} beyond $\Deltat =0$.}.

\subsection{Quark}

The quark diffractive TMD is defined as~\cite{Hatta:2022lzj}:
\begin{equation}
\begin{split}
&f_q^{\coh}( \kt, \Deltat, \xpom, \beta; \lambda) \\
=&
\frac{1}{16\pi^2}
 \sumint[X]
 \sum_{\lambda'}
\int \frac{\dd{r^-} \dd[2]{\rt}}{(2\pi)^3} e^{i x r^- p^+ - i \rt \vdot \kt} \\
&\times
    \mel{p}{\bar \psi \qty(-\frac{r}{2})\mathcal{W}_\neg^\dag\qty[- \frac{r}{2}]}{p' X}
    \frac{\gamma^+}{2}
    \mel{p'X}{\mathcal{W}_\neg\qty[\frac{r}{2}] \psi \qty(\frac{r}{2})}{p} \\
=& 
\frac{(p^+)^2}{4\pi^2}
(1 - \xpom) 
 \sumint[X] \sum_{\lambda'}
\int \dd{x^-} \dd[2]{\xt} 
\int \dd{y^{-}} \dd[2]{\yt}\\
& \times e^{
- i x p^+(x^- -y^- ) + i (\xt-\yt) \vdot \kt
} 
\delta(  p_X^+ - \xpom \qty[1-\beta] p^+)  \delta^{(2)}(\Deltat + \pt_X + \kt)
\\
&\times
    \expval{\bar\psi \qty(x) \mathcal{W}_\neg^\dag\qty[x]}_{\lambda' \lambda}
  \ket{X}
  \frac{\gamma^+}{2}
  \bra{X}
    \expval{\mathcal{W}_\neg\qty[y] \psi \qty(y)}_{\lambda \lambda'}
\end{split}
\end{equation}
This can be computed from the same diagram as the quark PDF,  Fig.~\ref{fig:quark_PDF}. 
This yields:
\begin{equation}
\label{eq:quark_DTMD_1}
\begin{split}
  f_q^{\coh} =& 
\frac{(p^+)^2}{8\pi^2}
(1 - \xpom) 
\int \dd[2]{\xt} \dd{x^-} \dd[2]{\yt} \dd{y^-}
e^{-i (x^- - y^-)x p^+ + i (\xt - \yt) \vdot \kt}
\times
\theta(- x^-) \theta( - y^-)
\\
&\times
 \int \frac{\dd[4]{q}\dd[4]{q'} \dd[4]{l}}{(2\pi)^{12}}
\Tr[
\gamma^+ \frac{i\slashed{q}}{q^2 + i \varepsilon} \gamma^- \slashed{l} \gamma^- \frac{-i \slashed{q'}}{q'^2 - i\varepsilon}
]
e^{-i q^+ y^- + i \qt \vdot \yt + i q^{\prime +}  x^- - i\qt' \vdot \xt}
\\
&\times 
\delta(  l^+ - \xpom \qty[1-\beta] p^+)  \delta^{(2)}(\Deltat + \lt + \kt) 
\\
&
\times
 2\pi  \delta(l^2) \theta(l^-) \times
(2\pi)^2 \delta(l^- + q^-) \delta(l^- + q'^-)
\int \dd[2]{\ut} \dd[2]{\vt} e^{-i (\lt +\qt)\vdot \vt + i (\lt + \qt')\vdot \ut} \\
&\times
\sum_{\lambda'}
\Tr \expval{-V(\ut) V^\dag(\xt) + 1}_{\lambda' \lambda}
\expval{
 -V(\yt)V^\dag(\vt) + 1 }_{\lambda \lambda'}.
\end{split}
\end{equation}
Evaluating the trace and integrating over $l$, $q^-$, $q'^{-}$, $x^-$, and $y^-$ this becomes
\begin{equation}
\label{eq:quark_DTMD_2}
\begin{split}
  f_q^{\coh} =& 
\frac{(p^+)^2}{8\pi^2}
\frac{1 - \xpom}{(2\pi)^3} 
\frac{l^-}{l^+}
\int \dd[2]{\xt}\dd[2]{\yt} \dd[2]{\ut} \dd[2]{\vt}
 \int \frac{ \dd{q^+} \dd{q'^+} \dd[2]{\qt}\dd[2]{\qt'} }{(2\pi)^{6}}
\\
&\times
\frac{4 \qt \vdot \qt'}{
\qty[-2l^- q^+ -\qt^2 + i\varepsilon ]  
\qty[-2l^- q'^+ -\qt'^2 - i\varepsilon ]  
}
\frac{-1}{
\qty[x p^+ -q^+ - i\varepsilon ]
\qty[-xp^+ + q'^+ - i\varepsilon ]
}
\\
&\times
e^{i (\xt - \yt) \vdot \kt
+i \qt \vdot \yt  - i\qt' \vdot \xt
-i (\lt +\qt)\vdot \vt + i (\lt + \qt')\vdot \ut} 
\times
\nc 
\sum_{\lambda'}
N_{\lambda' \lambda}(\ut,\xt)
N_{\lambda \lambda'}(\yt,\vt).
\end{split}
\end{equation}
We have also used Eq.~\eqref{eq:diffractive_trace} to write the trace in terms of dipole amplitudes, and we have used the delta function to set the components of the momentum $l$ as
 $l^+ = \xpom \qty[1-\beta] p^+$, $\lt = - \kt - \Deltat$, and $l^- =  \lt^2/(2l^+)$.
Integrating now over $q^+$ and $q'^+$ we get
\begin{equation}
\label{eq:quark_DTMD_3}
\begin{split}
  f_q^{\coh} =& 
\frac{(p^+)^2}{8\pi^2}
\frac{1 - \xpom}{(2\pi)^3} 
\frac{l^-}{l^+}
\int \dd[2]{\xt}\dd[2]{\yt} \dd[2]{\ut} \dd[2]{\vt}
 \int \frac{ \dd[2]{\qt}\dd[2]{\qt'} }{(2\pi)^{4}}
\frac{4 \qt \vdot \qt'}{
\qty[2l^- x p^+ +\qt^2  ]  
\qty[2l^- x p^+ +\qt'^2  ]  
}
\\
&\times
e^{i (\xt - \yt) \vdot \kt
+i \qt \vdot \yt  - i\qt' \vdot \xt
-i (\lt +\qt)\vdot \vt + i (\lt + \qt')\vdot \ut} 
\times
\nc 
\sum_{\lambda'}
N_{\lambda' \lambda}(\ut,\xt)
N_{\lambda \lambda'}(\yt,\vt).
\end{split}
\end{equation}
Finally, the Fourier transforms over $\qt$ and $\qt'$ can be done using Eq.~\eqref{eq:K1_FT}:
\begin{equation}
\label{eq:quark_DTMD_4}
\begin{split}
  f_q^{\coh} =& 
\frac{\nc}{(2\pi)^7}
\frac{(1-\xpom)\beta}{\xpom^2(1-\beta)^3}
(\kt + \Deltat)^4
\int \dd[2]{\xt}\dd[2]{\yt} \dd[2]{\ut} \dd[2]{\vt}
e^{-i (\xt - \yt) \vdot \Deltat
 - i  (\ut -\xt + \yt - \vt  ) \vdot (\kt + \Deltat)} 
\\
&\times
\frac{(\yt -\vt)\vdot (\xt -\ut) }{\abs{\yt -\vt} \abs{\xt -\ut}}
 K_1\qty(
 \abs{\yt -\vt}
\abs{\kt + \Deltat }
 \sqrt{\frac{\beta}{1-\beta}}
 )
 K_1\qty(
 \abs{\xt -\ut}
 \abs{\kt + \Deltat }
 \sqrt{\frac{\beta}{1-\beta}}
 )\\
 &
\times
\sum_{\lambda'}
N_{\lambda' \lambda}(\ut,\xt)
N_{\lambda \lambda'}(\yt,\vt).
\end{split}
\end{equation}
We again find exact matching with Ref.~\cite{Hatta:2022lzj} when $\Deltat= 0$, but in the general case our expressions are different.
This difference is similar to that in the diffractive gluon distribution case.

Compared to the quark GPD and PDF, the diffractive PDF will now be finite at leading order without renormalization.
This is because the diffractive gluon distribution is not enhanced by a factor $1/\alpha_s$, unlike in the gluon GPD and PDF cases.
For this reason, we can work at $D=4$ also for the quark diffractive PDF, and therefore the result can be obtained from the diffractive TMD using Eq.~\eqref{eq:DPDF} by integrating over $\kt$.

Similarly to standard TMDs, we can define the antiquark diffractive TMD as
\begin{equation}
\begin{split}
&f_{\bar q}^{\coh}( \kt, \Deltat, \xpom, \beta; \lambda) \\
=&
\frac{1}{16\pi^2}
 \sumint[X]
 \sum_{\lambda'}
\int \frac{\dd{r^-} \dd[2]{\rt}}{(2\pi)^3} e^{i x r^- p^+ - i \rt \vdot \kt} \\
&\times
\Tr \frac{\gamma^+}{2}
    \mel{p}{ \psi \qty(-\frac{r}{2})\mathcal{W}_\neg\qty[- \frac{r}{2}]}{p' X}
    \mel{p'X}{\mathcal{W}^\dag_\neg\qty[\frac{r}{2}] \bar \psi \qty(\frac{r}{2})}{p} \\
=& 
\frac{(p^+)^2}{4\pi^2}
(1 - \xpom) 
 \sumint[X] \sum_{\lambda'}
\int \dd{x^-} \dd[2]{\xt} 
\int \dd{y^{-}} \dd[2]{\yt}\\
& \times e^{
- i x p^+(x^- -y^- ) + i (\xt-\yt) \vdot \kt
} 
\delta(  p_X^+ - \xpom \qty[1-\beta] p^+)  \delta^{(2)}(\Deltat + \pt_X + \kt)
\\
&\times
\Tr
  \frac{\gamma^+}{2}
    \expval{\psi \qty(x) \mathcal{W}_\neg\qty[x]}_{\lambda' \lambda}
  \ket{X}
  \bra{X}
    \expval{\mathcal{W}^\dag_\neg\qty[y] \bar\psi \qty(y)}_{\lambda \lambda'}.
\end{split}
\end{equation}
This can be computed using the same diagram as the antiquark PDF, Fig.~\ref{fig:antiquark_PDF}.
We end up with the expression:
\begin{equation}
\label{eq:antiquark_DTMD_1}
\begin{split}
  f_{\bar q}^{\coh} =& 
\frac{(p^+)^2}{8\pi^2}
(1 - \xpom) 
\int \dd[2]{\xt} \dd{x^-} \dd[2]{\yt} \dd{y^-}
e^{-i (x^- - y^-)x p^+ + i (\xt - \yt) \vdot \kt}
\times
\theta(- x^-) \theta( - y^-)
\\
&\times
 \int \frac{\dd[4]{q}\dd[4]{q'} \dd[4]{l}}{(2\pi)^{12}}
\Tr[
\gamma^+ \frac{-i\slashed{q'}}{q'^2 - i \varepsilon} \gamma^- \slashed{l} \gamma^- \frac{i \slashed{q}}{q^2 + i\varepsilon}
]
e^{i q^+ y^- - i \qt \vdot \yt - i q^{\prime +}  x^- + i\qt' \vdot \xt}
\\
&\times 
\delta(  l^+ - \xpom \qty[1-\beta] p^+)  \delta^{(2)}(\Deltat + \lt + \kt) 
\\
&
\times
 2\pi  \delta(l^2) \theta(l^-) \times
(2\pi)^2 \delta(l^- - q^-) \delta(l^- - q'^-)
\int \dd[2]{\ut} \dd[2]{\vt} e^{-i (\lt -\qt)\vdot \vt + i (\lt - \qt')\vdot \ut} \\
&\times
\sum_{\lambda'}
\Tr \expval{-V(\xt) V^\dag(\ut) + 1}_{\lambda' \lambda}
\expval{
 -V(\vt)V^\dag(\yt) + 1 }_{\lambda \lambda'}.
\end{split}
\end{equation}
Similarly to the antiquark TMD and PDF, we can notice that the difference to the quark case corresponds to changing the Wilson-line structure to its complex conjugate.
Therefore, we can read the antiquark diffractive TMD from the quark case as:
\begin{equation}
\label{eq:antiquark_DTMD_2}
\begin{split}
  f_{\bar q}^{\coh} =& 
\frac{\nc}{(2\pi)^7}
\frac{(1-\xpom)\beta}{\xpom^2(1-\beta)^3}
(\kt + \Deltat)^4
\int \dd[2]{\xt}\dd[2]{\yt} \dd[2]{\ut} \dd[2]{\vt}
e^{-i (\xt - \yt) \vdot \Deltat
 - i  (\ut -\xt + \yt - \vt  ) \vdot (\kt + \Deltat)} 
\\
&\times
\frac{(\yt -\vt)\vdot (\xt -\ut) }{\abs{\yt -\vt} \abs{\xt -\ut}}
 K_1\qty(
 \abs{\yt -\vt}
\abs{\kt + \Deltat }
 \sqrt{\frac{\beta}{1-\beta}}
 )
 K_1\qty(
 \abs{\xt -\ut}
 \abs{\kt + \Deltat }
 \sqrt{\frac{\beta}{1-\beta}}
 )\\
 &
\times
\sum_{\lambda'}
N_{\lambda' \lambda}(\xt,\ut)
N_{\lambda \lambda'}(\vt,\yt).
\end{split}
\end{equation}

\section{Conclusions} \label{sec:Conclusion}

In this work, we have shown how to calculate parton distributions in the shockwave limit starting from the operator-level definition.
This is done by following the set of Feynman rules outlined in Sec.~\ref{sec:theory}.
Our results match those found in the literature, except for diffractive distributions with nonzero transverse momentum exchange $\Deltat$. 
We have also calculated, for the first time, the GTMDs and TMDs with general gauge links, and the quark GPD and PDF including their finite pieces.
The quark GPD and PDF are especially interesting, as we already need to renormalize these results at leading order.
This perhaps surprising result follows from the fact that in the shockwave approximation gluon distributions are enhanced by a factor of $1/\as$ compared to the quark ones.

When deriving these results, we did not assume the relevant $x$-variable to be small.
Instead, we only employed the shockwave approximation, such that the interaction with the target is instantaneous in the light-cone time, treating the power counting in the width of the shockwave separately from small $x$.
However, as discussed in Sec.~\ref{sec:theory}, for the shockwave approximation to be valid we will in practice need to be in the small-$x$ limit.
This limit can easily be taken from our results, as it only affects the overall coefficients in the GTMDs and GPDs where we can set $\sqrt{1-\xi^2} \to 1$;
we note that this is the \textit{only} place where further approximations from the small-$x$ limit can be taken.

Our result for the quark GPD also clarifies the role of non-zero skewness $\xi$ for exclusive processes.
In many phenomenological studies at small $x$, the non-zero skewness has been taken into account by adding an additional multiplicative factor~\cite{Kowalski:2006hc} corresponding to a maximally skewed gluon distribution~\cite{Shuvaev:1999ce}.
We do not see the need for this factor in our results.
This skewness correction essentially arises from running the GPD evolution to large momentum scales and,
as such, it is possible that a similar effect could be achieved by using our result for the quark GPD as an initial condition for the GPD evolution and evolving it to the relevant scale in the process of interest.

For the diffractive TMDs, we find that our results differ from those found in the literature~\cite{Hatta:2022lzj} when $\Deltat \neq 0$.
For diffractive PDFs, corresponding to the diffractive TMDs integrated over the momentum $\kt$, we find agreement with the results found in Ref.~\cite{Hauksson:2024bvv}.
However, we note that factorization for diffractive TMDs is currently poorly known, as the factorization theorems have been properly established only for diffractive PDFs~\cite{Collins:2001ga,Lee:2025fml}, and
the form of our results for quark and gluon diffractive TMDs suggests that it might be more natural to use variables $(\kt +\Deltat, \Deltat)$ instead of $(\kt,\Deltat)$ for defining the diffractive TMDs.
Whereas $\kt$ is the transverse momentum of the outgoing particle relative to the initial proton, $\kt + \Deltat$ is the momentum of the outgoing particle relative to the outgoing diffractive state $X$ instead.
Future work on factorization for diffractive TMDs, along with their evolution, will give us insight into the proper choice of the relative transverse momentum.

The combined framework presented in this work serves as a starting point for future calculations of parton distributions in the shockwave approximation or, more precisely, in the high-energy limit where the CGC approach is valid. 
As the next step, it will be natural to consider this matching of parton distributions to the CGC at NLO, where we expect to see directly the interplay between the high-energy evolution and the collinear or TMD evolution.
Such studies will serve as important cross-checks for a consistent framework including both high-energy factorization and collinear or TMD factorization. 
Additionally, we note that we have restricted ourselves to the eikonal limit in this work, and the inclusion of sub-eikonal corrections will be an important direction for a future study, which will allows us to consider e.g. helicity distributions that capture spin-dependent effects~\cite{Kovchegov:2015pbl,Kovchegov:2015zha,Kovchegov:2016zex,Kovchegov:2017lsr,Kovchegov:2018zeq,Kovchegov:2018znm,Kovchegov:2019rrz,Kovchegov:2020hgb,Kovchegov:2021iyc,Kovchegov:2022kyy,Kovchegov:2023yzd,Kovchegov:2024wjs,Kovchegov:2025gcg,Borden:2024bxa,Adamiak:2023okq,Altinoluk:2024tyx,Altinoluk:2024zom,Agostini:2024xqs}. 
Furthermore, the matching developed here can be used directly to calculate higher-twist distributions in a similar manner~\cite{Fu:2023jqv,Fu:2024sba}.
We expect that these future directions will improve our understanding of the interplay between partonic and CGC descriptions of the hadron structure.

\section*{Acknowledgment}
We thank Edmond Iancu, Kyle Lee and Feng Yuan for discussions on diffractive distributions.
\update{We would also like to thank Yuri Kovchegov for discussions that have led to improving the clarity of the manuscript.}
Z.K., D.P., and J.P. are supported National Science Foundation under grant No.~PHY-2515057. This work is also supported by the U.S. Department of Energy, Office of Science, Office of Nuclear Physics, within the framework of the Saturated Glue (SURGE) Topical Theory Collaboration.

\bibliographystyle{JHEP-2modlong.bst}
\bibliography{references}

\end{document}

%% file: packages.tex
\usepackage[utf8]{inputenc}
\usepackage{xcolor}

\definecolor{lcolor}{rgb}{0.5,0,0}
\definecolor{citcolor}{rgb}{0,0.3,0.0}

\usepackage{amssymb}
\usepackage{amsmath}
\usepackage{tikz}
\usepackage{siunitx}
\usepackage{physics}

\usepackage{slashed}

\usepackage{overpic}

\usepackage{subcaption}

\ExplSyntaxOn
\msg_redirect_name:nnn{siunitx}{physics-pkg}{none}
\ExplSyntaxOff

\usepackage{graphicx}

\usepackage{mathtools}

%% file: definitions.tex
\newcommand{\lt}{\mathbf{l}}
\newcommand{\Pt}{\mathbf{P}}
\newcommand{\pt}{\mathbf{p}}
\newcommand{\rt}{\mathbf{r}}
\newcommand{\xt}{\mathbf{x}}
\newcommand{\yt}{\mathbf{y}}

\newcommand{\bt}{\mathbf{b}}
\newcommand{\kt}{\mathbf{k}}
\newcommand{\qt}{\mathbf{q}}
\newcommand{\ut}{\mathbf{u}}
\newcommand{\vt}{\mathbf{v}}

\newcommand{\nc}{N_c}
\newcommand{\xpom}{x_\mathbb{P}}

\newcommand{\li}{\text{Li}_2}

\newcommand{\as}{\alpha_s}

\newcommand{\Deltat}{\mathbf{\Delta}_\perp}

\newcommand{\coh}{\text{coh}}

\newcommand{\msbar}{\overline{\text{MS}}}

\newcommand{\cl}{\text{cl}}

\newcommand{\sumint}[1][]{\mathrlap{\sum_{#1} } \int \;}
 
\newcommand{\ineg}{\text{\reflectbox{$\neg$}}}

%% file: references.bib
@article{Kovchegov:2023yzd,
    author = "Kovchegov, Yuri V. and Manley, Brandon",
    title = "{Orbital angular momentum at small x revisited}",
    eprint = "2310.18404",
    archivePrefix = "arXiv",
    primaryClass = "hep-ph",
    doi = "10.1007/JHEP02(2024)060",
    journal = "JHEP",
    volume = "02",
    pages = "060",
    year = "2024",
    note = "[Erratum: JHEP 08, 140 (2024)]"
}

@article{Zhou:2018lfq,
    author = "Zhou, Jian",
    title = "{Scale dependence of the small x transverse momentum dependent gluon distribution}",
    eprint = "1807.00506",
    archivePrefix = "arXiv",
    primaryClass = "hep-ph",
    doi = "10.1103/PhysRevD.99.054026",
    journal = "Phys. Rev. D",
    volume = "99",
    number = "5",
    pages = "054026",
    year = "2019"
}

@article{Brodsky:2002cx,
    author = "Brodsky, Stanley J. and Hwang, Dae Sung and Schmidt, Ivan",
    title = "{Final state interactions and single spin asymmetries in semiinclusive deep inelastic scattering}",
    eprint = "hep-ph/0201296",
    archivePrefix = "arXiv",
    reportNumber = "SLAC-PUB-9135, USM-TH-121",
    doi = "10.1016/S0370-2693(02)01320-5",
    journal = "Phys. Lett. B",
    volume = "530",
    pages = "99--107",
    year = "2002"
}

@article{Qiu:2023mrm,
    author = "Qiu, Jian-Wei and Yu, Zhite",
    title = "{Extraction of the Parton Momentum-Fraction Dependence of Generalized Parton Distributions from Exclusive Photoproduction}",
    eprint = "2305.15397",
    archivePrefix = "arXiv",
    primaryClass = "hep-ph",
    reportNumber = "JLAB-THY-23-3828, JLAB-THY-23-3828, MSUHEP-23-015",
    doi = "10.1103/PhysRevLett.131.161902",
    journal = "Phys. Rev. Lett.",
    volume = "131",
    number = "16",
    pages = "161902",
    year = "2023"
}

@article{Qiu:2022pla,
    author = "Qiu, Jian-Wei and Yu, Zhite",
    title = "{Single diffractive hard exclusive processes for the study of generalized parton distributions}",
    eprint = "2210.07995",
    archivePrefix = "arXiv",
    primaryClass = "hep-ph",
    reportNumber = "MSUHEP-22-032, JLAB-THY-22-3742, JLAB-THY-22-3742, MSUHEP-22-032",
    doi = "10.1103/PhysRevD.107.014007",
    journal = "Phys. Rev. D",
    volume = "107",
    number = "1",
    pages = "014007",
    year = "2023"
}

@book{Peskin:1995ev,
    author = "Peskin, Michael E. and Schroeder, Daniel V.",
    title = "{An Introduction to quantum field theory}",
    doi = "10.1201/9780429503559",
    isbn = "978-0-201-50397-5, 978-0-429-50355-9, 978-0-429-49417-8",
    publisher = "Addison-Wesley",
    address = "Reading, USA",
    year = "1995"
}

@article{Bomhof:2006dp,
    author = "Bomhof, C. J. and Mulders, P. J. and Pijlman, F.",
    title = "{The Construction of gauge-links in arbitrary hard processes}",
    eprint = "hep-ph/0601171",
    archivePrefix = "arXiv",
    doi = "10.1140/epjc/s2006-02554-2",
    journal = "Eur. Phys. J. C",
    volume = "47",
    pages = "147--162",
    year = "2006"
}

@article{Ji:2004wu,
    author = "Ji, Xiang-dong and Ma, Jian-ping and Yuan, Feng",
    title = "{QCD factorization for semi-inclusive deep-inelastic scattering at low transverse momentum}",
    eprint = "hep-ph/0404183",
    archivePrefix = "arXiv",
    reportNumber = "DOE-ER-40762-308, UM-PP-04-037, UM-PP-{\#}04-037",
    doi = "10.1103/PhysRevD.71.034005",
    journal = "Phys. Rev. D",
    volume = "71",
    pages = "034005",
    year = "2005"
}

@article{Belitsky:2002sm,
    author = "Belitsky, Andrei V. and Ji, X. and Yuan, F.",
    title = "{Final state interactions and gauge invariant parton distributions}",
    eprint = "hep-ph/0208038",
    archivePrefix = "arXiv",
    reportNumber = "DOE-ER-40762-263, UMD-PP-03-005",
    doi = "10.1016/S0550-3213(03)00121-4",
    journal = "Nucl. Phys. B",
    volume = "656",
    pages = "165--198",
    year = "2003"
}

@article{Collins:2002kn,
    author = "Collins, John C.",
    title = "{Leading twist single transverse-spin asymmetries: Drell-Yan and deep inelastic scattering}",
    eprint = "hep-ph/0204004",
    archivePrefix = "arXiv",
    doi = "10.1016/S0370-2693(02)01819-1",
    journal = "Phys. Lett. B",
    volume = "536",
    pages = "43--48",
    year = "2002"
}

@article{Achenbach:2023pba,
    author = "Achenbach, P. and others",
    title = "{The present and future of QCD}",
    eprint = "2303.02579",
    archivePrefix = "arXiv",
    primaryClass = "hep-ph",
    reportNumber = "JLAB-PHY-23-3808",
    doi = "10.1016/j.nuclphysa.2024.122874",
    journal = "Nucl. Phys. A",
    volume = "1047",
    pages = "122874",
    year = "2024"
}

@article{Manley:2024pcl,
    author = "Manley, Brandon",
    title = "{Orbital angular momentum small-x evolution: exact results in the large-N$_{c}$ limit}",
    eprint = "2401.05508",
    archivePrefix = "arXiv",
    primaryClass = "hep-ph",
    doi = "10.1007/JHEP04(2024)055",
    journal = "JHEP",
    volume = "04",
    pages = "055",
    year = "2024",
    note = "[Erratum: JHEP 09, 017 (2024)]"
}

@article{Kovchegov:2024wjs,
    author = "Kovchegov, Yuri V. and Manley, Brandon",
    title = "{Elastic dijet production in electron scattering on a longitudinally polarized proton at small x: A portal to orbital angular momentum distributions}",
    eprint = "2410.21260",
    archivePrefix = "arXiv",
    primaryClass = "hep-ph",
    doi = "10.1103/PhysRevD.111.054017",
    journal = "Phys. Rev. D",
    volume = "111",
    number = "5",
    pages = "054017",
    year = "2025"
}

@article{Bhattacharya:2018lgm,
    author = "Bhattacharya, Shohini and Metz, Andreas and Ojha, Vikash Kumar and Tsai, Jeng-Yuan and Zhou, Jian",
    title = "{Exclusive double quarkonium production and generalized TMDs of gluons}",
    eprint = "1802.10550",
    archivePrefix = "arXiv",
    primaryClass = "hep-ph",
    doi = "10.1016/j.physletb.2022.137383",
    journal = "Phys. Lett. B",
    volume = "833",
    pages = "137383",
    year = "2022"
}

@article{Bhattacharya:2017bvs,
    author = "Bhattacharya, Shohini and Metz, Andreas and Zhou, Jian",
    title = "{Generalized TMDs and the exclusive double Drell{\textendash}Yan process}",
    eprint = "1702.04387",
    archivePrefix = "arXiv",
    primaryClass = "hep-ph",
    doi = "10.1016/j.physletb.2017.05.081",
    journal = "Phys. Lett. B",
    volume = "771",
    pages = "396--400",
    year = "2017",
    note = "[Erratum: Phys.Lett.B 810, 135866 (2020)]"
}

@article{Bhattacharya:2022vvo,
    author = "Bhattacharya, Shohini and Boussarie, Renaud and Hatta, Yoshitaka",
    title = "{Signature of the Gluon Orbital Angular Momentum}",
    eprint = "2201.08709",
    archivePrefix = "arXiv",
    primaryClass = "hep-ph",
    doi = "10.1103/PhysRevLett.128.182002",
    journal = "Phys. Rev. Lett.",
    volume = "128",
    number = "18",
    pages = "182002",
    year = "2022"
}

@article{Bhattacharya:2023hbq,
    author = "Bhattacharya, Shohini and Zheng, Duxin and Zhou, Jian",
    title = "{Probing the Quark Orbital Angular Momentum at Electron-Ion Colliders Using Exclusive {\ensuremath{\pi}}0 Production}",
    eprint = "2312.01309",
    archivePrefix = "arXiv",
    primaryClass = "hep-ph",
    doi = "10.1103/PhysRevLett.133.051901",
    journal = "Phys. Rev. Lett.",
    volume = "133",
    number = "5",
    pages = "051901",
    year = "2024"
}

@article{Bhattacharya:2024sck,
    author = "Bhattacharya, Shohini and Boussarie, Renaud and Hatta, Yoshitaka",
    title = "{Exploring orbital angular momentum and spin-orbit correlations for gluons at the Electron-Ion Collider}",
    eprint = "2404.04209",
    archivePrefix = "arXiv",
    primaryClass = "hep-ph",
    doi = "10.1103/PhysRevD.111.034019",
    journal = "Phys. Rev. D",
    volume = "111",
    number = "3",
    pages = "034019",
    year = "2025"
}

@article{Accardi:2012qut,
    author = "Accardi, A. and others",
    editor = "Deshpande, A. and Meziani, Z. E. and Qiu, J. W.",
    title = "{Electron Ion Collider: The Next QCD Frontier}: {Understanding the glue that binds us all}",
    eprint = "1212.1701",
    archivePrefix = "arXiv",
    primaryClass = "nucl-ex",
    reportNumber = "BNL-98815-2012-JA, JLAB-PHY-12-1652",
    doi = "10.1140/epja/i2016-16268-9",
    journal = "Eur. Phys. J. A",
    volume = "52",
    number = "9",
    pages = "268",
    year = "2016"
}

@article{Boer:2011fh,
    author = "Boer, Daniel and others",
    title = "{Gluons and the quark sea at high energies: Distributions, polarization, tomography}",
    eprint = "1108.1713",
    archivePrefix = "arXiv",
    primaryClass = "nucl-th",
    reportNumber = "SLAC-R-995, INT-PUB-11-034, BNL-96164-2011, JLAB-THY-11-1373",
    month = "8",
    year = "2011"
}

@inproceedings{Lorce:2025aqp,
    author = "Lorc{\'e}, C{\'e}dric and Metz, Andreas and Pasquini, Barbara and Schweitzer, Peter",
    title = "{Parton Distribution Functions and their Generalizations}",
    eprint = "2507.12664",
    archivePrefix = "arXiv",
    primaryClass = "hep-ph",
    month = "7",
    year = "2025"
}

@article{Mankiewicz:1997bk,
    author = "Mankiewicz, L. and Piller, G. and Stein, E. and Vanttinen, M. and Weigl, T.",
    title = "{NLO corrections to deeply virtual Compton scattering}",
    eprint = "hep-ph/9712251",
    archivePrefix = "arXiv",
    reportNumber = "TUM-T39-97-31, DFTT-73-97",
    doi = "10.1016/S0370-2693(98)00190-7",
    journal = "Phys. Lett. B",
    volume = "425",
    pages = "186--192",
    year = "1998",
    note = "[Erratum: Phys.Lett.B 461, 423--423 (1999)]"
}

@article{Belitsky:1997rh,
    author = "Belitsky, Andrei V. and Mueller, Dieter",
    title = "{Predictions from conformal algebra for the deeply virtual Compton scattering}",
    eprint = "hep-ph/9709379",
    archivePrefix = "arXiv",
    reportNumber = "NTZ-23-97",
    doi = "10.1016/S0370-2693(97)01390-7",
    journal = "Phys. Lett. B",
    volume = "417",
    pages = "129--140",
    year = "1998"
}

@article{Ji:1997nk,
    author = "Ji, Xiang-Dong and Osborne, Jonathan",
    title = "{One loop QCD corrections to deeply virtual Compton scattering: The Parton helicity independent case}",
    eprint = "hep-ph/9707254",
    archivePrefix = "arXiv",
    reportNumber = "UMD-PP-97-001, DOE-ER-40762-124",
    doi = "10.1103/PhysRevD.57.R1337",
    journal = "Phys. Rev. D",
    volume = "57",
    pages = "1337--1340",
    year = "1998"
}

@article{Ji:2013dva,
    author = "Ji, Xiangdong",
    title = "{Parton Physics on a Euclidean Lattice}",
    eprint = "1305.1539",
    archivePrefix = "arXiv",
    primaryClass = "hep-ph",
    doi = "10.1103/PhysRevLett.110.262002",
    journal = "Phys. Rev. Lett.",
    volume = "110",
    pages = "262002",
    year = "2013"
}

@article{Lorce:2013pza,
    author = "Lorc\'e, C. and Pasquini, B.",
    title = "{Structure analysis of the generalized correlator of quark and gluon for a spin-1/2 target}",
    eprint = "1307.4497",
    archivePrefix = "arXiv",
    primaryClass = "hep-ph",
    doi = "10.1007/JHEP09(2013)138",
    journal = "JHEP",
    volume = "09",
    pages = "138",
    year = "2013"
}

@article{Radyushkin:1996nd,
    author = "Radyushkin, A. V.",
    title = "{Scaling limit of deeply virtual Compton scattering}",
    eprint = "hep-ph/9604317",
    archivePrefix = "arXiv",
    reportNumber = "CEBAF-TH-96-05",
    doi = "10.1016/0370-2693(96)00528-X",
    journal = "Phys. Lett. B",
    volume = "380",
    pages = "417--425",
    year = "1996"
}

@article{Ji:1996ek,
    author = "Ji, Xiang-Dong",
    title = "{Gauge-Invariant Decomposition of Nucleon Spin}",
    eprint = "hep-ph/9603249",
    archivePrefix = "arXiv",
    reportNumber = "MIT-CTP-2517",
    doi = "10.1103/PhysRevLett.78.610",
    journal = "Phys. Rev. Lett.",
    volume = "78",
    pages = "610--613",
    year = "1997"
}

@article{Burkardt:2000za,
    author = "Burkardt, Matthias",
    title = "{Impact parameter dependent parton distributions and off forward parton distributions for zeta ---{\ensuremath{>}} 0}",
    eprint = "hep-ph/0005108",
    archivePrefix = "arXiv",
    doi = "10.1103/PhysRevD.62.071503",
    journal = "Phys. Rev. D",
    volume = "62",
    pages = "071503",
    year = "2000",
    note = "[Erratum: Phys.Rev.D 66, 119903 (2002)]"
}

@article{Ji:2012sj,
    author = "Ji, Xiangdong and Xiong, Xiaonu and Yuan, Feng",
    title = "{Proton Spin Structure from Measurable Parton Distributions}",
    eprint = "1202.2843",
    archivePrefix = "arXiv",
    primaryClass = "hep-ph",
    doi = "10.1103/PhysRevLett.109.152005",
    journal = "Phys. Rev. Lett.",
    volume = "109",
    pages = "152005",
    year = "2012"
}

@article{Hatta:2011ku,
    author = "Hatta, Yoshitaka",
    title = "{Notes on the orbital angular momentum of quarks in the nucleon}",
    eprint = "1111.3547",
    archivePrefix = "arXiv",
    primaryClass = "hep-ph",
    doi = "10.1016/j.physletb.2012.01.024",
    journal = "Phys. Lett. B",
    volume = "708",
    pages = "186--190",
    year = "2012"
}

@article{Lorce:2011kd,
    author = "Lorce, C. and Pasquini, B.",
    title = "{Quark Wigner Distributions and Orbital Angular Momentum}",
    eprint = "1106.0139",
    archivePrefix = "arXiv",
    primaryClass = "hep-ph",
    doi = "10.1103/PhysRevD.84.014015",
    journal = "Phys. Rev. D",
    volume = "84",
    pages = "014015",
    year = "2011"
}

@article{Belitsky:2003nz,
    author = "Belitsky, Andrei V. and Ji, Xiang-dong and Yuan, Feng",
    title = "{Quark imaging in the proton via quantum phase space distributions}",
    eprint = "hep-ph/0307383",
    archivePrefix = "arXiv",
    doi = "10.1103/PhysRevD.69.074014",
    journal = "Phys. Rev. D",
    volume = "69",
    pages = "074014",
    year = "2004"
}

@article{Ji:2003ak,
    author = "Ji, Xiang-dong",
    title = "{Viewing the proton through 'color' filters}",
    eprint = "hep-ph/0304037",
    archivePrefix = "arXiv",
    doi = "10.1103/PhysRevLett.91.062001",
    journal = "Phys. Rev. Lett.",
    volume = "91",
    pages = "062001",
    year = "2003"
}

@article{Meissner:2009ww,
    author = "Meissner, Stephan and Metz, Andreas and Schlegel, Marc",
    title = "{Generalized parton correlation functions for a spin-1/2 hadron}",
    eprint = "0906.5323",
    archivePrefix = "arXiv",
    primaryClass = "hep-ph",
    reportNumber = "JLAB-THY-09-1018",
    doi = "10.1088/1126-6708/2009/08/056",
    journal = "JHEP",
    volume = "08",
    pages = "056",
    year = "2009"
}

@article{Collins:2001ga,
    author = "Collins, John C.",
    editor = "Grindhammer, Guenter and Kniehl, B. A. and Kramer, G. and Ochs, W.",
    title = "{Factorization in hard diffraction}",
    eprint = "hep-ph/0107252",
    archivePrefix = "arXiv",
    doi = "10.1088/0954-3899/28/5/327",
    journal = "J. Phys. G",
    volume = "28",
    pages = "1069--1078",
    year = "2002"
}

@article{Hauksson:2024bvv,
    author = "Hauksson, S. and Iancu, E. and Mueller, A. H. and Triantafyllopoulos, D. N. and Wei, S. Y.",
    title = "{TMD factorisation for diffractive jets in photon-nucleus interactions}",
    eprint = "2402.14748",
    archivePrefix = "arXiv",
    primaryClass = "hep-ph",
    doi = "10.1007/JHEP06(2024)180",
    journal = "JHEP",
    volume = "06",
    pages = "180",
    year = "2024"
}

@article{Berera:1995fj,
    author = "Berera, Arjun and Soper, Davison E.",
    title = "{Behavior of diffractive parton distribution functions}",
    eprint = "hep-ph/9509239",
    archivePrefix = "arXiv",
    reportNumber = "PSU-TH-163, OITS-581",
    doi = "10.1103/PhysRevD.53.6162",
    journal = "Phys. Rev. D",
    volume = "53",
    pages = "6162--6179",
    year = "1996"
}

@article{Diehl:2003ny,
    author = "Diehl, M.",
    title = "{Generalized parton distributions}",
    eprint = "hep-ph/0307382",
    archivePrefix = "arXiv",
    reportNumber = "DESY-THESIS-2003-018",
    doi = "10.1016/j.physrep.2003.08.002",
    journal = "Phys. Rept.",
    volume = "388",
    pages = "41--277",
    year = "2003"
}

@article{Golec-Biernat:2001gyl,
    author = "Golec-Biernat, Krzysztof J. and Wusthoff, M.",
    title = "{Diffractive parton distributions from the saturation model}",
    eprint = "hep-ph/0102093",
    archivePrefix = "arXiv",
    reportNumber = "DESY-00-180",
    doi = "10.1007/s100520100661",
    journal = "Eur. Phys. J. C",
    volume = "20",
    pages = "313--321",
    year = "2001"
}

@article{Buchmuller:1998jv,
    author = "Buchmuller, W. and Gehrmann, T. and Hebecker, Arthur",
    title = "{Inclusive and diffractive structure functions at small x}",
    eprint = "hep-ph/9808454",
    archivePrefix = "arXiv",
    reportNumber = "DESY-98-113, DAMTP-98-113",
    doi = "10.1016/S0550-3213(98)00682-8",
    journal = "Nucl. Phys. B",
    volume = "537",
    pages = "477--500",
    year = "1999"
}

@article{Hatta:2022lzj,
    author = "Hatta, Yoshitaka and Xiao, Bo-Wen and Yuan, Feng",
    title = "{Semi-inclusive diffractive deep inelastic scattering at small x}",
    eprint = "2205.08060",
    archivePrefix = "arXiv",
    primaryClass = "hep-ph",
    doi = "10.1103/PhysRevD.106.094015",
    journal = "Phys. Rev. D",
    volume = "106",
    number = "9",
    pages = "094015",
    year = "2022"
}

@article{Boussarie:2023izj,
    author = "Boussarie, Renaud and others",
    title = "{TMD Handbook}",
    eprint = "2304.03302",
    archivePrefix = "arXiv",
    primaryClass = "hep-ph",
    reportNumber = "JLAB-THY-23-3780, LA-UR-21-20798, MIT-CTP/5386",
    month = "4",
    year = "2023"
}

@book{Collins:2011zzd,
    author = "Collins, John",
    title = "{Foundations of Perturbative QCD}",
    doi = "10.1017/9781009401845",
    isbn = "978-1-009-40184-5, 978-1-009-40183-8, 978-1-009-40182-1",
    publisher = "Cambridge University Press",
    volume = "32",
    year = "2011"
}

@article{Boussarie:2023xun,
    author = "Boussarie, Renaud and Mehtar-Tani, Yacine",
    title = "{Low and moderate x gluon contribution to exclusive Compton scattering processes}",
    eprint = "2309.16576",
    archivePrefix = "arXiv",
    primaryClass = "hep-ph",
    doi = "10.1007/JHEP10(2024)056",
    journal = "JHEP",
    volume = "10",
    pages = "056",
    year = "2024"
}

@book {MR2360010,
    AUTHOR = {Gradshteyn, I. S. and Ryzhik, I. M.},
     TITLE = {Table of integrals, series, and products},
   EDITION = {Seventh},
 PUBLISHER = {Elsevier/Academic Press, Amsterdam},
      YEAR = {2007},
     PAGES = {xlviii+1171},
      ISBN = {978-0-12-373637-6; 0-12-373637-4},
   MRCLASS = {00A22 (33-00 65-00 65A05)},
  MRNUMBER = {2360010},
}

@article{Hanninen:2017ddy,
    author = {H{\"a}nninen, H. and Lappi, T. and Paatelainen, R.},
    title = "{One-loop corrections to light cone wave functions: the dipole picture DIS cross section}",
    eprint = "1711.08207",
    archivePrefix = "arXiv",
    primaryClass = "hep-ph",
    reportNumber = "HIP-2017-32-TH",
    doi = "10.1016/j.aop.2018.04.015",
    journal = "Annals Phys.",
    volume = "393",
    pages = "358--412",
    year = "2018"
}

@article{Lee:2025fml,
    author = "Lee, Kyle and Schindler, Stella T. and Stewart, Iain W.",
    title = "{Effective Field Theory Factorization for Diffraction}",
    eprint = "2508.10231",
    archivePrefix = "arXiv",
    primaryClass = "hep-ph",
    reportNumber = "MIT-CTP 5746, LA-UR-25-21807",
    month = "8",
    year = "2025"
}

@article{Hatta:2024vzv,
    author = "Hatta, Yoshitaka and Yuan, Feng",
    title = "{Angular dependence in transverse momentum dependent diffractive parton distributions at small-x}",
    eprint = "2403.19609",
    archivePrefix = "arXiv",
    primaryClass = "hep-ph",
    doi = "10.1016/j.physletb.2024.138738",
    journal = "Phys. Lett. B",
    volume = "854",
    pages = "138738",
    year = "2024"
}

@article{Huber:2005yg,
    author = "Huber, Tobias and Ma{\^\i}tre, Daniel",
    title = "{HypExp, a Mathematica package for expanding hypergeometric functions around integer-valued parameters}",
    eprint = "hep-ph/0507094",
    archivePrefix = "arXiv",
    reportNumber = "ZU-TH-13-05",
    doi = "10.1016/j.cpc.2006.01.007",
    journal = "Comput. Phys. Commun.",
    volume = "175",
    pages = "122--144",
    year = "2006"
}

@article{Huber:2007dx,
    author = "Huber, Tobias and Ma{\^\i}tre, Daniel",
    title = "{HypExp 2, Expanding hypergeometric functions about half-integer parameters}",
    eprint = "0708.2443",
    archivePrefix = "arXiv",
    primaryClass = "hep-ph",
    reportNumber = "SLAC-PUB-12748, PITHA-07-06",
    doi = "10.1016/j.cpc.2007.12.008",
    journal = "Comput. Phys. Commun.",
    volume = "178",
    pages = "755--776",
    year = "2008"
}

@article{McLerran:1993ni,
    author = "McLerran, Larry D. and Venugopalan, Raju",
    title = "{Computing quark and gluon distribution functions for very large nuclei}",
    eprint = "hep-ph/9309289",
    archivePrefix = "arXiv",
    reportNumber = "TPI-MINN-93-44-T, NUC-MINN-93-24-T, HEP-UMN-TH-1220-93",
    doi = "10.1103/PhysRevD.49.2233",
    journal = "Phys. Rev. D",
    volume = "49",
    pages = "2233--2241",
    year = "1994"
}

@article{McLerran:1994vd,
      author         = "McLerran, Larry D. and Venugopalan, Raju",
      title          = "{Green's functions in the color field of a large
                        nucleus}",
      journal        = "Phys. Rev.",
      volume         = "D50",
      year           = "1994",
      pages          = "2225-2233",
      doi            = "10.1103/PhysRevD.50.2225",
      eprint         = "hep-ph/9402335",
      archivePrefix  = "arXiv",
      primaryClass   = "hep-ph",
      reportNumber   = "TPI-MINN-94-7-T, NUC-MINN-94-2-T, HEP-MINN-94-1242-T",
      SLACcitation   = "%%CITATION = HEP-PH/9402335;%%"
}

@article{McLerran:1993ka,
      author         = "McLerran, Larry D. and Venugopalan, Raju",
      title          = "{Gluon distribution functions for very large nuclei at
                        small transverse momentum}",
      journal        = "Phys. Rev.",
      volume         = "D49",
      year           = "1994",
      pages          = "3352-3355",
      doi            = "10.1103/PhysRevD.49.3352",
      eprint         = "hep-ph/9311205",
      archivePrefix  = "arXiv",
      primaryClass   = "hep-ph",
      reportNumber   = "TPI-MINN-93-52-T, NUC-MINN-93-28-T, UMN-TH-1224-93",
      SLACcitation   = "%%CITATION = HEP-PH/9311205;%%"
}

@article{Moffat:2023svr,
    author = {Moffat, Eric and Freese, Adam and Clo{\"e}t, Ian and Donohoe, Thomas and Gamberg, Leonard and Melnitchouk, Wally and Metz, Andreas and Prokudin, Alexei and Sato, Nobuo},
    title = "{Shedding light on shadow generalized parton distributions}",
    eprint = "2303.12006",
    archivePrefix = "arXiv",
    primaryClass = "hep-ph",
    reportNumber = "JLAB-THY-23-3786",
    doi = "10.1103/PhysRevD.108.036027",
    journal = "Phys. Rev. D",
    volume = "108",
    number = "3",
    pages = "036027",
    year = "2023"
}

@article{Balitsky:1998ya,
    author = "Balitsky, Ian",
    title = "{Factorization and high-energy effective action}",
    eprint = "hep-ph/9812311",
    archivePrefix = "arXiv",
    reportNumber = "JLAB-THY-98-49",
    doi = "10.1103/PhysRevD.60.014020",
    journal = "Phys. Rev. D",
    volume = "60",
    pages = "014020",
    year = "1999"
}

@article{Balitsky:1995ub,
    author = "Balitsky, I.",
    title = "{Operator expansion for high-energy scattering}",
    eprint = "hep-ph/9509348",
    archivePrefix = "arXiv",
    reportNumber = "MIT-CTP-2470",
    doi = "10.1016/0550-3213(95)00638-9",
    journal = "Nucl. Phys. B",
    volume = "463",
    pages = "99--160",
    year = "1996"
}

@article{Marquet:2009ca,
    author = "Marquet, Cyrille and Xiao, Bo-Wen and Yuan, Feng",
    title = "{Semi-inclusive Deep Inelastic Scattering at small x}",
    eprint = "0906.1454",
    archivePrefix = "arXiv",
    primaryClass = "hep-ph",
    doi = "10.1016/j.physletb.2009.10.099",
    journal = "Phys. Lett. B",
    volume = "682",
    pages = "207--211",
    year = "2009"
}

@article{Dominguez:2011wm,
    author = "Dominguez, Fabio and Marquet, Cyrille and Xiao, Bo-Wen and Yuan, Feng",
    title = "{Universality of Unintegrated Gluon Distributions at small x}",
    eprint = "1101.0715",
    archivePrefix = "arXiv",
    primaryClass = "hep-ph",
    doi = "10.1103/PhysRevD.83.105005",
    journal = "Phys. Rev. D",
    volume = "83",
    pages = "105005",
    year = "2011"
}

@article{Boussarie:2018zwg,
    author = "Boussarie, Renaud and Hatta, Yoshitaka and Xiao, Bo-Wen and Yuan, Feng",
    title = {{Probing the Weizs{\"a}cker-Williams gluon Wigner distribution in $pp$ collisions}},
    eprint = "1807.08697",
    archivePrefix = "arXiv",
    primaryClass = "hep-ph",
    reportNumber = "YITP-18-67",
    doi = "10.1103/PhysRevD.98.074015",
    journal = "Phys. Rev. D",
    volume = "98",
    number = "7",
    pages = "074015",
    year = "2018"
}

@article{Xiao:2017yya,
    author = "Xiao, Bo-Wen and Yuan, Feng and Zhou, Jian",
    title = "{Transverse Momentum Dependent Parton Distributions at Small-x}",
    eprint = "1703.06163",
    archivePrefix = "arXiv",
    primaryClass = "hep-ph",
    doi = "10.1016/j.nuclphysb.2017.05.012",
    journal = "Nucl. Phys. B",
    volume = "921",
    pages = "104--126",
    year = "2017"
}

@article{Hatta:2017cte,
    author = "Hatta, Yoshitaka and Xiao, Bo-Wen and Yuan, Feng",
    title = "{Gluon Tomography from Deeply Virtual Compton Scattering at Small-x}",
    eprint = "1703.02085",
    archivePrefix = "arXiv",
    primaryClass = "hep-ph",
    reportNumber = "YITP-17-22",
    doi = "10.1103/PhysRevD.95.114026",
    journal = "Phys. Rev. D",
    volume = "95",
    number = "11",
    pages = "114026",
    year = "2017"
}

@article{Hatta:2016dxp,
    author = "Hatta, Yoshitaka and Xiao, Bo-Wen and Yuan, Feng",
    title = "{Probing the Small- x Gluon Tomography in Correlated Hard Diffractive Dijet Production in Deep Inelastic Scattering}",
    eprint = "1601.01585",
    archivePrefix = "arXiv",
    primaryClass = "hep-ph",
    reportNumber = "YITP-16-1",
    doi = "10.1103/PhysRevLett.116.202301",
    journal = "Phys. Rev. Lett.",
    volume = "116",
    number = "20",
    pages = "202301",
    year = "2016"
}

@article{Baier:1996sk,
    author = "Baier, R. and Dokshitzer, Yuri L. and Mueller, Alfred H. and Peigne, S. and Schiff, D.",
    title = "{Radiative energy loss and p(T) broadening of high-energy partons in nuclei}",
    eprint = "hep-ph/9608322",
    archivePrefix = "arXiv",
    reportNumber = "CU-TP-760, BI-TP-96-26, LPTHE-ORSAY-96-61",
    doi = "10.1016/S0550-3213(96)00581-0",
    journal = "Nucl. Phys. B",
    volume = "484",
    pages = "265--282",
    year = "1997"
}

@article{Mueller:1999wm,
    author = "Mueller, Alfred H.",
    title = "{Parton saturation at small x and in large nuclei}",
    eprint = "hep-ph/9904404",
    archivePrefix = "arXiv",
    reportNumber = "CU-TP-937",
    doi = "10.1016/S0550-3213(99)00394-6",
    journal = "Nucl. Phys. B",
    volume = "558",
    pages = "285--303",
    year = "1999"
}

@inproceedings{Mueller:2001fv,
    author = "Mueller, Alfred H.",
    title = "{Parton saturation: An Overview}",
    booktitle = "{Cargese Summer School on QCD Perspectives on Hot and Dense Matter}",
    eprint = "hep-ph/0111244",
    archivePrefix = "arXiv",
    reportNumber = "CU-TH-1035",
    pages = "45--72",
    month = "11",
    year = "2001"
}

@inbook{Iancu:2003xm,
    author = "Iancu, Edmond and Venugopalan, Raju",
    editor = "Hwa, Rudolph C. and Wang, Xin-Nian",
    title = "{The Color glass condensate and high-energy scattering in QCD}",
    booktitle = "{Quark-gluon plasma 4}",
    eprint = "hep-ph/0303204",
    archivePrefix = "arXiv",
    doi = "10.1142/9789812795533_0005",
    pages = "249--3363",
    month = "3",
    year = "2003"
}

@article{Penttala:2025tmp,
    author = "Penttala, Jani",
    title = "{Color-glass condensate beyond the Gaussian approximation}",
    eprint = "2507.18711",
    archivePrefix = "arXiv",
    primaryClass = "hep-ph",
    month = "7",
    year = "2025"
}

@article{Mantysaari:2024zxq,
    author = {M{\"a}ntysaari, Heikki and Penttala, Jani and Salazar, Farid and Schenke, Bj{\"o}rn},
    title = "{Finite-size effects on small-x evolution and saturation in proton and nuclear targets}",
    eprint = "2411.13533",
    archivePrefix = "arXiv",
    primaryClass = "hep-ph",
    doi = "10.1103/PhysRevD.111.054033",
    journal = "Phys. Rev. D",
    volume = "111",
    number = "5",
    pages = "054033",
    year = "2025"
}

@article{Ji:1998pc,
    author = "Ji, Xiang-Dong",
    title = "{Off forward parton distributions}",
    eprint = "hep-ph/9807358",
    archivePrefix = "arXiv",
    reportNumber = "UMD-PP-98-092, DOE-ER-40762-144",
    doi = "10.1088/0954-3899/24/7/002",
    journal = "J. Phys. G",
    volume = "24",
    pages = "1181--1205",
    year = "1998"
}

@book{Kovchegov:2012mbw,
    author = "Kovchegov, Yuri V. and Levin, Eugene",
    title = "{Quantum Chromodynamics at High Energy}",
    doi = "10.1017/9781009291446",
    isbn = "978-1-009-29144-6, 978-1-009-29141-5, 978-1-009-29142-2, 978-0-521-11257-4, 978-1-139-55768-9",
    publisher = "Oxford University Press",
    volume = "33",
    year = "2013"
}

@article{Kowalski:2006hc,
    author = "Kowalski, H. and Motyka, L. and Watt, G.",
    title = "{Exclusive diffractive processes at HERA within the dipole picture}",
    eprint = "hep-ph/0606272",
    archivePrefix = "arXiv",
    reportNumber = "DESY-06-095",
    doi = "10.1103/PhysRevD.74.074016",
    journal = "Phys. Rev. D",
    volume = "74",
    pages = "074016",
    year = "2006"
}

@article{Shuvaev:1999ce,
    author = "Shuvaev, A. G. and Golec-Biernat, Krzysztof J. and Martin, Alan D. and Ryskin, M. G.",
    title = "{Off diagonal distributions fixed by diagonal partons at small x and xi}",
    eprint = "hep-ph/9902410",
    archivePrefix = "arXiv",
    reportNumber = "DTP-99-18",
    doi = "10.1103/PhysRevD.60.014015",
    journal = "Phys. Rev. D",
    volume = "60",
    pages = "014015",
    year = "1999"
}

@article{Kovchegov:2015zha,
    author = "Kovchegov, Yuri V. and Sievert, Matthew D.",
    title = "{Calculating TMDs of a Large Nucleus: Quasi-Classical Approximation and Quantum Evolution}",
    eprint = "1505.01176",
    archivePrefix = "arXiv",
    primaryClass = "hep-ph",
    doi = "10.1016/j.nuclphysb.2015.12.008",
    journal = "Nucl. Phys. B",
    volume = "903",
    pages = "164--203",
    year = "2016"
}

@article{Kovchegov:2017lsr,
    author = "Kovchegov, Yuri V. and Pitonyak, Daniel and Sievert, Matthew D.",
    title = "{Small-$x$ Asymptotics of the Gluon Helicity Distribution}",
    eprint = "1706.04236",
    archivePrefix = "arXiv",
    primaryClass = "nucl-th",
    reportNumber = "LA-UR-16-27995",
    doi = "10.1007/JHEP10(2017)198",
    journal = "JHEP",
    volume = "10",
    pages = "198",
    year = "2017"
}

@article{Kovchegov:2015pbl,
    author = "Kovchegov, Yuri V. and Pitonyak, Daniel and Sievert, Matthew D.",
    title = "{Helicity Evolution at Small-x}",
    eprint = "1511.06737",
    archivePrefix = "arXiv",
    primaryClass = "hep-ph",
    reportNumber = "RBRC-1159, BNL-111624-2015-JA",
    doi = "10.1007/JHEP01(2016)072",
    journal = "JHEP",
    volume = "01",
    pages = "072",
    year = "2016",
    note = "[Erratum: JHEP 10, 148 (2016)]"
}

@article{Kovchegov:2016zex,
    author = "Kovchegov, Yuri V. and Pitonyak, Daniel and Sievert, Matthew D.",
    title = "{Helicity Evolution at Small $x$: Flavor Singlet and Non-Singlet Observables}",
    eprint = "1610.06197",
    archivePrefix = "arXiv",
    primaryClass = "hep-ph",
    reportNumber = "LA-UR-16-27996",
    doi = "10.1103/PhysRevD.95.014033",
    journal = "Phys. Rev. D",
    volume = "95",
    number = "1",
    pages = "014033",
    year = "2017"
}

@article{Kovchegov:2022kyy,
    author = "Kovchegov, Yuri V. and Santiago, M. Gabriel",
    title = "{T-odd leading-twist quark TMDs at small x}",
    eprint = "2209.03538",
    archivePrefix = "arXiv",
    primaryClass = "hep-ph",
    doi = "10.1007/JHEP11(2022)098",
    journal = "JHEP",
    volume = "11",
    pages = "098",
    year = "2022"
}

@article{Kovchegov:2021iyc,
    author = "Kovchegov, Yuri V. and Santiago, M. Gabriel",
    title = "{Quark sivers function at small $x$: spin-dependent odderon and the sub-eikonal evolution}",
    eprint = "2108.03667",
    archivePrefix = "arXiv",
    primaryClass = "hep-ph",
    doi = "10.1007/JHEP11(2021)200",
    journal = "JHEP",
    volume = "11",
    pages = "200",
    year = "2021",
    note = "[Erratum: JHEP 09, 186 (2022)]"
}

@article{Kovchegov:2020hgb,
    author = "Kovchegov, Yuri V. and Tawabutr, Yossathorn",
    title = "{Helicity at Small $x$: Oscillations Generated by Bringing Back the Quarks}",
    eprint = "2005.07285",
    archivePrefix = "arXiv",
    primaryClass = "hep-ph",
    doi = "10.1007/JHEP08(2020)014",
    journal = "JHEP",
    volume = "08",
    pages = "014",
    year = "2020"
}

@article{Kovchegov:2019rrz,
    author = "Kovchegov, Yuri V.",
    title = "{Orbital Angular Momentum at Small $x$}",
    eprint = "1901.07453",
    archivePrefix = "arXiv",
    primaryClass = "hep-ph",
    reportNumber = "INT pre-print number INT-PUB-18-065",
    doi = "10.1007/JHEP03(2019)174",
    journal = "JHEP",
    volume = "03",
    pages = "174",
    year = "2019"
}

@article{Kovchegov:2018zeq,
    author = "Kovchegov, Yuri V. and Sievert, Matthew D.",
    title = "{Valence Quark Transversity at Small $x$}",
    eprint = "1808.10354",
    archivePrefix = "arXiv",
    primaryClass = "hep-ph",
    doi = "10.1103/PhysRevD.99.054033",
    journal = "Phys. Rev. D",
    volume = "99",
    number = "5",
    pages = "054033",
    year = "2019"
}

@article{Kovchegov:2018znm,
    author = "Kovchegov, Yuri V. and Sievert, Matthew D.",
    title = "{Small-$x$ Helicity Evolution: an Operator Treatment}",
    eprint = "1808.09010",
    archivePrefix = "arXiv",
    primaryClass = "hep-ph",
    doi = "10.1103/PhysRevD.99.054032",
    journal = "Phys. Rev. D",
    volume = "99",
    number = "5",
    pages = "054032",
    year = "2019"
}

@article{Kovchegov:2025gcg,
    author = "Kovchegov, Yuri V. and Li, Ming",
    title = {{Weizs{\"a}cker-Williams gluon helicity distribution and inclusive dijet production in longitudinally polarized electron-proton collisions}},
    eprint = "2504.12979",
    archivePrefix = "arXiv",
    primaryClass = "hep-ph",
    doi = "10.1007/JHEP08(2025)206",
    journal = "JHEP",
    volume = "08",
    pages = "206",
    year = "2025"
}

@article{Borden:2024bxa,
    author = "Borden, Jeremy and Kovchegov, Yuri V. and Li, Ming",
    title = "{Helicity evolution at small x: quark to gluon and gluon to quark transition operators}",
    eprint = "2406.11647",
    archivePrefix = "arXiv",
    primaryClass = "hep-ph",
    doi = "10.1007/JHEP09(2024)037",
    journal = "JHEP",
    volume = "09",
    pages = "037",
    year = "2024"
}

@article{Adamiak:2023okq,
    author = "Adamiak, Daniel and Kovchegov, Yuri V. and Tawabutr, Yossathorn",
    title = "{Helicity evolution at small x: Revised asymptotic results at large Nc and Nf}",
    eprint = "2306.01651",
    archivePrefix = "arXiv",
    primaryClass = "hep-ph",
    reportNumber = "JLAB-THY-23-3922",
    doi = "10.1103/PhysRevD.108.054005",
    journal = "Phys. Rev. D",
    volume = "108",
    number = "5",
    pages = "054005",
    year = "2023"
}

@article{Altinoluk:2024tyx,
    author = "Altinoluk, Tolga and Beuf, Guillaume and Blanco, Etienne and Mulani, Swaleha",
    title = "{Quark TMDs from back-to-back dijet production at forward rapidities in pA collisions beyond eikonal accuracy in the CGC}",
    eprint = "2412.08485",
    archivePrefix = "arXiv",
    primaryClass = "hep-ph",
    doi = "10.1007/JHEP06(2025)097",
    journal = "JHEP",
    volume = "06",
    pages = "097",
    year = "2025"
}

@article{Altinoluk:2024zom,
    author = "Altinoluk, Tolga and Beuf, Guillaume and Czajka, Alina and Marquet, Cyrille",
    title = "{Back-to-back dijet production in DIS at next-to-eikonal accuracy and twist-3 gluon TMDs}",
    eprint = "2410.00612",
    archivePrefix = "arXiv",
    primaryClass = "hep-ph",
    doi = "10.1103/PhysRevD.111.014010",
    journal = "Phys. Rev. D",
    volume = "111",
    number = "1",
    pages = "014010",
    year = "2025"
}

@article{Agostini:2024xqs,
    author = "Agostini, Pedro and Altinoluk, Tolga and Armesto, N{\'e}stor",
    title = "{Next-to-eikonal corrections to dijet production in Deep Inelastic Scattering in the dilute limit of the Color Glass Condensate}",
    eprint = "2403.04603",
    archivePrefix = "arXiv",
    primaryClass = "hep-ph",
    doi = "10.1007/JHEP07(2024)137",
    journal = "JHEP",
    volume = "07",
    pages = "137",
    year = "2024"
}

@article{Fu:2023jqv,
    author = "Fu, Yu and Kang, Zhong-Bo and Salazar, Farid and Wang, Xin-Nian and Xing, Hongxi",
    title = "{Correspondence between Color Glass Condensate and High-Twist Formalism}",
    eprint = "2310.12847",
    archivePrefix = "arXiv",
    primaryClass = "hep-ph",
    doi = "10.1103/PhysRevLett.135.032301",
    journal = "Phys. Rev. Lett.",
    volume = "135",
    number = "3",
    pages = "032301",
    year = "2025"
}

@article{Fu:2024sba,
    author = "Fu, Yu and Kang, Zhong-Bo and Salazar, Farid and Wang, Xin-Nian and Xing, Hongxi",
    title = "{Color glass condensate meets high twist expansion}",
    eprint = "2406.01684",
    archivePrefix = "arXiv",
    primaryClass = "hep-ph",
    reportNumber = "INT-PUB-24-024",
    doi = "10.1103/ckhv-5213",
    journal = "Phys. Rev. D",
    volume = "112",
    number = "1",
    pages = "014029",
    year = "2025"
}

@article{Diehl:1998sm,
    author = "Diehl, Markus and Gousset, Thierry",
    title = "{Time ordering in off diagonal parton distributions}",
    eprint = "hep-ph/9801233",
    archivePrefix = "arXiv",
    reportNumber = "DAPNIA-SPHN-98-01, CPT-S593-1297",
    doi = "10.1016/S0370-2693(98)00439-0",
    journal = "Phys. Lett. B",
    volume = "428",
    pages = "359--370",
    year = "1998"
}
